\documentclass[12pt]{article}

\usepackage{amssymb}
\usepackage{amsmath}
\usepackage{amscd}
\usepackage{latexsym}
\usepackage{graphicx}

\usepackage{cite}

\newcommand{\be}{\begin{equation}}
\newcommand{\ee}{\end{equation}}

\newcommand{\dlt}{\delta}

\newcommand{\bt}{\beta}
\newcommand{\vp}{\varphi}
\newcommand{\ep}{\varepsilon}
\newcommand{\al}{\alpha}
\newcommand{\ra}{\rightarrow}
\newcommand{\sgm}{\sigma}

\newcommand{\gm}{\gamma}

\newcommand{\lbd}{\lambda}

\newcommand{\cH}{{\cal H}}

\newcommand{\rgl}{\rangle}
\newcommand{\lgl}{\langle}

\begin{document}

\begin{center}

{\Large{\bf Role of collective information\\ in networks of quantum operating agents} \\ [5mm]

V.I. Yukalov$^{1,2,*}$, E.P. Yukalova$^{1,3}$, and
D. Sornette$^{1,4}$} \\ [3mm]

{\it
$^1$Department of Management, Technology and Economics, \\
ETH Z\"urich, Swiss Federal Institute of Technology, \\
Z\"urich CH-8032, Switzerland \\ [3mm]

$^2$Bogolubov Laboratory of Theoretical Physics, \\
Joint Institute for Nuclear Research, Dubna 141980, Russia \\ [3mm]

$^3$Laboratory of Information Technologies, \\
Joint Institute for Nuclear Research, Dubna 141980, Russia \\ [3mm]

$^4$ Institute of Risk Analysis, Prediction and Management (Risks-X),\\
Academy for Advanced Interdisciplinary Studies,\\ Southern University of Science 
and Technology (SUSTech), Shenzhen, 518055, China} \\ [3mm]

$^*${\it Corresponding author e-mail}: yukalov@theor.jinr.ru \\ [3mm]

\end{center}

\vskip 2cm

\begin{abstract}
A network of agents is considered whose decision processes are described by the quantum 
decision theory previously advanced by the authors. Decision making is done by evaluating 
the utility of alternatives, their attractiveness, and the available information, whose 
combinations form the probabilities to choose a given alternative. As a result of the 
interplay between these three contributions, the process of choice between several 
alternatives is multimodal. The agents interact by exchanging information, which can take 
two forms: information that an agent can directly receive from another agent and 
information collectively created by the members of the society. The information field 
common to all agents tends to smooth out sharp variations in the temporal behaviour of 
the probabilities and can even remove them. For agents with short-term memory, the 
probabilities often tend to their limiting values through strong oscillations and, for 
a range of parameters, these oscillations last for ever, representing an ever lasting 
hesitation of decision makers. Switching on the information field makes the amplitude 
of the oscillations smaller and even halt the oscillations forcing the probabilities 
to converge to fixed limits. The dynamic disjunction effect is described.  
\end{abstract}

\vskip 1cm

{\parindent=0pt


}

\newpage

\section{Introduction}

In recent years, great interest has been paid to the study of quantum 
information processing \cite{Williams_1,Nielsen_2,Vedral_3,Keyl_4,Guhne_5,Wilde_6}. 
A special attention is directed towards the investigation of quantum networks 
\cite{Albert_7,Meter_8}. The latter are usually based on the collections of 
elementary interacting quantum objects, such as photons, cold atoms, or spins.
These quantum objects form quantum registers that can be used for quantum computing
\cite{Bernien_33,Zhang_34,Wu_71,Zhong_72}. Promising candidates for the creation of 
quantum registers are atoms possessing electric or magnetic dipoles
\cite{Griesmaier_35,Baranov_36,Baranov_37,Boudjemaa_38,Yukalov_39}. 

In the present paper, we consider another kind of a network, a network composed 
not of elementary quantum objects, but consisting of intelligent agents operating
according to quantum rules. Such networks can model information processing in real 
human society or in artificial intelligence. The use of quantum techniques allows us 
to take account of rational-irrational duality in decision making. In order that 
the reader could better understand the meaning of the used terminology, it is necessary 
to introduce some explanations.  

Artificial intelligence, enhanced by quantum computing, is usually named quantum 
artificial intelligence \cite{Miakisz_40,Ying_41}. In that sense, the latter is 
understood as an ensemble of elementary quantum objects, e.g. photons, atoms, spins, 
or dipoles, that are organized in such a way that their functioning accomplishes 
intelligent computational actions. In other words, the considered system consists 
of elementary quantum objects, each of which does not possess intelligence and 
realizes simple quantum operations, while due to the wise organizational architecture 
of these operations the system as a whole accomplishes intelligent actions, for 
instance quantum computation, quantum machine learning, or similar achievements 
\cite{Wittek_42,Bhattacharyya_43,Schuld_44,Gangulu_45}. In our case, we consider
a more complicated situation, where the intelligence itself is composed of intelligent 
agents whose information processing employs quantum rules.   
 
Employing quantum rules does not mean to be quantum. Although some researchers assume 
that the brain's neurons act as miniature quantum devices, thus that the brain functions 
similarly to a quantum computer \cite{Penrose_44,Hameroff_45}, however others accept 
a different interpretation with regard to the functioning of the brain, whether human 
or animal. One does not assume that the brain is composed of quantum devices, but the 
overall functioning of such a complex system as the brain can be modeled in the frame 
of quantum techniques. Then one says that consciousness is not based on quantum processes, 
but just it proceeds in a quantum-like manner 
\cite{Khrennikov_46,Khrennikov_47,Yukalov_12,Yukalov_48}. We keep this point of view 
that the brain is not necessarily a quantum system, but its functioning can be 
conveniently described by the mathematics of quantum theory. The use of quantum 
techniques makes it possible to characterize the behavior of real agents, taking 
account of their rational and irrational sides of decision making.
 
Throughout the paper, we use the standard terminology generally accepted by the scientific 
community. In order to be precise, we formulate below the basic definitions that one 
has to keep in mind in order to avoid confusion on the meaning of the problem under 
consideration. 

{\it Artificial Intelligence} is intelligence demonstrated by machines, as opposed 
to natural intelligence displayed by animals including humans. Leading Artificial 
Intelligence textbooks define the field as the study of artificial intelligent 
systems that are understood as systems perceiving their environment and taking 
decisions and actions that define their chance for their goal attainment 
\cite{Nilsson_49,Poole_50,Luger_51,Rich_52,Russell_53}.

{\it An Intelligent Agent} is a system that, evaluating the available information, 
is able to take autonomous actions and decisions directed to the achievement of the 
desired goals and may improve its performance with learning or using obtained
knowledge \cite{Nilsson_49,Poole_50,Luger_51,Rich_52,Russell_53}. Often, the term
intelligent agent is applied to a system that possesses artificial intelligence.
However the intelligent agent paradigm is closely related to and employed with respect 
to agents in economics, in cognitive science, ethics, philosophy, as well as in 
many interdisciplinary socio-cognitive modeling and simulations. Generally, from the 
technical or mathematical point of view, the notion of intelligent agent can be 
associated with either real or artificial intelligence. An intelligent agent could 
be anything that makes decisions, as a person, firm, machine, or software.
 
{\it Quantum Artificial Intelligence} is artificial intelligence employing quantum 
computing for its functioning. This includes quantum computation itself and the related 
techniques, such as quantum machine learning 
\cite{Wittek_42,Bhattacharyya_43,Schuld_44,Gangulu_45,Wichert_54}. Quantum artificial 
intelligence can also use Hybrid Quantum/Classical Algorithms, when a quantum state 
preparation and measurement are combined with classical optimization \cite{Liang_55}. 

{\it Quantum Intelligence} is a field that focuses on building quantum algorithms 
for improving computational tasks within artificial intelligence, including the
related sub-fields, such as machine learning, data mining etc. 
\cite{Wittek_42,Bhattacharyya_43,Schuld_44,Gangulu_45,Wichert_54,Liang_55}.

{\it A Quantum Intelligent Agent} is a quantum device that interacts with the 
surrounding through sensors and actuators, contains a learning algorithm that 
correlates the sensor and actuator results by learning features \cite{Kewming_56}. 
A sensor is a physical device that the agent can use to read in information about 
the world. An actuator is a physical device that the agent can use to write 
information out into the world. In other words, one says that a quantum intelligent 
agent is a system possessing quantum artificial intelligence. Since the latter assumes 
the use of quantum computation, a quantum intelligent agent is also called a quantum 
computing agent \cite{Wittek_42,Bhattacharyya_43,Schuld_44,Gangulu_45,Wichert_54,Liang_55}.

{\it A Quantum Operating Agent} is a system whose operation can be described by employing
the mathematical techniques of quantum theory, without assuming that this system 
is composed of or includes quantum parts. A quantum operating agent is principally 
different from quantum agent, as it is not reliant on the hypothesis that there is 
something quantum mechanical in the agent structure, but it is merely the operation 
of the agent that can formally be described in terms of quantum techniques
\cite{Yukalov_12,Yukalov_48}. Sometimes, one calls this kind of agents ``quantum-like''
\cite{Khrennikov_46,Khrennikov_47}. 

The present paper studies the behavior of a system of quantum operating agents 
solving the standard problem of decision making by choosing between several 
alternatives. These agents can be called decision makers. The functioning of 
quantum operating agents can model both quantum intelligent agents equipped with 
quantum artificial intelligence as well as real humans, provided their goal is the 
choice between several alternatives.    
    
The process of taking decisions evolves in time, which requires to consider the 
related dynamics. This dynamics for a single decision maker is based on quantum 
evolution equations for a closed system \cite{Khrennikov_46,Khrennikov_47,Busemeyer_49}. 
To take into account the surrounding environment, one has to deal with evolution 
equations for open systems \cite{Asano_50,Asano_51,Bagarello_52}. 

In the present paper, we consider a situation that is principally different from the
cases discussed above. We are interested not in the process of decision making of 
a single agent, but in the behavior of a society composed of intelligent decision 
makers interacting with each other through information exchange \cite{Yukalov_9}. 
Thus the society as a whole is a closed system, while the agents interact with 
each other, so that the surroundings of an agent is formed by other agents.   

Each decision maker takes decisions by employing quantum rules. The goal of a decision 
maker is to perform a choice among the set of given alternatives. The process of 
making decisions is described in a realistic way taking into account the duality of 
decision processes related to rational reasoning as well as to irrational feelings. 
The choice is not deterministic as in classical utility theory \cite{Neumann_10}, 
but probabilistic. In the choice process, behavioural effects are included and the 
influence of available information is taken into account. Thus the choice between 
alternatives is multimodal, considering the evaluation of the alternative utility, 
the alternative attractiveness, and the information associated with the attitude of 
the society to different alternatives. In that way, decision making takes account 
of rational and irrational features associated with the considered alternatives. 
The rational part of decision making allows for the evaluation of the given 
alternatives following explicitly prescribed rules, for example the Luce rule 
\cite{Luce_11}. The irrational part, although being random, can nevertheless be 
evaluated by means of noninformative priors. Technically, it has been shown 
\cite{Yukalov_12,Yukalov_13,Yukalov_14,Yukalov_15,Yukalov_16} that the 
rational-irrational duality of decision making can be effectively characterized 
by resorting to the mathematical tools of quantum theory, particularly, to the 
language of quantum measurement theory. 

The ability of each member of the network to make decisions is what distinguishes our
approach from the models representing social systems as laser-like physical objects
composed of finite-level atoms \cite{Khrennikov_83,Khrennikov_84,Tsarev_85}.
   
In that way, three major points principally distinguish the content of the present 
paper from the previously studied cases: (i) We consider not a single decision maker, 
but a society of intelligent quantum-operating agents. Thus we study the behavior 
of a network of agents, each of which takes decisions following the rules of quantum 
decision theory \cite{Yukalov_12,Yukalov_13,Yukalov_14,Yukalov_15,Yukalov_16}. 
(ii) This network of intelligent agents is different from the networks of simple 
quantum objects, like photons, spins or dipoles accomplishing some actions, such as 
quantum computation, which can be treated as intelligent only for the system as a 
whole. (iii) The network of intelligent agents employing quantum decision theory is 
different from the networks representing classical multi-agent societies 
\cite{Jackson_17,Perc_18,Perc_19}.

In our previous paper \cite{Yukalov_9}, only the direct exchange of information 
between each pair of agents has been considered. However, in any real society, 
there always exists an information background that is common for all members 
of the society. This information background is formed by common social and cultural 
biases, memory of past events, general habits, and so on. This background effectively 
acts on all agents of the society and it can be manipulated by mass media, be it 
government controlled, in private hands or more delocalised (but still under some 
supervising control) via various social media channels. The information background, 
while being formed by the society itself, at the same time certainly can strongly 
influence the choices made by the members of the society. In the present paper, we 
suggest a model taking fully into account the two sides of the information processing 
by an intelligence network. First, there exists a direct exchange of information between 
pairs of agents. And, second, the society forms a common information field acting on 
all society members. We show that regulating the intensity of the common information 
field makes it possible to cover a large variety of behaviours exhibited by 
the society. 

The presentation is organised as follows.  Section 2 recalls the operation of a single 
intelligent agent, adapting the formulation of quantum decision theory to include the 
information dimension. Section 3 describes how to account for possible entanglements 
between the presented alternatives, the feelings of the decision maker and the available 
information. Section 4 formulates the decision maker process of a society of interacting 
intelligent agents, assuming a single-shot decision making accomplished once and in a 
short time. Section 5 goes further by considering that a dynamical decision making 
corresponds to a possibly lengthly temporal process occurring over a finite time 
allowing for repeated interactions. Section 6 applies the above formalism to the choice 
between two competing alternatives. Section 7 considers the case of a society populated 
by two types of agents whose difference is in the initial estimation of utility and 
attraction of the considered alternatives. Section 8 (respectively 9) studies the case 
of a society with long-term (resp. short-term) memory. Section 10 applies the above 
results to a description of the dynamic disjunction effect, which violates the sure-thing 
principle. Section 11 concludes.

\section{Single intelligent agent}     
   
Before considering a network of quantum operating agents, it is necessary to describe 
the operation of a single intelligent agent. For this, we build on the quantum decision 
theory \cite{Yukalov_12,Yukalov_13,Yukalov_14,Yukalov_15,Yukalov_16} and adapt it to 
add the information dimension.

Suppose the goal of a decision maker is to select an alternative from the set of $N_A$ 
alternatives $A_n$ enumerated by the index $n=1,2,\ldots,N_A$. An alternative is 
characterized by a vector $|A_n\rangle$ in a Hilbert space $\cH_A$. The latter is defined 
as the closed linear envelope over the basis formed by the vectors of alternatives,
\be
\label{1}
\cH_A = {\rm span}_n \{ \; | \; A_n \; \rgl \; \} \; .
\ee
This is the {\it space of alternatives}. The basis, as usual, is assumed to 
be orthonormalized. 

An intelligent subject possesses a collection of feelings portraying him/her 
as an individual, with the associated {\it subject space of mind} that can be 
defined as a closed linear envelope over the vectors of elementary feelings,
\be
\label{2}
\cH_S = {\rm span}_\al \{ \; | \; \al \; \rgl \; \} \;   .
\ee
Irrational characteristics, such as the attractiveness of an $n$-th alternative, 
are generally composed of elementary feelings and are represented as composite 
vectors
\be
\label{3}
 | \; z_n \; \rgl = \sum_\al \; b_{n \al}\; | \; \al \; \rgl \; ,
\ee
describing superpositions of elementary feelings. Contrary to the basis, formed 
by the vectors $|\alpha\rangle$, that is orthonormalized, the vectors $|z_n\rangle$ 
are not necessarily mutually orthogonal and normalized. The appropriate normalization 
conditions will be imposed later on.  
     
The {\it information space} is a Hilbert space
\be
\label{4}
\cH_I = {\rm span}_\sgm \{ \; | \; \sgm \; \rgl \; \}
\ee
that is a closed linear envelope of the orthonormalized vectors $|\sgm\rangle$ 
representing elementary pieces of information. A vector, associated with the 
information about an $n$-th alternative, generally, is a composite vector
\be
\label{5}
| \; \vp_n \; \rgl = \sum_\sgm \; c_{n\sgm}\; | \; \sgm \; \rgl
\ee
being a superposition of several elementary pieces of information. Vectors (\ref{5}) 
do not need to be necessarily orthonormalized. 

The total space, where decisions are made, is the {\it decision space}
\be
\label{6}
 \cH = \cH_A \bigotimes \cH_S \bigotimes \cH_I \;  .
\ee
A decision with respect to an alternative $A_n$ involves the associated feelings 
$z_n$ and uses the available information $\varphi_n$, so that actually the choice 
is multimodal and is characterized by a {\it prospect}
\be
\label{7}
 \pi_n =  A_n \bigcap z_n \bigcap \vp_n \; .
\ee
This prospect corresponds to the vector
\be
\label{8}
|\; \pi_n \; \rgl = | \; A_n z_n \vp_n \; \rgl =  
| \; A_n \; \rgl \; \bigotimes \; | \; z_n \; \rgl \; \bigotimes \;
| \; \vp_n \; \rgl
\ee
in the decision space (\ref{6}). Expression (\ref{8}) corresponds to constructing 
$|\; \pi_n \; \rgl$ by taking the tensorial product of a vector among possible 
alternatives, by a vector representing the attractiveness or feelings associated with 
this alternative, and by a vector of the information associated with this alternative.

In quantum theory, observable quantities are represented by self-adjoint 
operators \cite{Neumann_20}. In quantum decision theory 
\cite{Yukalov_12,Yukalov_15,Yukalov_16,Yukalov_21,Yukalov_22,Yukalov_23,Yukalov_24}, 
as well as in the theory of quantum measurements, the role of operators of 
observables is played by the prospect operators
\be
\label{9}
 \hat P(\pi_n) \equiv | \; \pi_n \; \rgl  \lgl \; \pi_n \; |
\ee
that can be written as the product of three operators,
\be
\label{10}
\hat P(\pi_n) =  \hat P(A_n) \bigotimes \hat P(z_n) \bigotimes \hat P(\vp_n) \; .
\ee
Here, the operator 
\be
\label{11}
\hat P(A_n) = | \; A_n \; \rgl  \lgl \; A_n \; |
\ee
is a projector, while the operators  
\be
\label{12}
\hat P(z_n) = | \; z_n \; \rgl  \lgl \; z_n \; | \; , \qquad
\hat P(\vp_n) = | \; \vp_n \; \rgl  \lgl \; \vp_n \; |
\ee
are not necessarily projectors, since the vectors (\ref{3}) and (\ref{5}), generally,  
are not orthonormalized. 

The state of a decision maker is associated with a statistical operator $\hat\rho$ 
that is a semi-positive trace-one operator acting on $\cH$. The pair $\{\cH,\hat\rho\}$ 
is a {\it decision ensemble}. The probability of selecting a prospect $\pi_n$ is 
given by the average
\be
\label{13}
 p(\pi_n) = {\rm Tr}_\cH \; \hat\rho \; \hat P(\pi_n) \;  .
\ee
The probability is normalized, so that
\be
\label{14}
 \sum_{n=1}^{N_A} \; p(\pi_n) = 1 \; , \qquad 0 \leq \; p(\pi_n) \; \leq 1 \; .
\ee
According to the frequentist statistical interpretation, probability (\ref{13}) 
describes either the chance of choosing a prospect $\pi_n$ or the fraction of 
times when the prospect $\pi_n$ has been chosen. When comparing theoretical 
probabilities with empirical data related to a pool of homogeneous agents, probability 
(\ref{13}) should be compared with the fraction of subjects choosing a prospect 
$\pi_n$. When more specific characteristics are available on individual decision makers,
the theory can be calibrated via maximum likelihood methods, for instance following
\cite{Favreetal16,Kovaetal16,Ferroetal2020}.

Substituting into probability (\ref{13}) the explicit expressions for the vectors 
entering the prospect operator (\ref{10}), it is straightforward to represent the 
probability as a sum of three terms,
\be
\label{15}
p(\pi_n) = f(\pi_n) + q(\pi_n) + h(\pi_n)   
\ee
normalized in the following way.  By definition, the first term is semi-positive 
and normalized as
\be
\label{16}
 \sum_{n=1}^{N_A} \; f(\pi_n) = 1 \; , \qquad 
0 \; \leq f(\pi_n) \; \leq 1 \; .
\ee
This term corresponds to the classical probability of choosing a prospect $\pi_n$, 
being based on rational rules of evaluating the utility of the alternative $A_n$. 
Hence the term $f(\pi_n)$ can be called the {\it utility factor}. 

The second term, defined so that
\be
\label{17}
 \sum_{n=1}^{N_A} \; q(\pi_n) = 0 \; , \qquad 
-1 \leq \; q(\pi_n) \; \leq 1 \; ,
\ee
characterizes the hesitation of the agent being subject to irrational feelings 
and balancing the attractiveness of the alternatives. Thence the term (\ref{17}) 
can be named the {\it attraction factor}. 

The third term, satisfying the conditions
\be
\label{18}
\sum_{n=1}^{N_A} \; h(\pi_n) = 0 \; , \qquad 
-1 \leq \; h(\pi_n) \; \leq 1 \; ,
\ee 
accounts for the impact of the available information on the decision, and it is thus
called the {\it information factor}. 

Expression (\ref{15}) generalises the previous general expression 
$p(\pi_n) = f(\pi_n) + q(\pi_n)$ derived in 
\cite{Yukalov_12,Yukalov_15,Yukalov_16,Yukalov_21,Yukalov_22,Yukalov_23,Yukalov_24}
to decision making in the presence of information coming from various sources, which 
contributes an additional influence term $h(\pi_n)$ to the decision making process.

\section{Alternatives-feelings-information entanglement}

As explained in the previous section, the decision space (\ref{6}) is formed 
by the direct product of three spaces, the space of alternatives $\mathcal{H}_A$, the
subject space $\mathcal{H}_S$, and the information space $\mathcal{H}_I$. In the 
process of defining the prospect probabilities, the three modes corresponding to 
alternatives, feelings, and information, become entangled. The entanglement is 
produced by the decision-maker state $\hat{\rho}$. The general form of the latter 
can be represented as
\be
\label{E1}
\hat\rho = \sum_{mn} \; \sum_{\al\bt} \; \sum_{\sgm\gm} \;
\rho_{mn}^{\al\bt}(\sgm\gm) \; |\; 
A_m \al\sgm \; \rgl \lgl \; \gm\bt A_n \; | \; .
\ee
This operator, acting on disentangled vectors of the decision space, generally,  
transforms them into entangled vectors. This is clearly seen by acting on the 
simplest disentangled vector that is a basis vector,
\be
\label{E2}
\hat\rho\; |\; A_n \al \sgm \; \rgl = \sum_{m\bt\gm} \; 
\rho_{mn}^{\bt\al} \; |\; A_m \bt\gm \; \rgl \; ,   
\ee
which results in an entangled vector.

It is necessary to distinguish entangled states $\hat{\rho}$ and the states $\hat{\rho}$ 
generating entanglement. Thus the decision maker state $\hat{\rho}$ can be separable, 
that is nonentangled, but at the same time generating entanglement 
\cite{Yukalov_68,Yukalov_69,Yukalov_70}. For instance, let us consider a separable 
decision maker state
\be
\label{E3}
 \hat\rho_{sep} = \sum_i \lbd_i \hat\rho_{Ai} \; \bigotimes \; 
 \hat\rho_{Si} \; \bigotimes \; \hat\rho_{Ii} \; ,
\ee
where
$$
  \sum_i \lbd_i  = 1 \; , \qquad 0 \leq \lbd_i \leq 1 \;  .
$$
This state comprises three factors, a state in the space of alternatives
$$
\hat\rho_{Ai} = \sum_{mn} \rho_{mni} \; | \; A_m \; \rgl \lgl \; A_n \; |\; ,
$$
a state in the subject space
$$
\hat\rho_{Si} = \sum_{\al\bt} \rho_{i}^{\al\bt} \; | \; \al \; \rgl \lgl \; \bt \; |\;   ,
$$
and a state in the information space
$$
\hat\rho_{Ii} = \sum_{\sgm\gm} 
\rho_{i}(\sgm\gm) \; | \; \sgm \; \rgl \lgl \; \gm \; |\;   .
$$
In this case of a separable state, the coefficients in the state (\ref{E1}) become
$$
\rho_{mn}^{\al\bt}(\sgm\gm) = 
\sum_i \lbd_i \;\rho_{mni} \;\rho_i^{\al\bt}\; \rho_i(\sgm\gm) \; .
$$
The action of the separable state (\ref{E3}) on a disentangled basis vector, 
\be
\label{E4}
 \hat\rho_{sep} \;| \; A_n \al \sgm \; \rgl  = \sum_i \; \sum_{m\bt\gm}
\lbd_i \; \rho_{mni} \; \rho_i^{\bt\al} \; \rho_i(\gm\sgm) \; | \; 
A_m \bt \gm \; \rgl \;  ,
\ee
produces an entangled vector, if at least two $\lambda_i$ are not zero. 

This shows that in the process of decision making all three modes corresponding to 
alternatives, feelings, and information, generally, become entangled and cannot be 
separated from each other.

\section{Society of intelligent agents}

The structure of the prospect probabilities can be straightforwardly generalized 
to the family of $N$ intelligent agents enumerated by the index $j=1,2,\ldots,N$. 
The decision space for the society of $N$ agents is
\be
\label{19}
 \cH = \bigotimes_{j=1}^N \cH_j \; ,
\ee
where 
\be
\label{20}
 \cH_j = \cH_A \bigotimes \cH_{jS} \bigotimes \cH_{jI}
\ee
is the decision space of a $j$-th agent, as derived in (\ref{6}).

The prospect of choosing an alternative $A_n$, in the presence of feelings
accompanying this choice, and of the related information, becomes
\be
\label{21}
 \pi_{nj}= A_n \bigcap z_{nj} \bigcap \vp_{nj} \; .
\ee
The corresponding prospect vector becomes
\be
\label{22}
 |\; \pi_{nj} \; \rgl =  |\; A_n \; \rgl \; \bigotimes \; |\; z_{nj} \; \rgl  
\; \bigotimes \; |\; \vp_{nj} \; \rgl \; .
\ee
And the prospect operator is
\be
\label{23}
 \hat P( \pi_{nj}) = |\; \pi_{nj} \; \rgl  \lgl \; \pi_{nj} \; | \; .
\ee

Each $j$-th agent is characterized by a state $\hat\rho_j$ acting on the decision 
space (\ref{20}). The probability that a $j$-th agent selects a prospect $\pi_{nj}$ 
reads as
\be
\label{24}
p( \pi_{nj}) = {\rm Tr}_{\cH_j} \;\hat\rho_j \; \hat P(\pi_{nj}) \; ,
\ee
with the properties
\be
\label{25}
 \sum_{n=1}^{N_A} \; p(\pi_{nj}) = 1 \; , \qquad 
0 \leq \; p(\pi_{nj}) \; \leq 1 \; .
\ee

Similarly to the previous section, the probability can be separated into three terms 
with the same meanings as earlier,
\be
\label{26}
p(\pi_{nj}) = f(\pi_{nj}) + q(\pi_{nj}) + h(\pi_{nj}) \; ,
\ee
and with the same normalization conditions for each agent,
$$
\sum_{n=1}^{N_A} \; f(\pi_{nj}) = 1 \; , \qquad 
0 \leq \; f(\pi_{nj}) \; \leq 1 \; ,
$$
$$
\sum_{n=1}^{N_A} \; q(\pi_{nj}) = 0 \; , \qquad 
-1 \leq \; q(\pi_{nj}) \; \leq 1 \; ,
$$
\be
\label{27}
\sum_{n=1}^{N_A} \; h(\pi_{nj}) = 0 \; , \qquad 
-1 \leq \; h(\pi_{nj}) \; \leq 1 \; .
\ee
The ultimate question that is the most interesting for applications is: What is 
the probability $p(\pi_{nj})$ that a $j$-th agent selects a prospect $\pi_{nj}$?

\section{Dynamics of intelligent agents network}

In the above sections, a single-shot decision making is treated, assumed to be 
accomplished once and in a short time, practically immediately. However a realistic 
process of decision making is not static but requires some period of time, say of 
length $\tau$. In addition, the agents can repeatedly interact by exchanging 
information. The repeated process of decision making leads to the appearance of 
dynamics in the values of the probabilities. Then the dynamical decision making 
corresponds to the temporal process occurring at steps $t=\tau,2\tau,\ldots$, taking 
the time as discrete multiples of $\tau$. Assuming the formation of the utility factor 
$f(\pi_{nj},t)$, attraction factor $q(\pi_{nj},t)$, and information factor $h(\pi_{nj},t)$ 
at time $t$, we accept a causal decision time $\tau$ to realise the choice at $t+\tau$
with the probability thus denoted as
\be
\label{28}
 p(\pi_{nj},t+\tau) =  f(\pi_{nj},t) + q(\pi_{nj},t) + h(\pi_{nj},t) \; .
\ee
The utility of alternatives usually changes slowly, as compared to the typical 
time of information exchange. Really, utility can be understood as price. Prices 
do not change for years, or at least for months, while information exchange can 
happen several times a day. I that sense, the utility factor can be kept constant, 
\be
\label{29}
  f(\pi_{nj},t) =  f(\pi_{nj}) \;  .
\ee

The attraction factor essentially depends on the information $M_j(t)$ available 
to a $j$-th subject and can be modeled \cite{Yukalov_9} as follows:
\be
\label{30}
q(\pi_{nj},t) = q(\pi_{nj},0) \exp\{ - M_j(t) \} \; .
\ee
Note that this form can be derived \cite{Yukalov_25+3,Yukalov_26+3} considering 
nondestructive repeated measurements over a quantum system. The available 
information, received by a $j$-th agent at time $t$, comes from other members 
of the society due to the exchange interactions of intensity $J_{ij}(t,s)$ 
acting during the periods of time before the present time $t$,
\be
\label{31}
 M_j(t) = \sum_{s=0}^t \; \sum_{i=1}^N J_{ij}(t,s) \mu_{ji}(s)  \; .
\ee
Here the information gain received by a $j$-th agent from an $i$-th agent is 
given by the Kullback-Leibler \cite{Kullback_27+3,Kullback_28+3} relative information
\be
\label{32}
 \mu_{ji}(t) = \sum_{n=1}^{N_A} \; p(\pi_{nj},t) \; 
\ln\; \frac{p(\pi_{nj},t)}{p(\pi_{ni},t)} \; .
\ee
It is straightforward to check that $\mu_{ij}(t) \geq 0$. At the initial moment of 
time, before the process of decision making has started, no additional information 
has yet been transferred, which implies  
$$
M_j(0) \equiv 0.
$$ 

The information exchange can be classified into two end-member cases, namely 
short-range and long-range interactions. The short-range interactions can be taken for instance
to be just nearest-neighbour interactions. In the case 
of short-range interactions, the net properties strongly depend of the interaction 
range and the net topology. In contrast, {\it long-range interactions} having 
the form
\be
\label{33}
J_{ij}(t,s) = \frac{1}{N-1} \; J(t,s) \;  ,
\ee
do not depend on the net geometry.

We consider a modern society composed of intelligent agents, such as of humans, where 
the society members are not nailed to fixed locations in space, but can move anywhere 
they wish and are able to communicate at any distance through such modern means as phones, 
Skype, WhatsApp, Twitter, Telegram, and so on. Since the members are not nailed to some 
nodes of a geometric net, there is no net at al. For this case, interactions do not depend 
on the distance between the members exchanging information. Therefore the long-range 
interactions are not an approximation, but the sole possible form of information exchange 
in the modern society.

Then for an $j$-th agent, the available information can be represented 
as
\be
\label{34}
 M_j(t) = \sum_{s=0}^t \; \sum_{i=1}^N \frac{J(t,s)}{N-1} \; \mu_{ji}(s) \; .
\ee
This corresponds to a ``mean-field'' approximation according to which every agent 
can be connected to every other agent.

The available information also depends on the type of memory typical of the net 
agents. There are two limiting types of memory, long-term and short-term memory. 
The {\it long-term memory} does not disappear with time, so that
\be
\label{35}
 J(t,s) = J \; ,
\ee
where we recall that the long-range interactions are assumed. Then the available 
information takes the form   
\be
\label{36}
M_j(t) = \frac{J}{N-1} \sum_{s=0}^t \; \sum_{i=1}^N \mu_{ji}(s) \; .
\ee

In the extreme case of {\it short-term memory}, only the nearest past is remembered, 
that is
\be
\label{37}
  J(t,s) = J \dlt_{ts} \; .
\ee
As a result, the available information becomes
\be
\label{38}
M_j(t) = \frac{J}{N-1} \sum_{i=1}^N \mu_{ji}(t) \;  .
\ee
  
In addition to the direct information exchange between agents, there exists 
a common information background created by the whole society. This background is 
formed by the agent activity not directed towards particular individuals, but rather 
intended to the society in total. Examples include literature, 
paper and journal articles, public lectures, and any kind of general information. 
This general information background can be represented as the information field 
described by analogy with the balance of fitness in biological and social evolution 
equations \cite{Sandholm_29+3}. Explicitly, the information field in the net of $N>1$ 
agents is modeled by the expression
\be
\label{39}
h(\pi_{nj},t) = \ep_j \left\{ \; \frac{1}{N-1} \sum_{i(\neq j)}^N 
[\; f(\pi_{ni}) + q(\pi_{ni},t)\; ] - [\; f(\pi_{nj}) + q(\pi_{nj},t)\; ] 
\right\} \; .
\ee
This expression means that the influence of the whole of society on an individual $j$
is proportional to the difference between the average of the sum
of the utility and attraction factors over the $N-1$ other agents
and the sum of the utility and attraction factors of agent $j$.
The parameter $\varepsilon_j$ defines the intensity with which the common information 
field acts on the $j$-th agent. Agents for which $\ep_j=0$ are not subjected to 
the common information field. Because of normalization (\ref{27}), the value of $\ep_j$
has to be bounded from above by $1$. So everywhere below we keep in mind that
$$
 0 \leq \; \ep_j \; \leq 1 \; .
$$
           
Without loss of generality,  as already mentioned, time is measured in units of $\tau$. Then, 
substituting the information field (\ref{39}) into relation (\ref{28}), we come to 
the evolution equation
\be
\label{40}
p(\pi_{nj},t+1) = ( 1 - \ep_j) [\; f(\pi_{nj}) + q(\pi_{nj},t)\; ] +
\frac{\ep_j}{N-1} \sum_{i(\neq j)}^N  [\; f(\pi_{ni}) + q(\pi_{ni},t)\; ] \; .
\ee
At the initial time $t = 0$, expression (\ref{26}) serves as the initial condition
\be
\label{41}
p(\pi_{nj},0) = ( 1 - \ep_j) [\; f(\pi_{nj}) + q(\pi_{nj},0)\; ] +
\frac{\ep_j}{N-1} \sum_{i(\neq j)}^N  [\; f(\pi_{ni}) + q(\pi_{ni},0)\; ] \; .
\ee
Expression (\ref{40}) shows that $\varepsilon_j$ plays the role of the weight with which 
the average decision of a society influences the personal evaluation of a $j$-th 
individual.

\section{Choice between two competing alternatives}

It is instructive to study the often met situation of choosing between two 
alternatives, i.e., $N_A=2$. Then it is sufficient to consider only the quantities 
corresponding to one of the alternatives, while the quantities related to the second 
alternative can be obtained from normalization (\ref{27}). It is useful to simplify 
the notation for the probabilities
\be 
\label{42}
p(\pi_{1j},t) \equiv p_j(t)\;  , \qquad p(\pi_{2j},t) = 1 - p_j(t)\;  , 
\ee
for the utility factors
\be
\label{43}
f(\pi_{1j}) \equiv f_j \;  , \qquad f(\pi_{2j}) = 1 - f_j \;   ,
\ee
for the attraction factors
\be
\label{44}
q(\pi_{1j},t) \equiv q_j(t)\;  , \qquad q(\pi_{2j},t) =  - q_j(t)\; ,
\ee
and for the information fields
\be
\label{45}
h(\pi_{1j},t) \equiv h_j(t)\;  , \qquad h(\pi_{2j},t) = - h_j(t)\;    .
\ee
Then equation (\ref{28}) for the probability that a $j$-th agent selects the first 
alternative becomes
\be
\label{46}
 p_j(t+1) = f_j + q_j(t) + h_j(t)  \; .
\ee
The attraction factor (\ref{30}) is
\be
\label{47}
q_j(t) = q_j(0) \exp\{ - M_j(t) \}  \;  ,
\ee
where $M_j(t) = J  \sum_{s=0}^t \; \sum_{i=1}^N \mu_{ji}(s)$
for the long memory case and $M_j(t) = J  \mu_{ji}(t)$ for the short memory case, 
with the information gain (\ref{32}) reading
\be
\label{48}
\mu_{ij} = p_i(t) \; \ln \; \frac{p_i(t)}{p_j(t)} + 
[ \; 1 - p_i(t) \; ] \; \ln \; \frac{1-p_i(t)}{1-p_j(t)}  \; .
\ee
And the information field (\ref{39}) turns to
\be
\label{49}
 h_j(t) = \ep_j \left\{ \; \frac{1}{N-1} \sum_{i(\neq j)}^N 
[\; f_i + q_i(t) \; ] - [\; f_j + q_j(t) \; ] \right\} \;  .
\ee
Finally, for the evolution equation (\ref{40}), we obtain
\be
\label{50}
 p_j(t+1) = ( 1 - \ep_j) [\; f_j + q_j(t) \; ] +  
\frac{\ep_j}{N-1} \sum_{i(\neq j)}^N  [\; f_i + q_i(t) \; ] \;  .
\ee

\section{Two types of agents}

Another typical situation corresponds to the case where the society is separated 
into two groups, so that inside each group the agent opinions are close to each 
other, while they are essentially different for the representatives of different 
groups. In that case, the treatment of a group of similar agents is equivalent 
to the consideration of one superagent. While the initial opinions of 
superagents of different groups may be very dissimilar, they can evolve significantly
as agents exchange information with each other, and the superagents
representing different groups even can come to a consensus.      

For one of the groups that can be marked with index $1$, we have
$$
p_1(t+1) = f_1 + q_1(t) + h_1(t)  \; , \qquad 
q_1(t) = q_1(0) \exp\{ - M_1(t)\} \; ,
$$
where $M_1(t) = J  \sum_{s=0}^t  \mu_{12}(s)$
for the long memory case and $M_1(t) = J  \mu_{12}(t)$ for the short memory case. 
The information factor for group 1 reads
\be
\label{51}
 h_1(t) = \ep_1 [\; (\; f_2 + q_2(t)\; ) - (\; f_1 + q_1(t)\;) \; ] \;  .
\ee
Respectively, for the second group, we get
$$
p_2(t+1) = f_2 + q_2(t) + h_2(t)  \; , \qquad 
q_2(t) = q_2(0) \exp\{ - M_2(t)\} \; ,
$$
where $M_2(t) = J  \sum_{s=0}^t  \mu_{21}(s)$
for the long memory case and $M_2(t) = J  \mu_{21}(t)$ for the short memory case. 
The information factor for group 2 reads
\be
\label{52}
 h_2(t) = \ep_2 [\; (\; f_1 + q_1(t)\; ) - (\; f_2 + q_2(t)\;) \; ] \;   .
\ee
Then the evolution equations (\ref{50}) take the form
$$
p_1(t+1) = (1 - \ep_1 )  [\; f_1 + q_1(t)\; ] + 
\ep_1 [\; f_2 + q_2(t) \; ] \; ,
$$
\be
\label{53}
p_2(t+1) = (1 - \ep_2 )  [\; f_2 + q_2(t)\; ] + 
\ep_2 [\; f_1 + q_1(t) \; ] \; .
\ee

The values $f_1$, $f_2$, $q_1(0)$, and $q_2(0)$ serve as initial conditions for
the evolution equations. It is easy to show that, if the initial conditions satisfy 
the normalization conditions (\ref{25}) and (\ref{27}), then equations (\ref{53}) 
guarantee that these normalization conditions are valid for all times $t > 0$. 
From equations (\ref{53}), the following general property follows. 

\vskip 2mm

{\bf Proposition 1}. If $\ep_1+\ep_2=1$, then $p_1(t)=p_2(t)$ for all $t\geq 1$. 

\vskip 2mm
{\it Proof}. Subtracting the second of equations (\ref{53}) from the first 
equation gives
\be
\label{54}
 p_1(t+1) - p_2(t+1) = 
(1 - \ep_1 - \ep_2) [\; f_1 + q_1(t) - f_2 - q_2(t)\; ] \; .
\ee
Hence, when $\ep_1+\ep_2=1$, then 
$$
p_1(t+1) - p_2(t+1) = 0 \qquad ( \ep_1 + \ep_2 = 1 )\; ,
$$
which proves the proposition. $\square$

\vskip 2mm
The overall dynamics depends on the memory type of the agents of the society, which defines 
the available information (\ref{31}). In the case of long-term memory, the available 
information is given by equation (\ref{36}) and for short-term memory, it acquires 
the form (\ref{38}). In both the cases, without loss of generality, we can set 
$J=1$.

\section{Agents with long-term memory}

Let all agents have long-term memory, so that the information they receive at 
different steps of repeated decision making is accumulated,
\be
\label{55}
 M_j(t) = \sum_{s=1}^t \mu_{ji}(s) \; , \qquad M_j(0) \equiv 0 \; ,
\ee
where $i \neq j$ and the information gain $\mu_{ij}(t)$ at a time $t$ is given by the 
Kullback-Leibler form (\ref{48}).

\subsection{General properties}

{\bf Proposition 2}. For agents with long-term memory, if their probabilities 
$p_1(t_0)$ and $p_2(t_0)$ of choosing alternative 1 coincide at some time $t_0\in\mathbb{N}$, 
then they coincide for all later times $t\geq t_0$.

\vskip 2mm

{\it Proof}. Suppose there is a point of time $t_0 \in \mathbb{N}$, when 
$p_1(t_0)=p_2(t_0)$. Then, $\mu_{ij}(t_0)=0$ from expression (\ref{48}). 
From definition (\ref{55}), we have
$$
 M_j(t_0) = \sum_{s=1}^{t_0-1} \mu_{ji}(s) + \mu_{ji}(t_0) \; ,
$$
with $j\neq i$. This yields
$$
  M_j(t_0) =  M_j(t_0-1)\; .
$$
Then the attraction factors possess the property
$$
q_j(t_0) = q_j(0) \exp\{ - M_j(t_0-1)\} = q_j(t_0-1) \;   .
$$
For probabilities (\ref{53}), we find
$$
p_j(t_0+1) = 
( 1 - \ep_j) [\; f_j + q_j(t_0)\; ] + \ep_j [\; f_i + q_i(t_0)\; ] =
$$
$$
= ( 1 - \ep_j) [\; f_j + q_j(t_0-1)\; ] + \ep_j [\; f_i + q_i(t_0-1)\; ] = 
p_j(t_0) \;,
$$
where $j\neq i$. Therefore
$$
p_1(t_0+1) = p_1(t_0) = p_2(t_0) = p_2(t_0+1) \;  .
$$
Repeating the same chain of arguments for the times $t_0+k$, where $k$ is a positive 
integer, we obtain
$$
 p_1(t_0+k) = p_2(t_0+k) \qquad ( k = 1, 2, \ldots ) \; ,
$$
which proves the proposition. $\square$

\vskip 2mm
The classification of possible dynamical regimes depends of the relation between the 
initial probabilities
$$
p_1(0) = 
( 1 - \ep_1) [\; f_1 + q_1(t_0)\; ] + \ep_1 [\; f_2 + q_2(t_0)\; ] \; ,
$$
\be
\label{56}
p_2(0) = 
( 1 - \ep_2) [\; f_2 + q_2(t_0)\; ] + \ep_2 [\; f_1 + q_1(t_0)\; ] \; ,
\ee
and the values
\be
\label{57}
 p_1^* \equiv ( 1 - \ep_1)  f_1 + \ep_1 f_2 \qquad 
 p_2^* \equiv ( 1 - \ep_2)  f_2 + \ep_2 f_1 \;  .
\ee

By assumption, the agents from the two different groups are different in the sense 
that their initial conditions $p_1(0)$ and $p_2(0)$ differ from each other. For 
concreteness, let us accept that
\be
\label{58}
p_1(0) > p_2(0) \; .
\ee
We exclude the case $p_1(0)=p_2(0)$ since, by Proposition 2, this leads to $p_1(t)=p_2(t)$
for all $t>0$ ($t=1,2,\ldots$). 

From expressions (\ref{56}), it follows 
\be
\label{59}
 p_1(0) - p_2(0) = 
( 1 - \ep_1 - \ep_2) [\; f_1 + q_1(0) - f_2 - q_2(0) \; ] \; .
\ee
Therefore inequality (\ref{58}) is valid when either
\be
\label{60}
f_1 + q_1(0) > f_2 + q_2(0) \; , \qquad  0 \leq \ep_1 + \ep_2 < 1 \; ,
\ee
or when
\be
\label{61}
f_1 + q_1(0) < f_2 + q_2(0) \;  ,  \qquad  1 < \ep_1 + \ep_2 \leq 2 \; .
\ee
Notice that the case $\ep_1+\ep_2>1$ reduces to the case $\ep_1+\ep_2<1$ by 
replacing the initial conditions $f_1$ and $q_1(0)$ by $f_2$ and $q_2(0)$. 
Therefore in what follows, it is sufficient to consider only the case
\be
\label{62}
 \ep_1 + \ep_2 < 1 \;  .
\ee

The main aim of the paper is to study how the common information in the society 
influences the opinion dynamics.  
For this purpose, we consider different dynamic regimes of the agents in 
the absence of the common information field, when $\ep_j=0$, and then switch on the 
parameters $\ep_j$ to nonzero values, thus coming to equations (\ref{53}).     

When $\ep_j=0$, two situations can develop with respect to the initial 
conditions. One situation corresponds to {\it rational initial preference}, when 
the inequality for the initial probabilities $p_1(0)>p_2(0)$ is accompanied by the 
similar relation between the rational utility factors:  
\be
\label{63}
p_1(0) > p_2(0) \; , \qquad f_1 > f_2 \qquad ( \ep_j = 0 ) \;  .
\ee
Then each probability tends with time to the related utility factor,
\be
\label{64}
\lim_{t\ra\infty} p_j(t) = f_j \qquad (\ep_j = 0 ) \; .
\ee
Switching on the common information field, under $0<\ep_1+\ep_2<1$, results in 
the limiting probabilities
\be
\label{65}
\lim_{t\ra\infty} p_j(t) = p_j^* \;  ,
\ee
where $p_j^*$ are defined in equation (\ref{57}).

If $\ep_1+\ep_2$ becomes larger then one, the probabilities $p_1$ and $p_2$ 
interchange their places, so that $p_2$ turns larger then $p_1$.  

The other rather nontrivial situation corresponds to {\it irrational initial 
preferences}, when at the initial time the inequality between the probabilities 
is opposite to that between the rational utility factors:
\be
\label{66}
p_1(0) > p_2(0) \; , \qquad f_1 < f_2 \qquad ( \ep_j = 0 ) \; ,
\ee
which is caused by the presence of the attraction factors. Then, two different dynamical 
behaviours can occur. 

One possibility, for $\ep_j=0$, is when the probabilities tend to the {\it common 
consensual limit}
\be
\label{67}
 \lim_{t\ra\infty} p_j(t) = p^* \cong 
\frac{f_1q_2(0)-f_2q_1(0)}{q_2(0)-q_1(0)} ~~~~~  ( \forall ~j~) \; .
\ee
Interestingly, switching on the common information field, while this changes the temporal 
behaviour, the probabilities still converge to the same consensual limit (\ref{67}) 
independent of the parameters $\ep_j$. 

The other possibility is the occurrence of {\it dynamic preference reversal}, when 
at the initial moment of time and at the first step one has
\be
\label{68}
p_1(0) > p_2(0) \; , \qquad p_1(1) > p_2(1) \;  ,
\ee
but at the second step the relation between the preferences reverses:
\be
\label{69}
  p_1(2) < p_2(2) \;  .
\ee
After this reversal, the probabilities tend to the limiting values (\ref{57}). 
Switching on the common information field removes this preference reversal, so 
that both probabilities tend to the common consensus (\ref{67}). 

Below, we present the results of numerical calculations describing in detail the 
role of the information field on the dynamics of preferences. For short, we use 
the notation $q_j(0)\equiv q_j$.

\subsection{Monotonically diminishing probabilities}

The regime, where both probabilities monotonically decrease with time, happens for 
$\ep_1=\ep_2=0$ when 
\be
\label{70}
 q_1 > 0 \; , \qquad q_2 > 0 \qquad (\ep_j = 0 \; , ~~\forall~ j ) \;  .
\ee
Switching on the information field leaves this behaviour unchanged, since under condition 
(\ref{70}) and $\ep_j<1$ , the inequalities 
\be
\label{71}
(1 - \ep_1) q_1 + \ep_1 q_2 > 0 \; , \qquad (1 - \ep_2) q_2 + \ep_2 q_1 > 0
\ee
follow. This is illustrated in Fig.~1 for the rational as well as irrational initial 
preferences. The information field decreases the difference between the probabilities 
at finite times, but they reach the same limiting values $p_j(\infty)$ at long times.

\subsection{Monotonically increasing probabilities}

In the case $\ep_1=\ep_2=0$, monotonically increasing probabilities occur if
\be
\label{72}
 q_1 < 0 \; , \qquad q_2 < 0 \qquad (\ep_j = 0 \; , ~~\forall~ j ) \;   .
\ee
From this, the inequalities
\be
\label{73}
(1 - \ep_1) q_1 + \ep_1 q_2 < 0 \; , \qquad (1 - \ep_2) q_2 + \ep_2 q_1 < 0 \; ,
\ee
for $0 <\varepsilon_j< 1$, follow. Therefore, the qualitative behaviour of the 
probabilities remains the same, they monotonically grow with time, as is shown 
in Fig. 2 for the rational and irrational initial preferences.

\subsection{Probabilities diverging from each other}

In the absence of information field, the probabilities diverge form each other 
when $p_1(t)$ increases while $p_2(t)$ decreases, which happens when 
\be
\label{74}
q_1 < 0 \; , \qquad q_2 > 0 \qquad (\ep_j = 0 ) \;   .
\ee
Switching on the information field can essentially change the temporal behaviour 
of the probabilities. Thus choosing the parameters $\ep_j$ so that 
\be
\label{75}
- \; \frac{q_1}{q_2-q_1} < \ep_1 < 1 \; , \qquad
0 < \ep_2 < \frac{q_2}{q_2-q_1} 
\ee
makes both probabilities diminishing. And for the case 
\be
\label{76}
0 < \ep_1 < -\; \frac{q_1}{q_2-q_1} \; , \qquad 
\frac{q_2}{q_2-q_1} < \ep_2 < 1
\ee
both probabilities increase, as is demonstrated in Fig. 3. In all cases, the 
probabilities diverge from each other.

\subsection{Probabilities approaching each other}

The probabilities approach each other if $p_1(t)$ decreases, while $p_2(t)$ 
increases, which occurs when
\be
\label{77}
q_1 > 0 \; , \qquad q_2 < 0 \qquad (\ep_j = 0 ) \;    .
\ee
Choosing the parameters of the information field such that
\be
\label{78}
0 < \ep_1 < \frac{q_1}{q_1-q_2} \; , \qquad 
-\; \frac{q_2}{q_1-q_2} < \ep_2 < 1
\ee
makes both probabilities decreasing. And for the parameters in the range
\be
\label{79}
\frac{q_1}{q_1-q_2} < \ep_1 < 1 \; ,   \qquad
0 < \ep_2 < -\; \frac{q_2}{q_1-q_2}
\ee
both probabilities increase. The role of the information field is illustrated 
in Fig. 4 for the case of the rational initial preference and in Fig. 5 for the 
irrational initial preference. While the information field changes the temporal 
behaviour of the probabilities, they continue approaching each other.

\subsection{Dynamic preference reversal}

Under a special choice of parameters, the following interesting effect occurs. 
The parameters are taken such that $\ep_1=\ep_1=0$, with the initial conditions $f_1<f_2$, 
$q_1>0$, $q_2<0$, $f_1+q_1$ close to one, and $f_2+q_2\ll 1$. This corresponds to 
the case of the irrational initial preference, since
\be
\label{80}
p_1(0) \gg p_2(0) \; , \qquad f_1 < f_2 \;   .
\ee
At the first time step, the condition $p_1(1) > p_2(1)$ is still preserved. However, at 
the second step, the initial inequality reverses to $p_1(2) < p_2(2)$. After this, 
the probabilities tend to their limits 
\be
\label{81}
 \lim_{t\ra\infty} p_j(t) = f_j \qquad ( \ep_j = 0 ) \;  ,
\ee
with different $f_j$. However, switching on the information field again drastically 
changes the behaviour of the probabilities that now tend to the common consensual 
limit
\be
\label{82}
 \lim_{t\ra\infty} p_j(t) = p^* \qquad ( \ep_j > 0 ) 
\ee
defined in equation (\ref{67}). In that way, the existence of the common information 
in the society removes the effect of preference reversal, but leads to the consensual 
limit. This situation is shown in Fig. 6.

\section{Agents with short-term memory}

In the case of short-term memory, the available information reads
\be
\label{83}
M_j(t) = \mu_{ji}(t) \qquad ( i \neq j ) \;  .
\ee

\subsection{General properties}

{\bf Proposition 3}. If the initial probabilities coincide,
\be
\label{84}
 p_1(0) = p_2(0) \;  ,
\ee
then they coincide for all times:
\be
\label{85}
 p_1(t) = p_2(t) \qquad ( t = 1,2,3,\ldots ) \;  .
\ee

\vskip 2mm

{\it Proof}. Because of the initial condition (\ref{84}), we have 
$$
\mu_{12}(0) = \mu_{21}(0) = 0 \;  .
$$
Therefore
$$
p_1(1) = ( 1 - \ep_1) [\; f_1 + q_1(0)\exp\{-\mu_{12}(0) \} \; ] +
\ep_1 [\; f_2 + q_2(0)\exp\{-\mu_{21}(0) \} \; ]  =
$$
$$
= 
( 1 - \ep_1) [\; f_1 + q_1(0)\; ] + \ep_1 [\; f_2 + q_2(0)\; ] =
p_1(0) \;  .
$$
Similarly, we find
$$
p_2(1) = p_2(0)  \;  .
$$
Taking into account condition (\ref{84}) gives
$$
 p_1(1) = p_1(0) = p_2(0) = p_2(1) \; .
$$
From here
$$
\mu_{12}(1) = \mu_{21}(1) = 0 \;   .
$$
Using this, we obtain
$$
p_1(2) = ( 1 - \ep_1) [\; f_1 + q_1(0)\exp\{ -\mu_{12}(1) \} \; ] +
\ep_1 [\; f_2 + q_2(0)\exp\{ -\mu_{21}(1) \} \; ] =
$$
$$
= 
( 1 - \ep_1) [\; f_1 + q_1(0)\; ] + \ep_1 [\; f_2 + q_2(0)\; ] =
p_1(0) \;  ,
$$
which yields
$$
p_1(2) = p_1(0) = p_2(0) = p_2(2) \;   .
$$
Continuing this chain of arguments, we come to equality (\ref{85}). $\square$

\subsection{Monotonic dynamics of probabilities}

In the absence of the common information field, there are three main temporal 
regimes for the probabilities. One is the monotonic tendency of the probabilities 
to different limits. Switching on the information field does not change the 
monotonicity of the dynamics, but makes the probabilities closer to each other, 
as is clear from Fig. 7. The larger the parameters $\ep_j$, characterizing the 
intensity of the information field, the closer to each other the probabilities. 
It looks rather natural that the common information in a society reduces the 
difference between the agent opinions.

\subsection{Oscillatory dynamics of probabilities}

The convergence to the different limits $p_j(\infty)$ can be oscillatory in the 
absence of the common information field. Switching on the information field smoothes 
out the oscillations (but does not remove them completely), as is shown in Fig. 8. 
The stronger the influence of the common information, the smaller the oscillations. 
The oscillations can even disappear for sufficiently large parameters $\ep_j$. This 
effect of smoothing the opinion oscillations caused by the common information in a 
strongly correlated society seems quite natural.

\subsection{Everlasting probability oscillations}

When there is no common information field, the agents with a short-term memory can 
have their choice probabilities exhibiting a permanent oscillatory behaviour, without 
convergence to a fixed limit. This can be interpreted as agents who are 
hesitating without hope to converge to a clear decision.
However switching on the information field reduces the amplitude of 
such oscillations. For sufficiently large information fields,  
the probabilities do not oscillate anymore and they converge to well defined limits.
Again, such a smoothing effect of the common information in a society is 
understandable and very reasonable. These behaviours are shown in Fig. 9.

\section{Dynamic disjunction effect}

In the above sections, we have presented a complete classification of qualitatively 
different regimes of dynamic decision making for two groups of subjects choosing one 
of two given alternatives. In the real world, there exist various societies with rather 
different features. To our mind, the developed theory allows us to characterize 
practically any of the possible situations by fixing the appropriate society parameters. 
For concreteness, we study below the case where the so-called disjunction effect occurs, which
violates the sure-thing principle \cite{Tversky_53,Croson_54,Lambdin_55,Li_56}.

The disjunction effect was discovered by Tversky and Shafir \cite{Tversky_53} and confirmed 
in numerous studies (e.g. \cite{Croson_54,Lambdin_55,Li_56}). This effect is typical for 
two-step composite games of the following structure. At the first step, a group of subjects
takes part in a game, where each member with equal probability can either win (event $B_1$) 
or lose (event $B_2$). At the second step, the agents are invited to participate in a 
second game, having the right either to accept the second game (event $A_1$) or to refuse
it (event $A_2$). The second stage is realized in different variants: One can either accept 
or decline the second game under the condition of knowing the result of the first game. 
Or one can either accept or decline the second game without knowing the result of the first 
game. 

In their experimental studies, Tversky and Shafir \cite{Tversky_53} find that the fraction 
of people accepting the second game, under the condition that the first game was won, and  
the fraction of those accepting the second game, under the condition that the first game 
was lost, respectively, are
\be
\label{85a}
f(A_1|B_1) = 0.69 \; , \qquad  f(A_1|B_2) = 0.59 \;  .
\ee
Here and below we denote classical probabilities by the letter $f$ to distinguish them 
from quantum probabilities denoted by the letter $p$. From the normalization condition 
for the conditional probabilities, we have
\be
\label{86}
f(A_2|B_1) = 0.31 \; , \qquad  f(A_2|B_2) = 0.41 \;    .
\ee
Hence the related joint probabilities are
\be
\label{87}
f(A_1B_1) = 0.345 \; , \qquad  f(A_1B_2) = 0.295 \; , \qquad 
f(A_2B_1) = 0.155 \; , \qquad  f(A_2B_2) = 0.205 \; ,
\ee
using the fact that each member has equal probability to either winning (event $B_1$) 
or losing (event $B_2$), so that $f(B_1) = f(B_2)=0.5$.

The question of interest is: How many subjects will accept or decline the second game,
when the result of the first game is not known to them? That is, one needs to define
the behavioral probabilities of two alternative events
\be
\label{88}
 A_1 B = A_1 \bigcap \left( B_1 \bigcup B_2 \right) \; , \qquad 
A_2 B = A_2 \bigcap \left( B_1 \bigcup B_2 \right) \;  .
\ee
The classical probabilities for these events are
\be
\label{89}
 f(A_1B) = f(A_1B_1) + f(A_1B_2) = 0.64 \qquad
 f(A_2B) = f(A_2B_1) + f(A_2B_2) = 0.36 \; ,
\ee
which implies that, by classical theory, the probability of accepting the second 
game, not knowing the results of the first one, is larger than that of refusing the 
second game, not knowing the result of the first game. This, actually, is nothing 
but the sure-thing principle \cite{Savage_57} telling that if $f(A_1B_1)>f(A_2B_1)$ 
and $f(A_1B_2) > f(A_2B_2)$, then $f(A_1B) > f(A_2B)$. To great surprise, Tversky 
and Shafir in their experiments \cite{Tversky_53} observe that human decision makers 
behave opposite to the prescription of the sure-thing principle, with the majority 
refusing the second game, if the result of the first one is not known, 
\be
\label{90}
p_{exp}(A_1B) = 0.36 \; , \qquad p_{exp}(A_2B) = 0.64 \;  .
\ee
The magnitude of the disjunction effect is characterized by the difference between 
the fraction of agents choosing the alternative $A_nB$ and the corresponding utility 
factor,
$$
p_{exp}(A_1B) - f(A_1B) = 0.36 - 0.64 = - 0.28 \; ,
$$
\be
\label{91} 
p_{exp}(A_2B) - f(A_2B) = 0.64 - 0.36 =  0.28 \; .
\ee
This difference measures the magnitude of the attraction factor. 
 
The resolution of the disjunction effect in the frame of the quantum decision theory 
has been given in \cite{Yukalov_9,Yukalov_15}, taking into account that the behavioral 
probability, according to quantum rules, includes a utility factor and an attraction 
factor. The latter was estimated using a noninformative prior, which resulted in the 
magnitude of the attraction factor $0.25$, close to the empirical value of $0.28$.

The above discussion of the disjunction effect keeps in mind decision making performed
in a short interval of time and without information exchange between the agents. But when
the agents are provided sufficiently long time during which they are allowed to exchange 
information between them, then the disjunction effect attenuates, in the sense that the initial
difference 
\be
\label{92}
| \; p(A_nB) - f(A_nB) \; | = 0.28 \qquad ( n = 1,2 )
\ee
becomes smaller with time. The disjunction-effect attenuation has been proved by 
numerous empirical observations 
\cite{Kuhberger_58,Charness_59,Blinder_60,Cooper_61,Tsiporkova_62,Charness_63,
Charness_64,Chen_65,Liu_66,Charness_67}. 

In order to study the dynamics of the disjunction effect, we have to resort to the 
equations of the theory developed above. The setup of Tversky and Shafir 
\cite{Tversky_53} can be captured by the evolution equations for the probability $p_j(t)$ of 
choosing the alternative $A_1B$ by the $j$-th agent at time $t$, which take the form
$$
 p_1(t+1) = ( 1 - \ep_1) [ \; f + q_1(t) \; ] + \ep_1 [ \; f + q_2(t) \; ] \; ,
$$
\be
\label{93}
 p_2(t+1) = ( 1 - \ep_2) [ \; f + q_2(t) \; ] + \ep_2 [ \; f + q_1(t) \; ] \; ,
\ee
with the initial conditions
$$
 p_1(0) = ( 1 - \ep_1) ( f + q_1 ) + \ep_1 (f + q_2 )  \; ,
$$
\be
\label{94}
 p_2(0) = ( 1 - \ep_2) ( f + q_2 )  + \ep_2 ( f + q_1 ) \;  ,
\ee
where $q_j \equiv q_j(0)$ and $f = 0.64$. As is seen from the difference
\be
\label{95}
 p_1(0) - p_2(0) = ( 1 - \ep_1 - \ep_2) ( q_1 - q_2) \;  ,
\ee
switching on the information field, so that $0 \leq \ep_1 + \ep_2 \leq 1$,
makes the probabilities $p_1(t)$ and $p_2(t)$ closer to each other.

We solve equations (\ref{93}) in the long-term memory case, where
agents accumulate the information from the previous interactions with other 
agents. The numerical solution   
shows that both probabilities $p_1(t)$ and $p_2(t)$ always tend to the same consensual 
limit:
\be
\label{96}
\lim_{t\ra\infty} p_1(t) = \lim_{t\ra\infty} p_2(t) = f \;  ,
\ee
while the attraction factors decrease with time,
\be
\label{97}
 \lim_{t\ra\infty} q_1(t) = \lim_{t\ra\infty} q_2(t) = 0 \;  ,
\ee
thus diminishing the disjunction effect. This explains the disjunction effect 
attenuation observed in experiments 
\cite{Kuhberger_58,Charness_59,Blinder_60,Cooper_61,Tsiporkova_62,Charness_63,
Charness_64,Chen_65,Liu_66,Charness_67}. The information provided to decision makers 
reduces the difference between $p_1(t)$ and $p_2(t)$ at the very beginning of the
process. See Fig. 10.

\section{Conclusion}

We have considered a network of agents taking decisions on the basis of quantum 
rules. This implies that the choice between several alternatives is characterized 
by the probabilities calculated using the techniques of quantum theory, which makes 
the process of choice multimodal. According to the quantum rules, decision 
makers take into account different features associated with the alternatives. The agents evaluate 
the utility of alternatives, characterized by utility factors as well as their 
attractiveness, described by attraction factors. They interact with other agents 
of the network by exchanging information and also they are influenced by the common 
information collectively created by the society members. 
  
We consider two ways by which decision makers consume the available information, 
either accumulating the information from the previous interactions with other 
agents, or using the information only from one previous time step. The first case 
represents the presence of long-term memory and the second, short-term memory. 
Both these cases have been treated.

In order to understand the influence of the common information available in the 
society, we started from the situation where this common information is absent. Then 
we analyzed how the dynamics of taking decisions changes when the information field 
is switched on and its intensity is increased. 

We have documented that the information field can essentially influence the temporal 
dynamics of the probabilities, whose values can be interpreted as the fraction of agents 
preferring one or the other alternative. Different variants of this influence have been 
carefully studied by varying initial conditions and different intensities of the common 
information field. We have treated the case of long-range interactions between the members
of the society, keeping in mind that the members of the modern societies are able to 
exchange information at any distance through a number of means, such as phone, internet, 
mass media, etc. 

In real life, there can exist various situations, quite diverse types of societies, and 
varying initial conditions. This is why our aim has been to present a complete analysis 
of different possible dynamic regimes and to examine the influence of the common 
information field on these regimes. 
   
Summarizing the most important impact of the information field on the decision making 
dynamics in the network, we found that it smoothes out sharp variations in the temporal 
behaviour of the probabilities and can even remove them. For example, in the case of 
agents with long-term memory, we have documented the existence of a dynamic preference 
reversal where, starting from $p_1(0)>p_2(0)$, the dynamics gives at the second step 
an abrupt reversal to $p_1(2)<p_2(2)$, after which the probabilities tend to different 
limits $f_j$. Switching on the information field removes this reversal and makes the 
probabilities converging to a common consensual limit $p^*$. 

For agents with short-term memory, the probabilities often tend to their limiting 
values through strong oscillations and, for a range of parameters, these oscillations 
last for ever, representing an ever lasting hesitation of the decision makers.
However switching on the information field makes the amplitude of the oscillations 
smaller and even can halt the everlasting oscillations forcing the probabilities to 
converge to fixed limits.  

The theory explains the attenuation of the disjunction effect with time, as a 
consequence of the exchange of information between the agents. This is embodied by 
the reduction of the magnitude of the attraction factors with increasing time. 
Additional common information further reduces the disjunction effect from the very 
beginning of the process. 

Concluding, a society sharing a sufficiently large flow of common available information 
exhibits smoother dynamical processes of decision taking compared with a society without 
this common information. The common information removes sudden preference reversals in 
a society with long-term memory and can stop permanent decision oscillations in a society 
with short-term memory.

\newpage

\begin{figure}[ht]
\centerline{
\hbox{ \includegraphics[width=7.5cm]{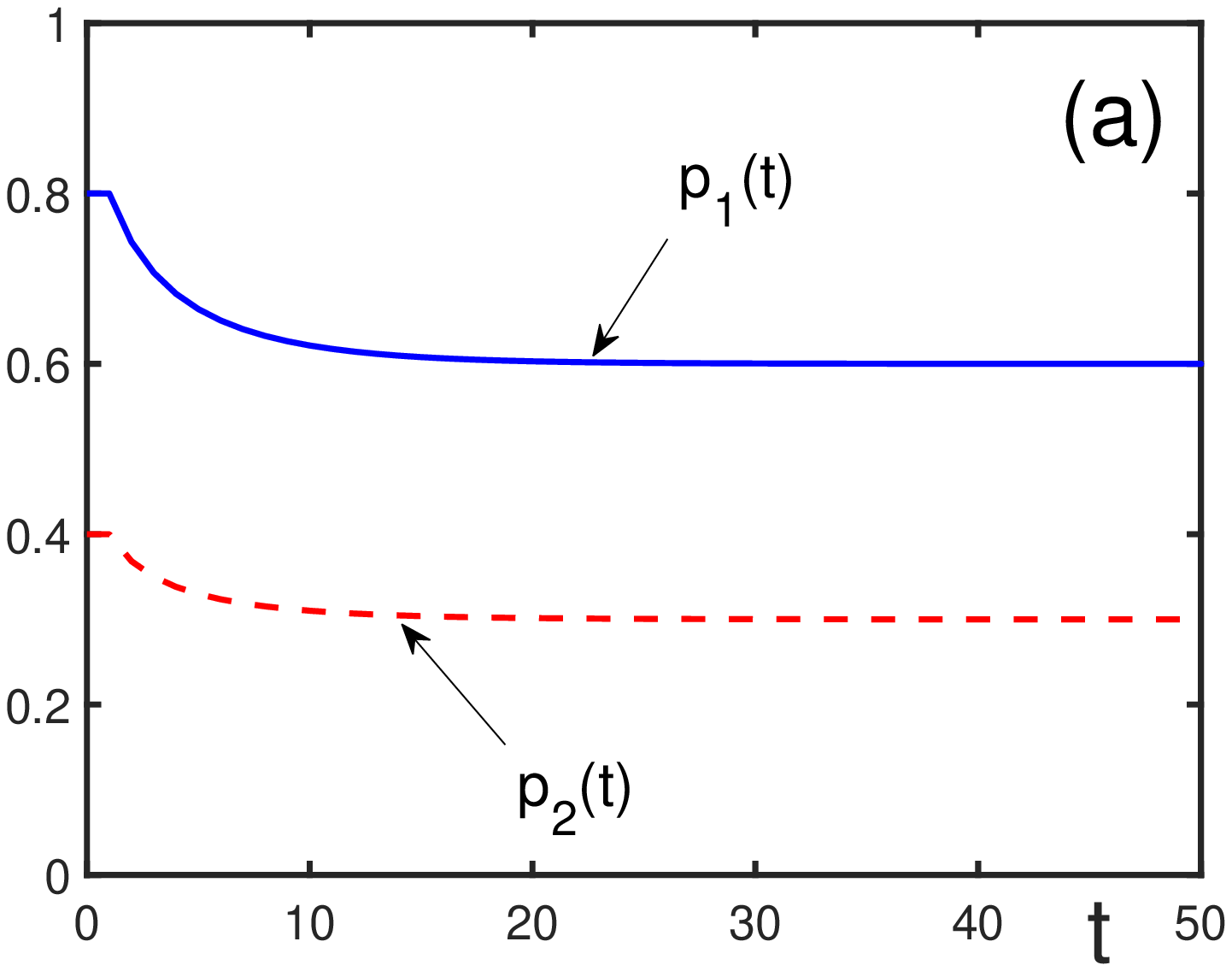} \hspace{1cm}
\includegraphics[width=7.5cm]{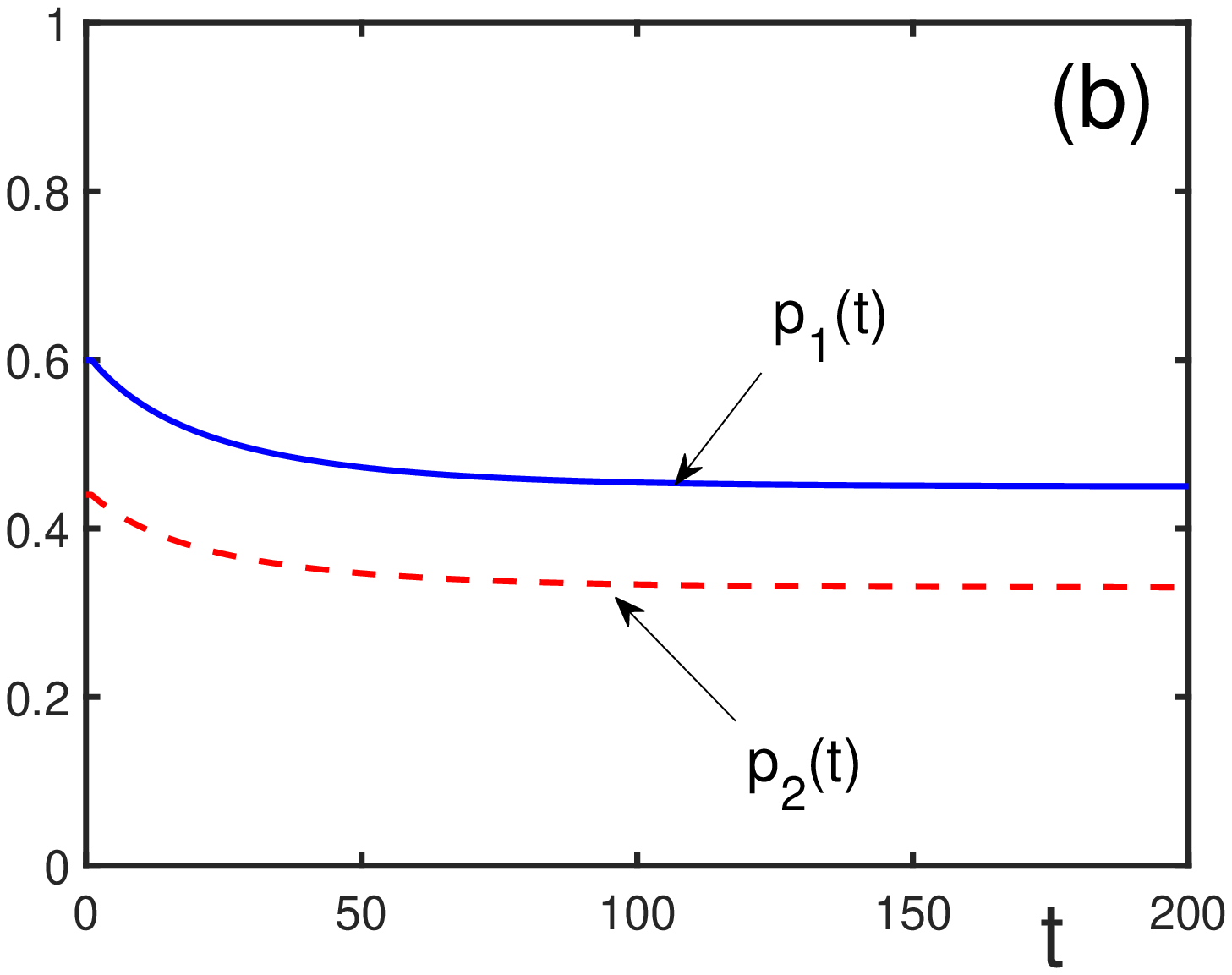}  } }
\vspace{12pt}
\centerline{
\hbox{ \includegraphics[width=7.5cm]{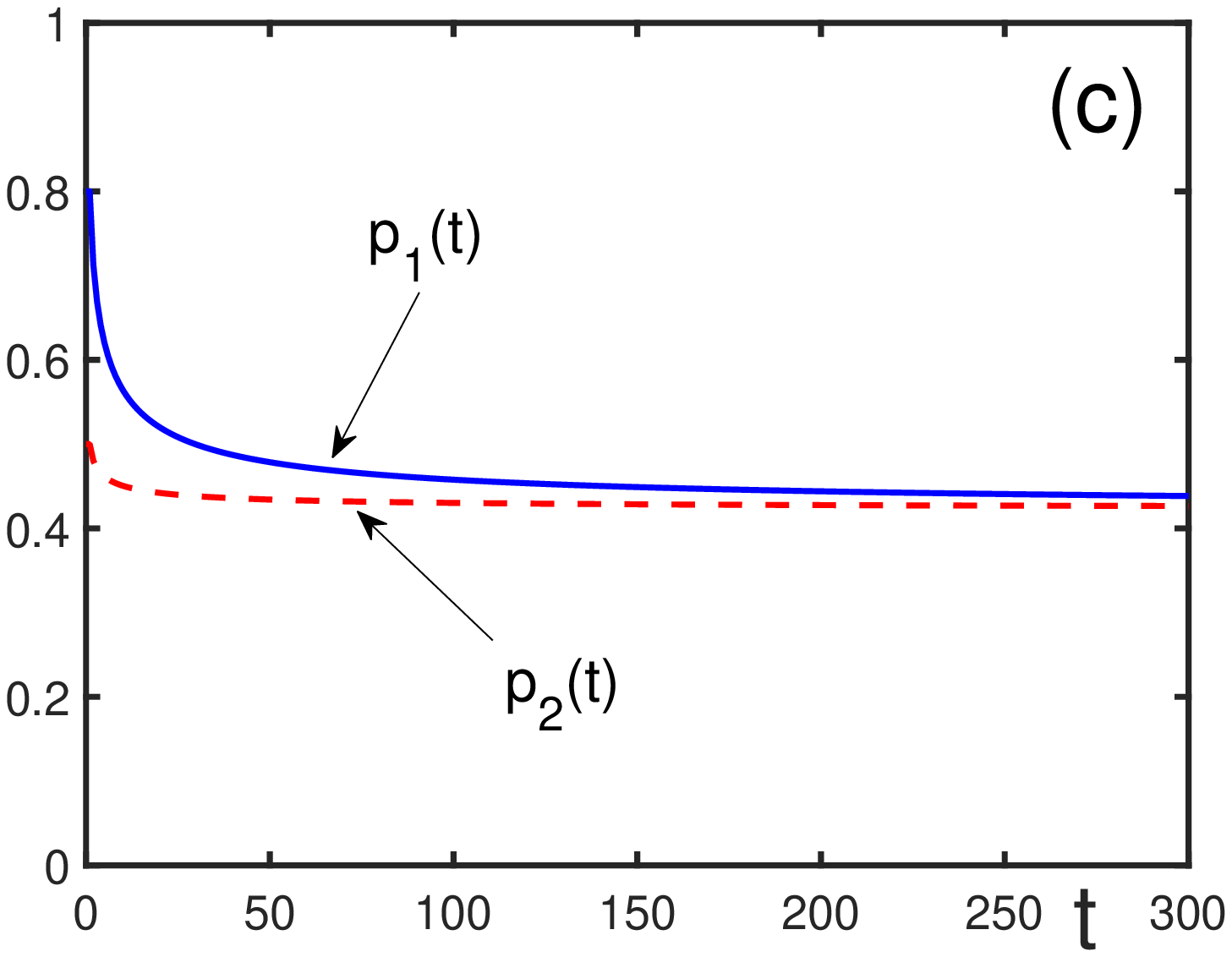} \hspace{1cm}
\includegraphics[width=7.5cm]{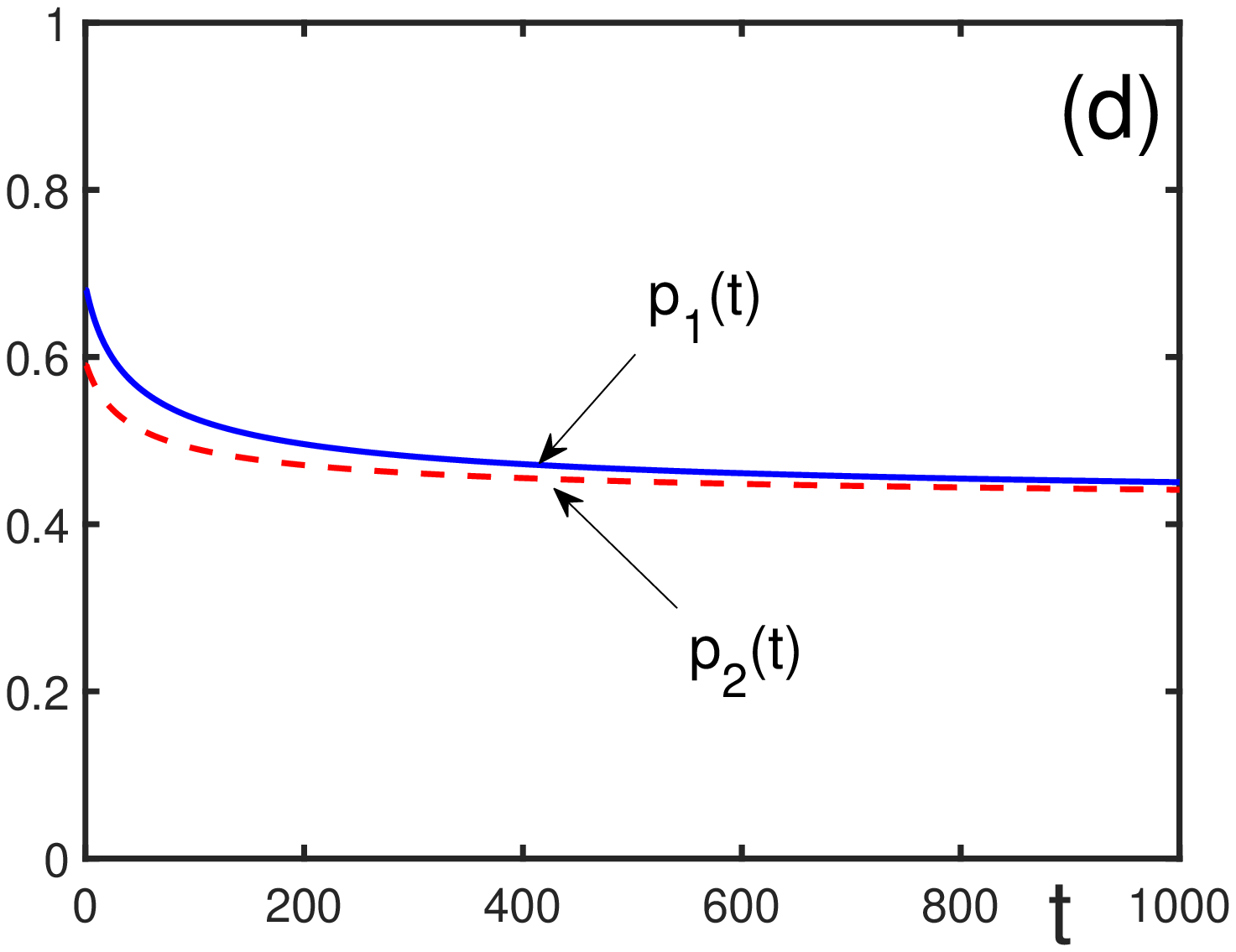} } }
\caption{
Long-term memory. Monotonically diminishing probabilities $p_1(t)$ (solid line) 
and $p_2(t)$ (dashed line) as functions of time: 
(a) Rational initial preference for $\ep_1=\ep_2=0$, the initial conditions are
$f_1=0.6$, $f_2=0.3$, $q_1=0.2$, $q_2=0.1$, which implies $p_1(0)=0.8$ and 
$p_2(0)=0.4$. The limiting probabilities are $p_1(\infty)=f_1=0.6$ and 
$p_2(\infty)=f_2=0.3$. 
(b) Rational initial preference for $\ep_1=0.5$, $\ep_2=0.1$, the initial 
conditions are the same as in Fig. 1(a), which gives $p_1(0)=0.6$ and 
$p_2(0)=0.44$. The limiting probabilities are $p_1(\infty)=p_1^*=0.45$ 
and $p_2(\infty)=p_2^*=0.33$.
(c) Irrational initial preference for $\ep_1=\ep_2=0$, the initial conditions are
$f_1=0.3$, $f_2=0.4$, $q_1=0.5$, $q_2=0.1$, which implies $p_1(0)=0.8$ and 
$p_2(0)=0.5$. The consensual probabilities are $p_1(\infty)=p_2(\infty)=p^*=0.42$. 
(d) Irrational initial preference for $\ep_1=0.4$, $\ep_2=0.3$, the initial 
conditions are the same as in Fig. 1(c), which implies $p_1(0)=0.68$ and 
$p_2(0)=0.59$. The consensual probabilities are $p_1(\infty)=p_2(\infty)=p^*=0.42$. 
}
\label{fig:Fig.1}
\end{figure}

\newpage

\begin{figure}[ht]
\centerline{
\hbox{ \includegraphics[width=7.5cm]{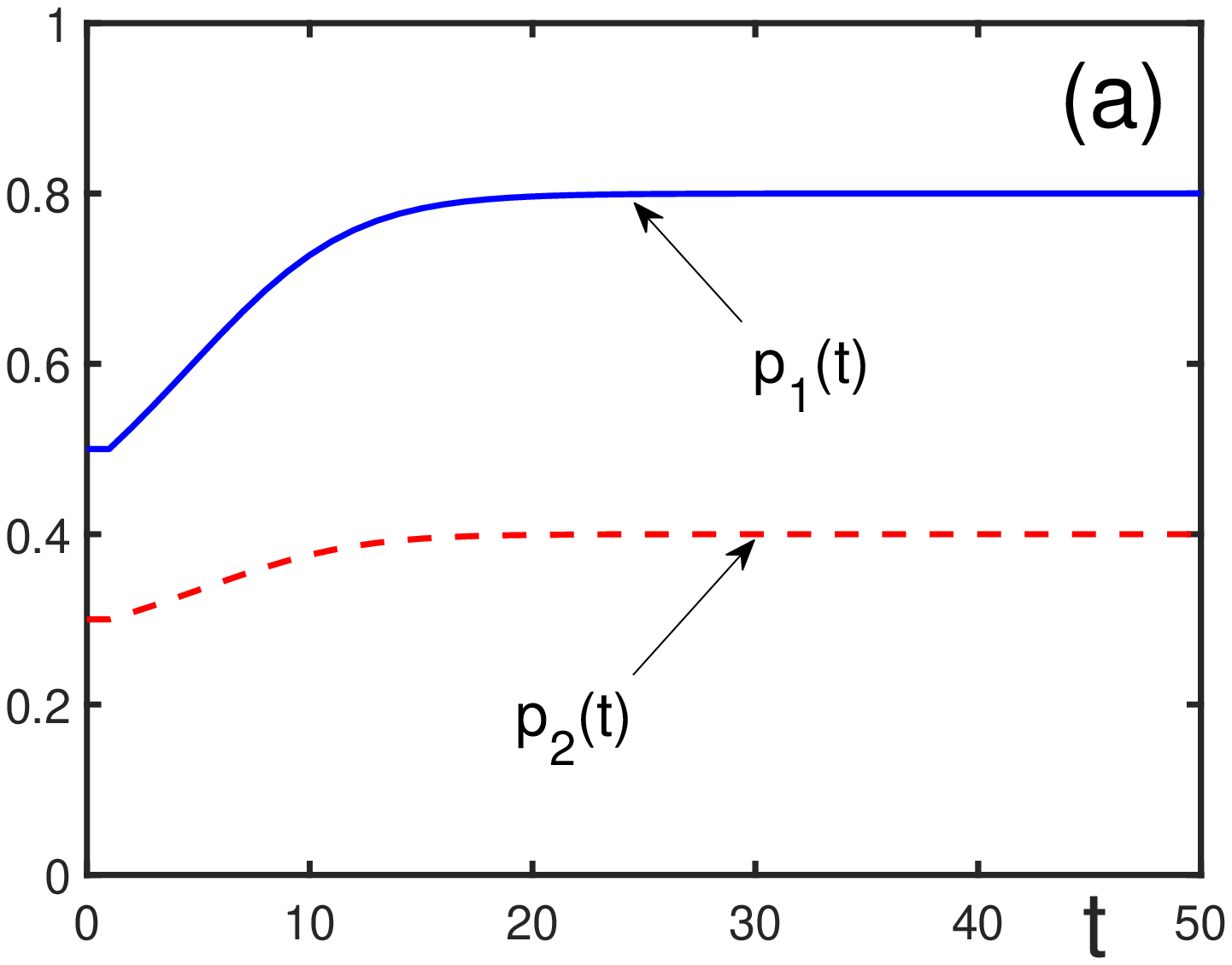} \hspace{1cm}
\includegraphics[width=7.5cm]{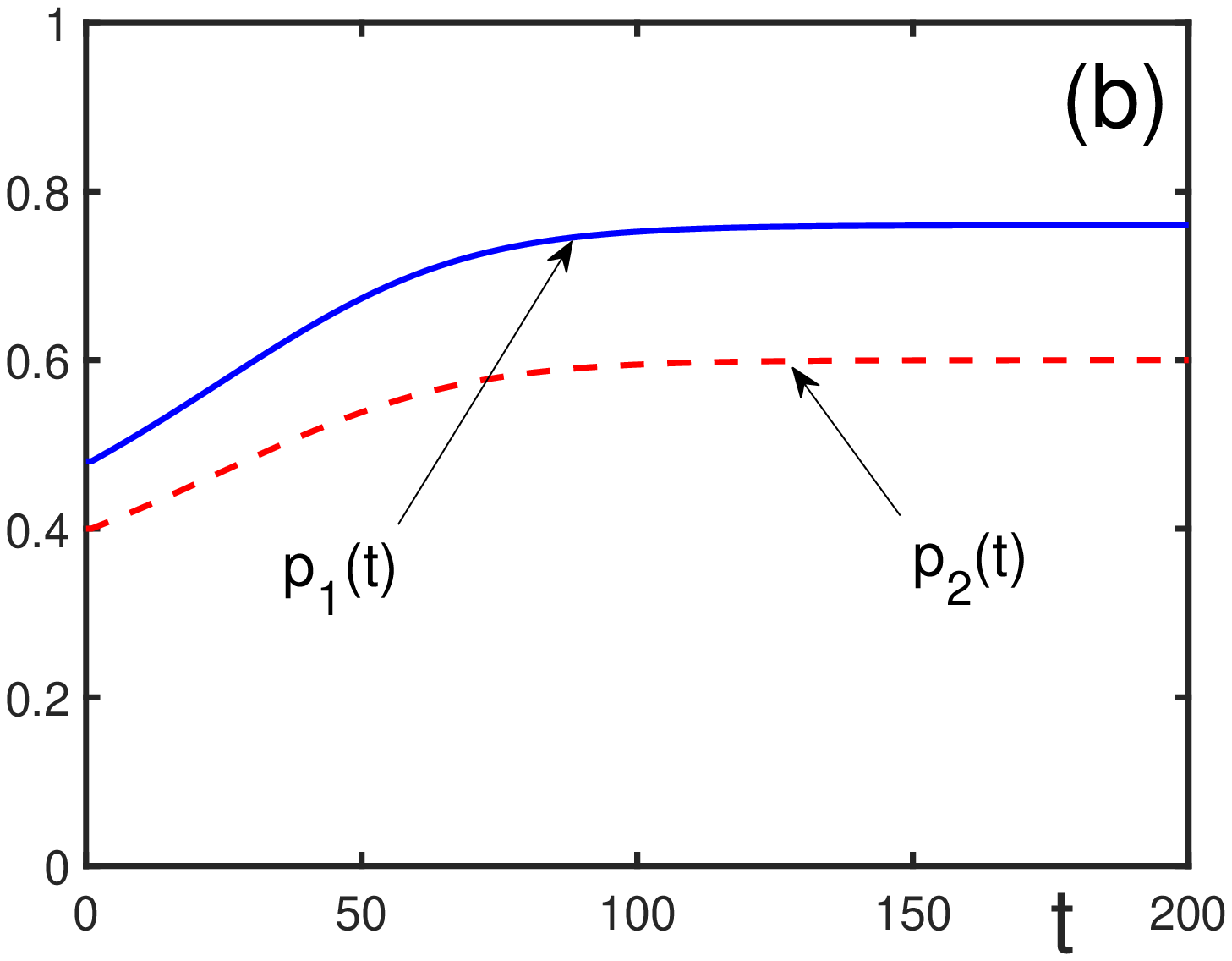}  } }
\vspace{12pt}
\centerline{
\hbox{ \includegraphics[width=7.5cm]{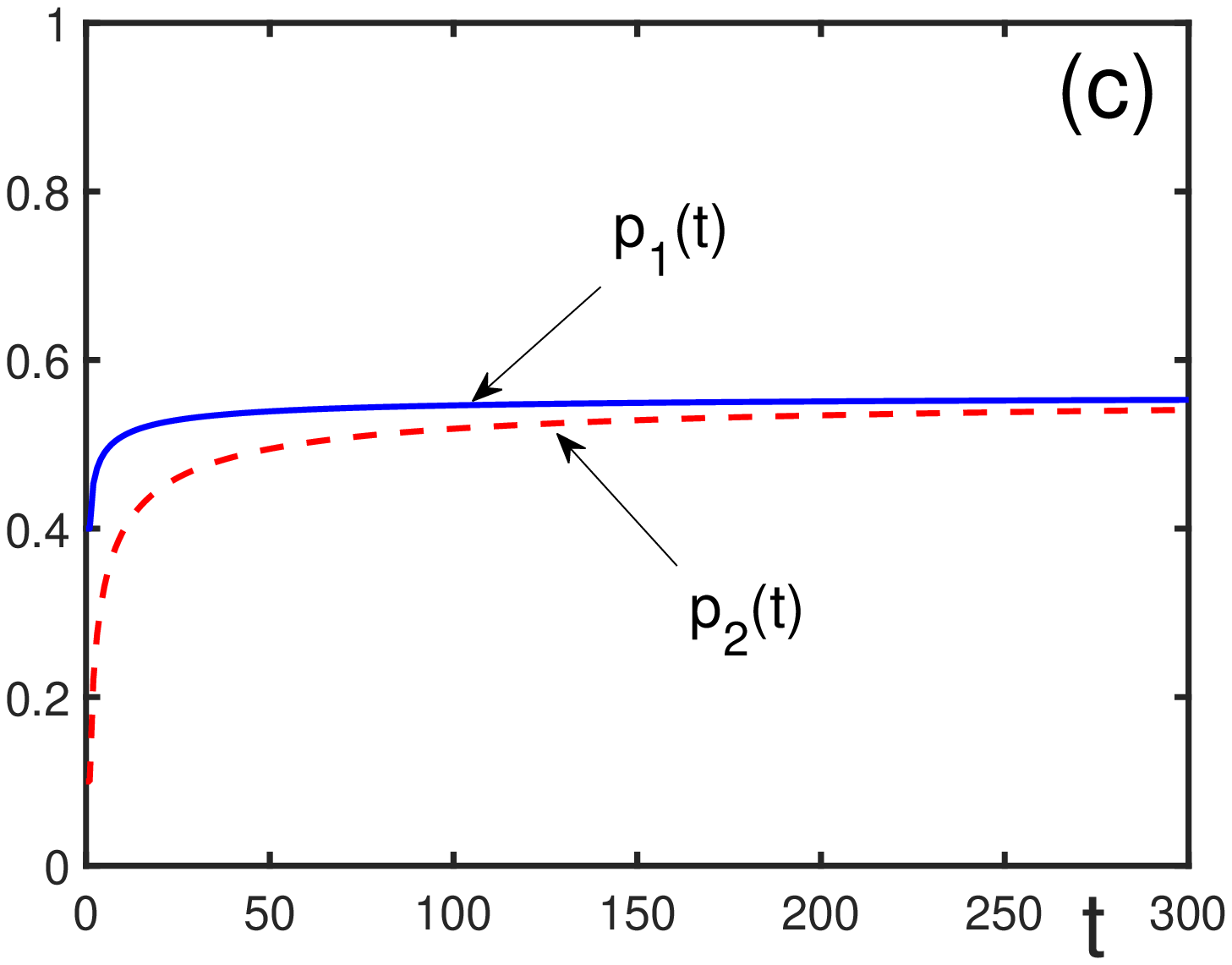} \hspace{1cm}
\includegraphics[width=7.5cm]{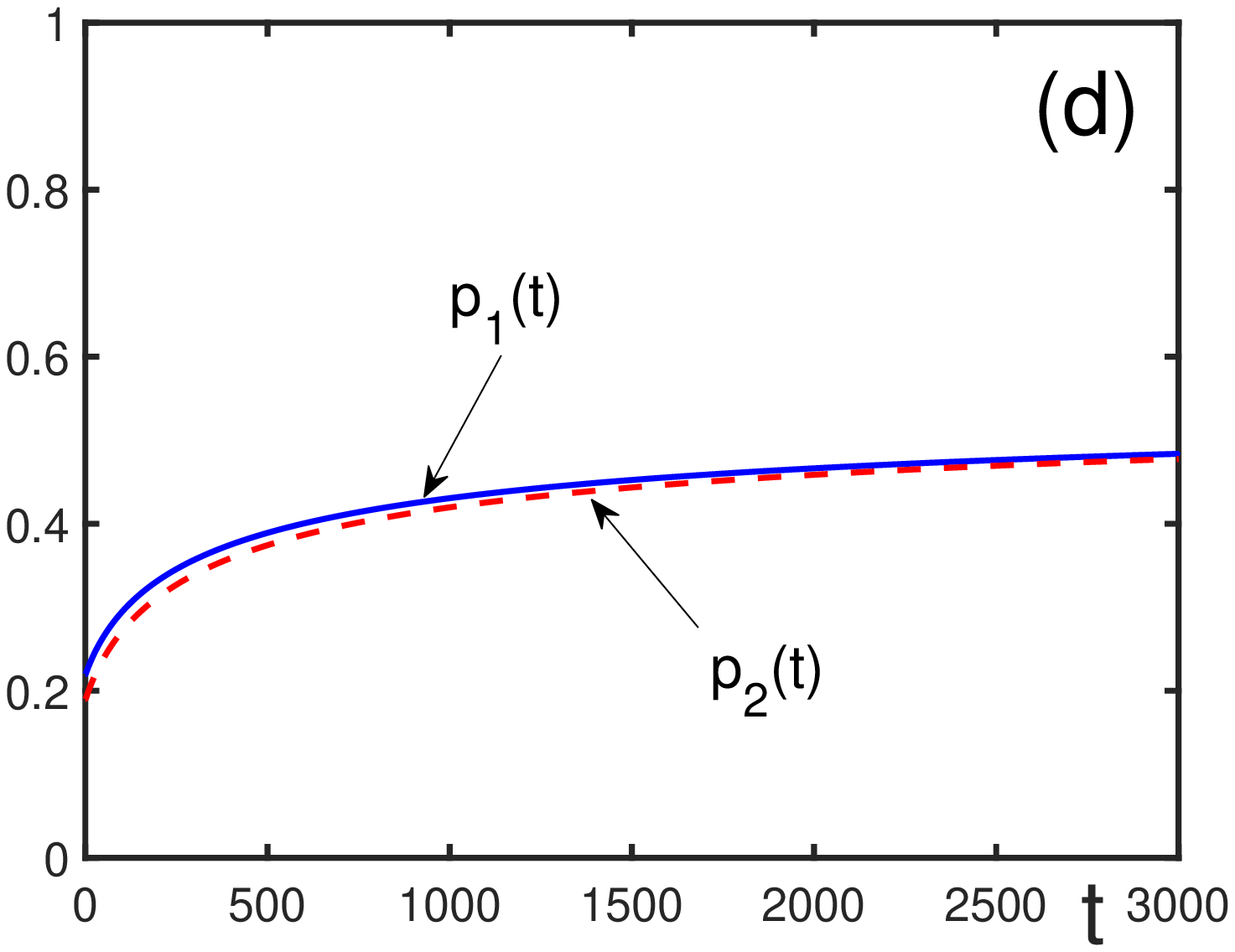} } }
\caption{
Long-term memory. Monotonically increasing probabilities $p_1(t)$ (solid line) 
and $p_2(t)$ (dashed line) as functions of time: 
(a) Rational initial preference for $\ep_1=\ep_2=0$, the initial conditions are
$f_1=0.8$, $f_2=0.4$, $q_1=-0.3$, $q_2=-0.1$, which implies $p_1(0)=0.5$ and 
$p_2(0)=0.3$. The limiting probabilities are $p_1(\infty)=f_1=0.8$ and 
$p_2(\infty)=f_2=0.4$. 
(b) Rational initial preference for $\ep_1=0.1$, $\ep_2=0.5$, initial conditions 
are the same as for Fig. 2(a), which gives $p_1(0)=0.48$ and $p_2(0)=0.4$. The 
limiting probabilities are $p_1(\infty)=p_1^*=0.76$ and $p_2(\infty)=p_2^*=0.6$.
(c) Irrational initial preference for $\ep_1=\ep_2=0$, initial conditions are
$f_1=0.6$, $f_2=0.7$, $q_1=-0.2$, $q_2=-0.6$, which implies $p_1(0)=0.4$ 
and $p_2(0)=0.1$. The consensual probabilities are 
$p_1(\infty)=p_2(\infty)=p^*=0.55$. 
(d) Irrational initial preference for $\ep_1=0.6$, $\ep_2=0.3$, initial conditions 
are the same as for Fig. 2(c), which implies $p_1(0)=0.22$ and $p_2(0)=0.19$. The 
consensual probabilities are $p_1(\infty)=p_2(\infty)=p^*=0.55$. 
}
\label{fig:Fig.2}
\end{figure}

\newpage

\begin{figure}[ht]
\centerline{
\hbox{ \includegraphics[width=7.5cm]{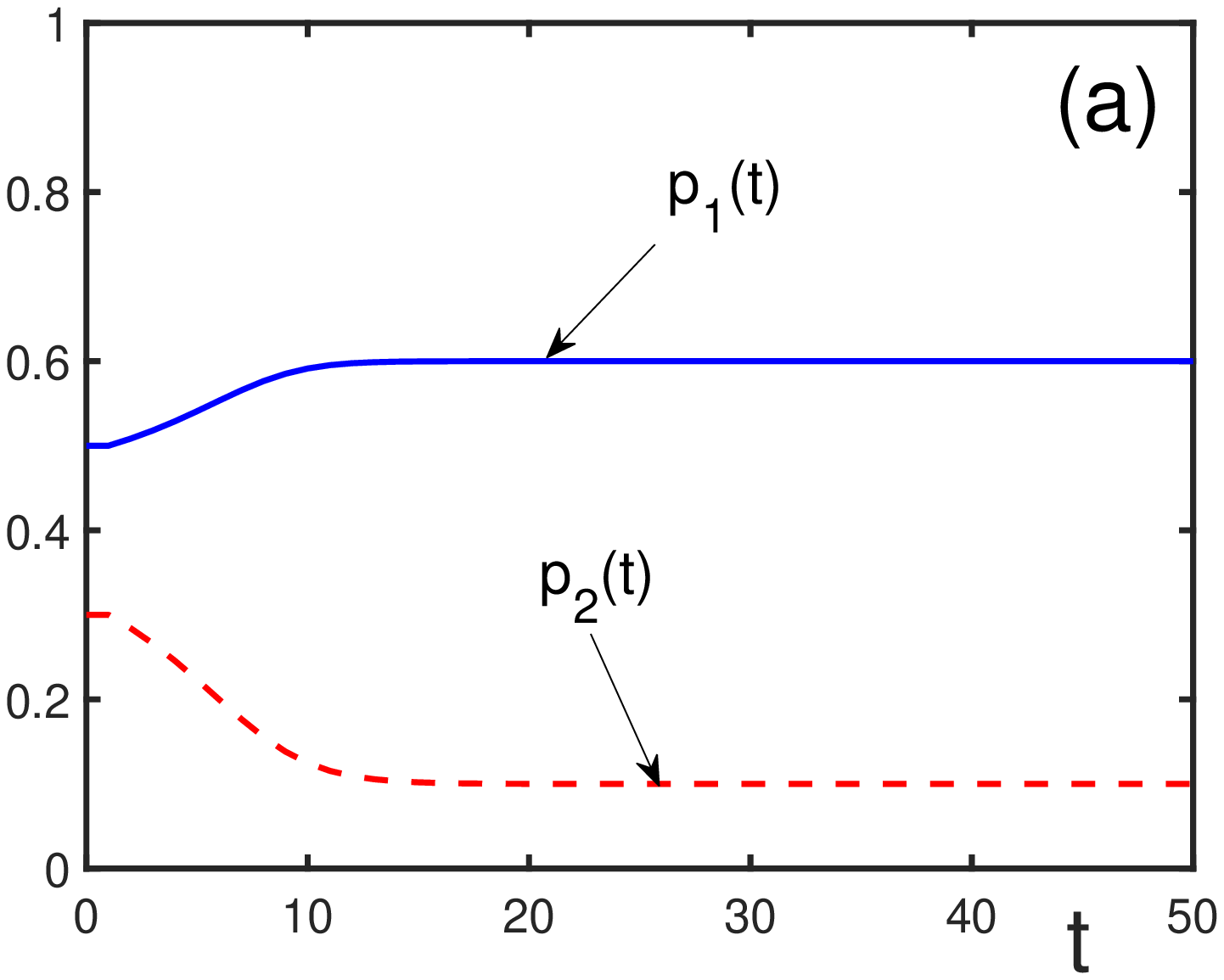} \hspace{1cm}
\includegraphics[width=7.5cm]{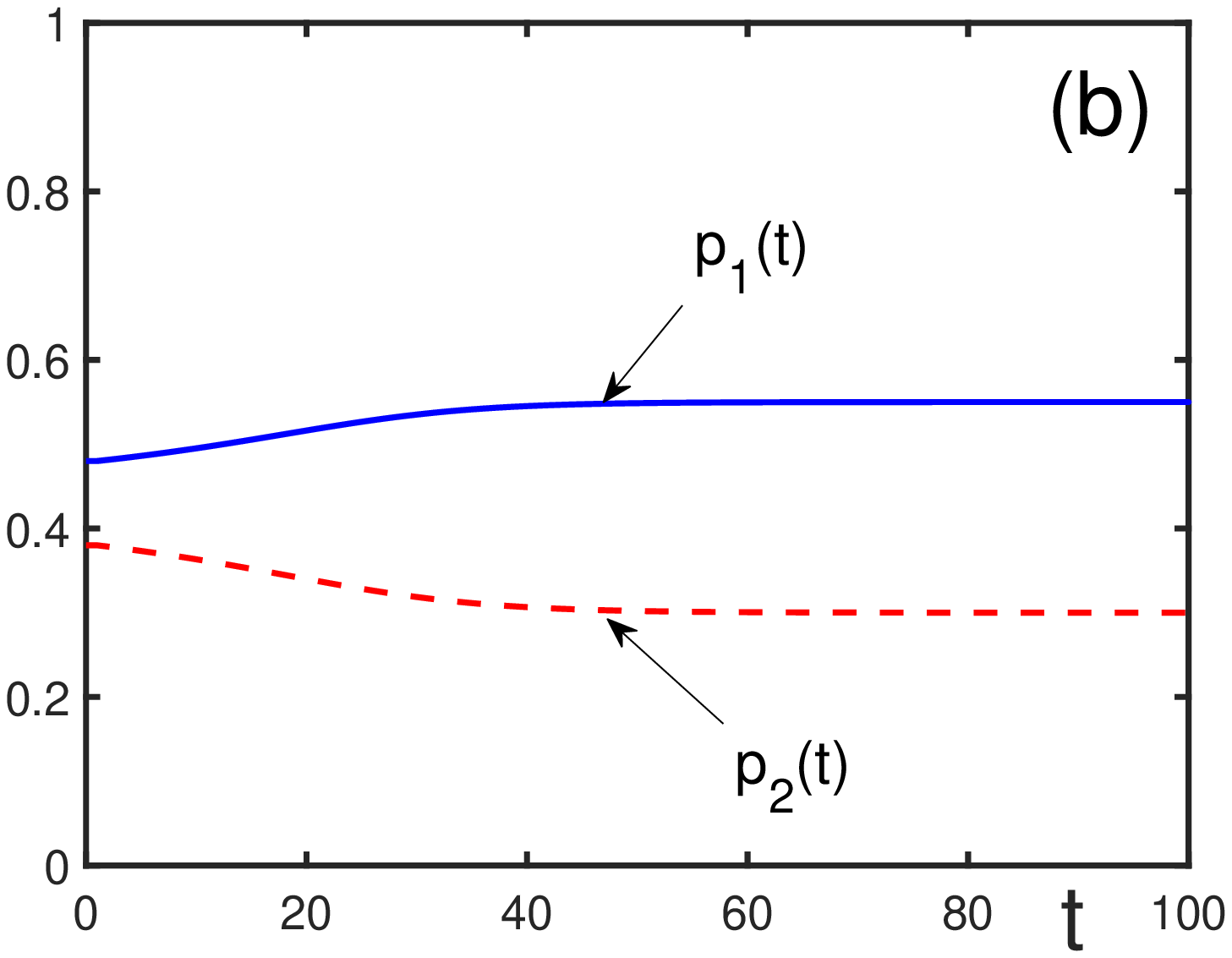}  } }
\vspace{12pt}
\centerline{
\hbox{ \includegraphics[width=7.5cm]{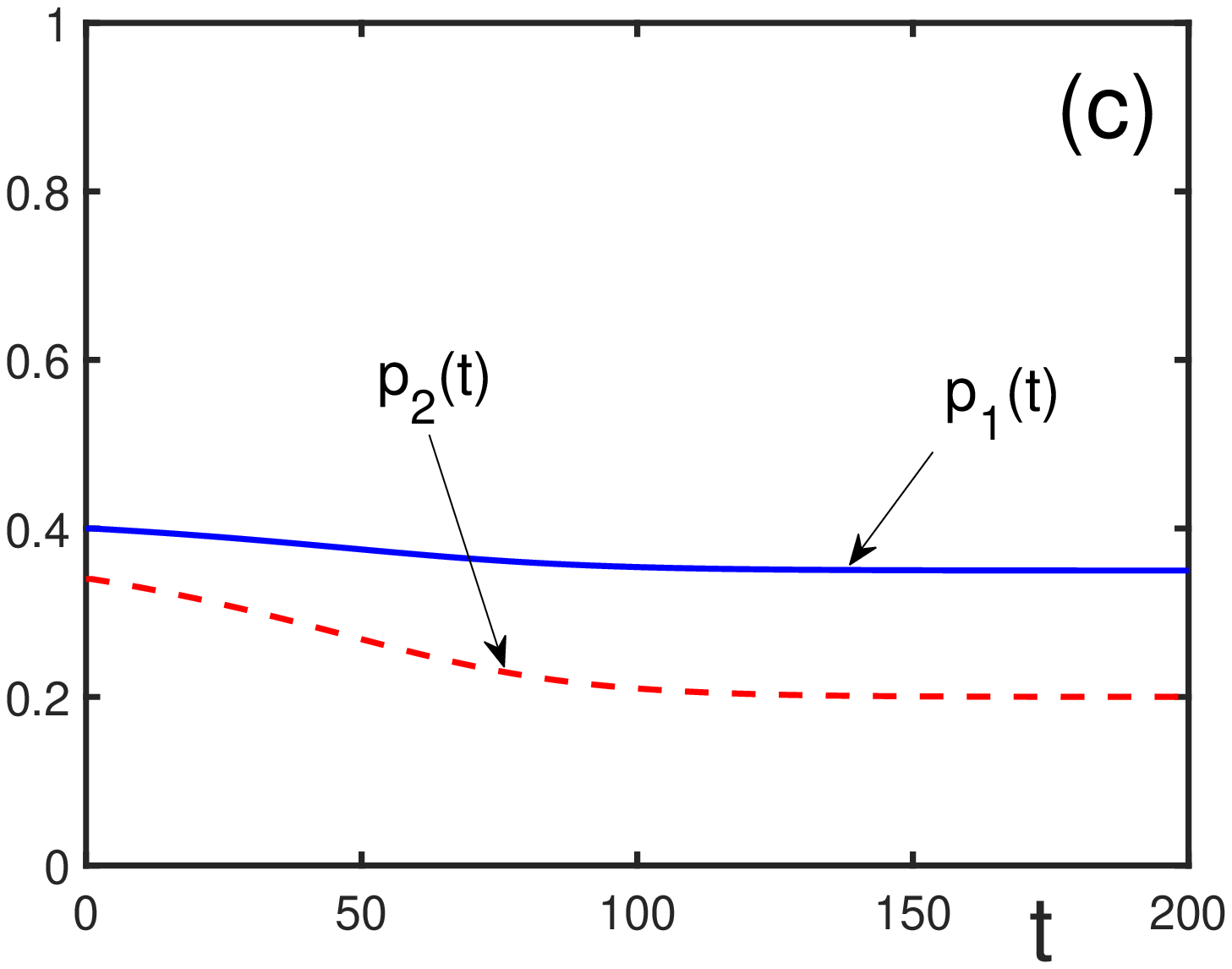} \hspace{1cm}
\includegraphics[width=7.5cm]{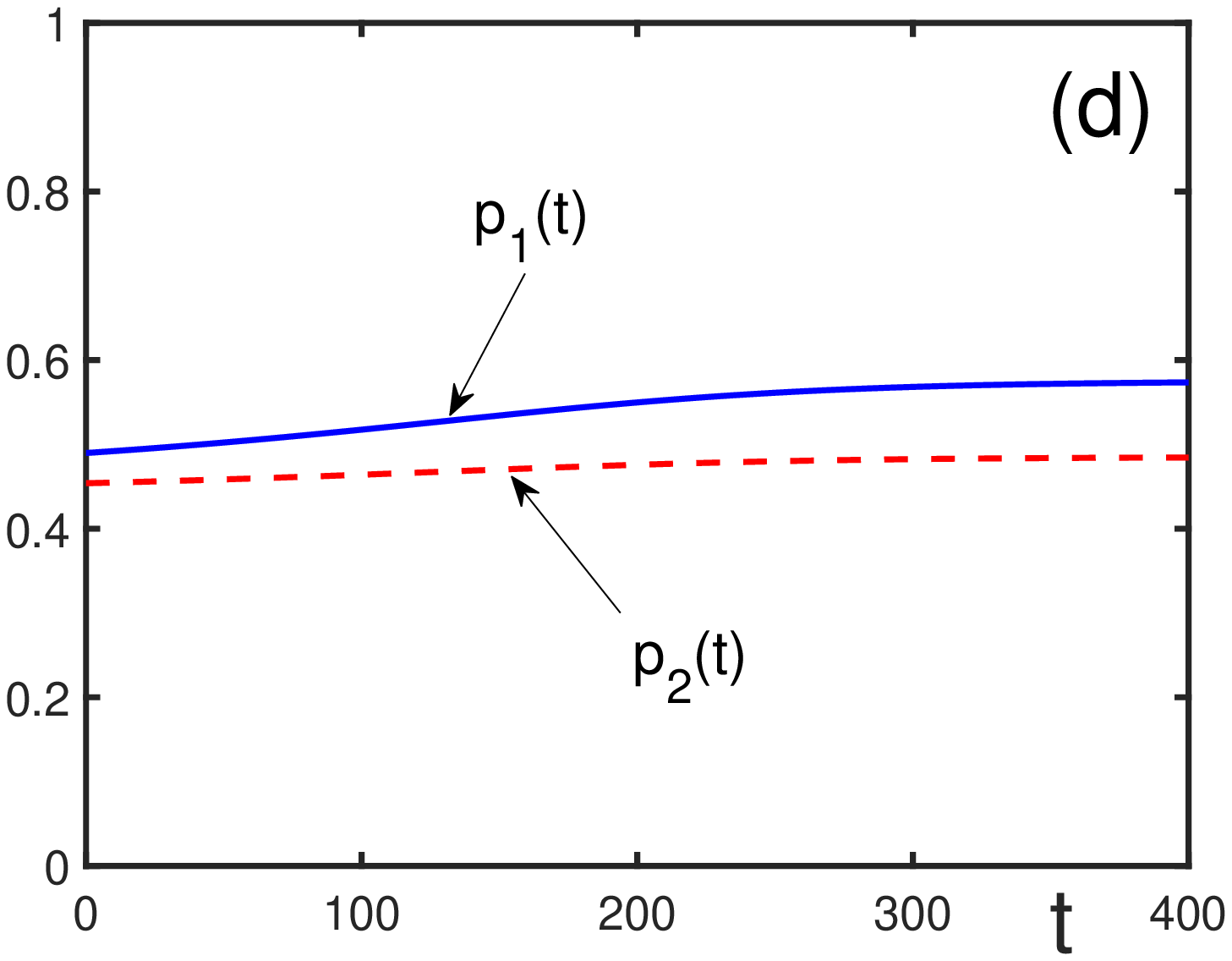} } }
\caption{
Long-term memory. Influence of the information field on probabilities
 $p_1(t)$ (solid line) and $p_2(t)$  (dashed line)
diverging from each other, with the 
same initial conditions $f_1=0.6$, $f_2=0.1$, $q_1=-0.1$, and $q_2=0.2$:  
(a) Rational initial preference for $\ep_1=\ep_2=0$, which gives $p_1(0)=0.5$ 
and $p_2(0)=0.3$. The limiting probabilities are $p_1(\infty)=f_1=0.6$ and 
$p_2(\infty)=f_2=0.1$.
(b) Rational initial preference for $\ep_1=0.1$ and $\ep_2=0.4$, which gives 
$p_1(0)=0.48$ and $p_2(0)=0.38$. The limiting probabilities are 
$p_1(\infty)=p_1^*=0.55$ and $p_2(\infty)=p^*=0.3$. 
(c) Rational initial preference for $\ep_1=0.5$ and $\ep_2=0.2$, which gives 
$p_1(0)=0.4$ and $p_2(0)=0.34$. The limiting probabilities are 
$p_1(\infty)=p_1^*=0.35$ and $p_2(\infty)=p_2^*=0.2$. 
(d) Rational initial preference for $\ep_1=0.05$ and $\ep_2=0.77$, which gives
$p_1(0)=0.49$ and $p_2(0)=0.45$. The limiting probabilities are 
$p_1(\infty)=p_1^*=0.58$ and $p_2(\infty)=p_2^*= 0.49$.
}
\label{fig:Fig.3}
\end{figure}

\newpage

\begin{figure}[ht]
\centerline{
\hbox{ \includegraphics[width=7.5cm]{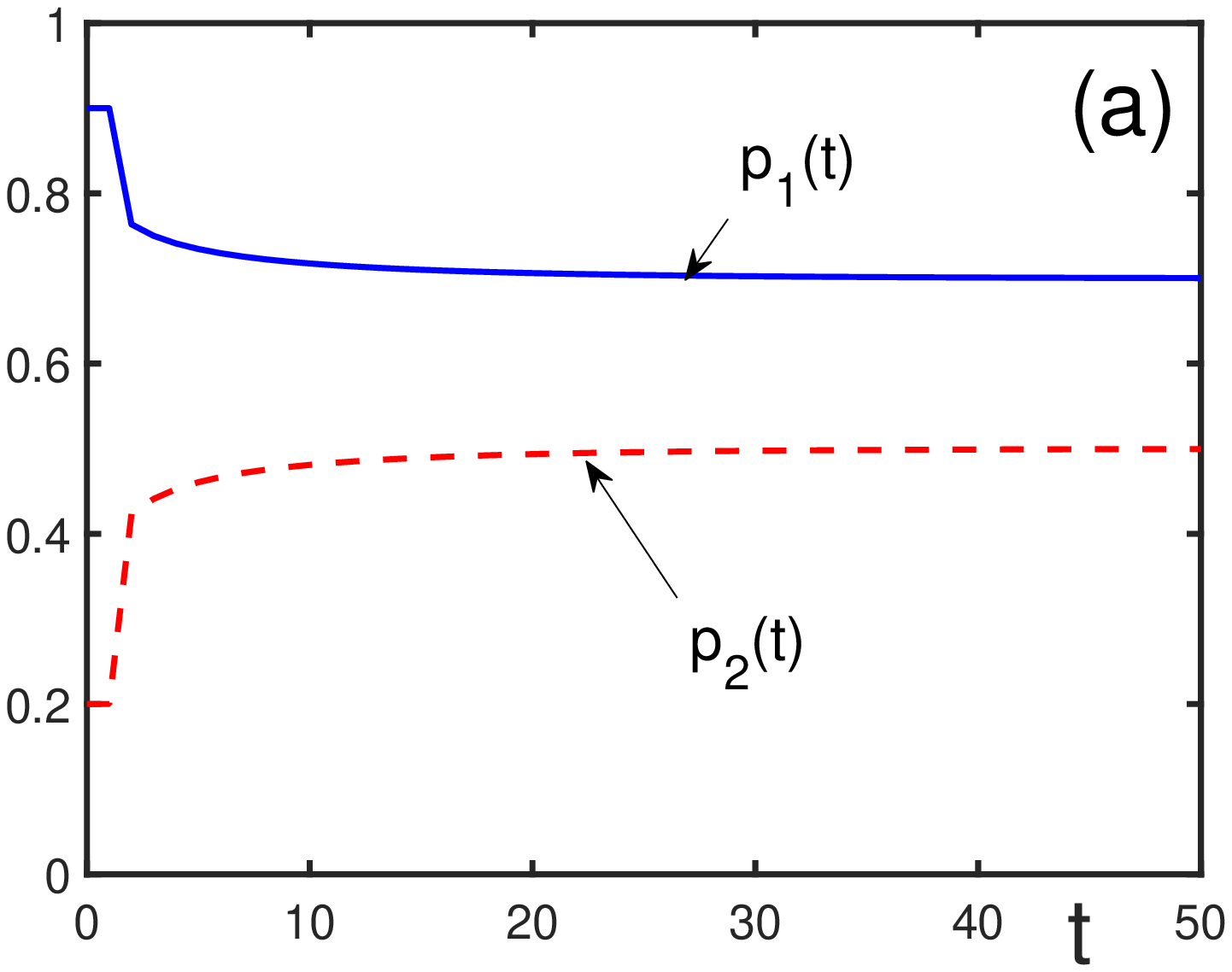} \hspace{1cm}
\includegraphics[width=7.5cm]{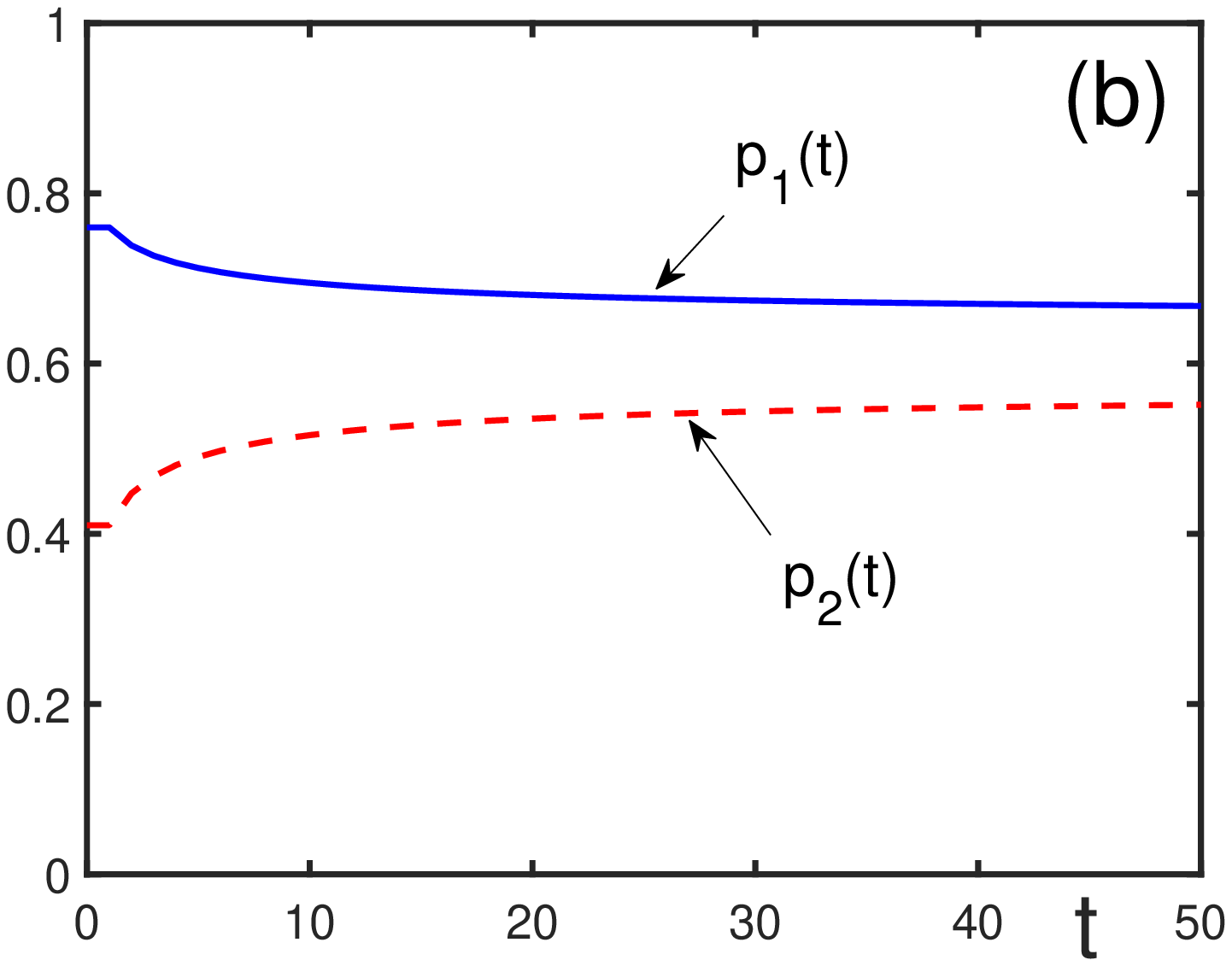}  } }
\vspace{12pt}
\centerline{
\hbox{ \includegraphics[width=7.5cm]{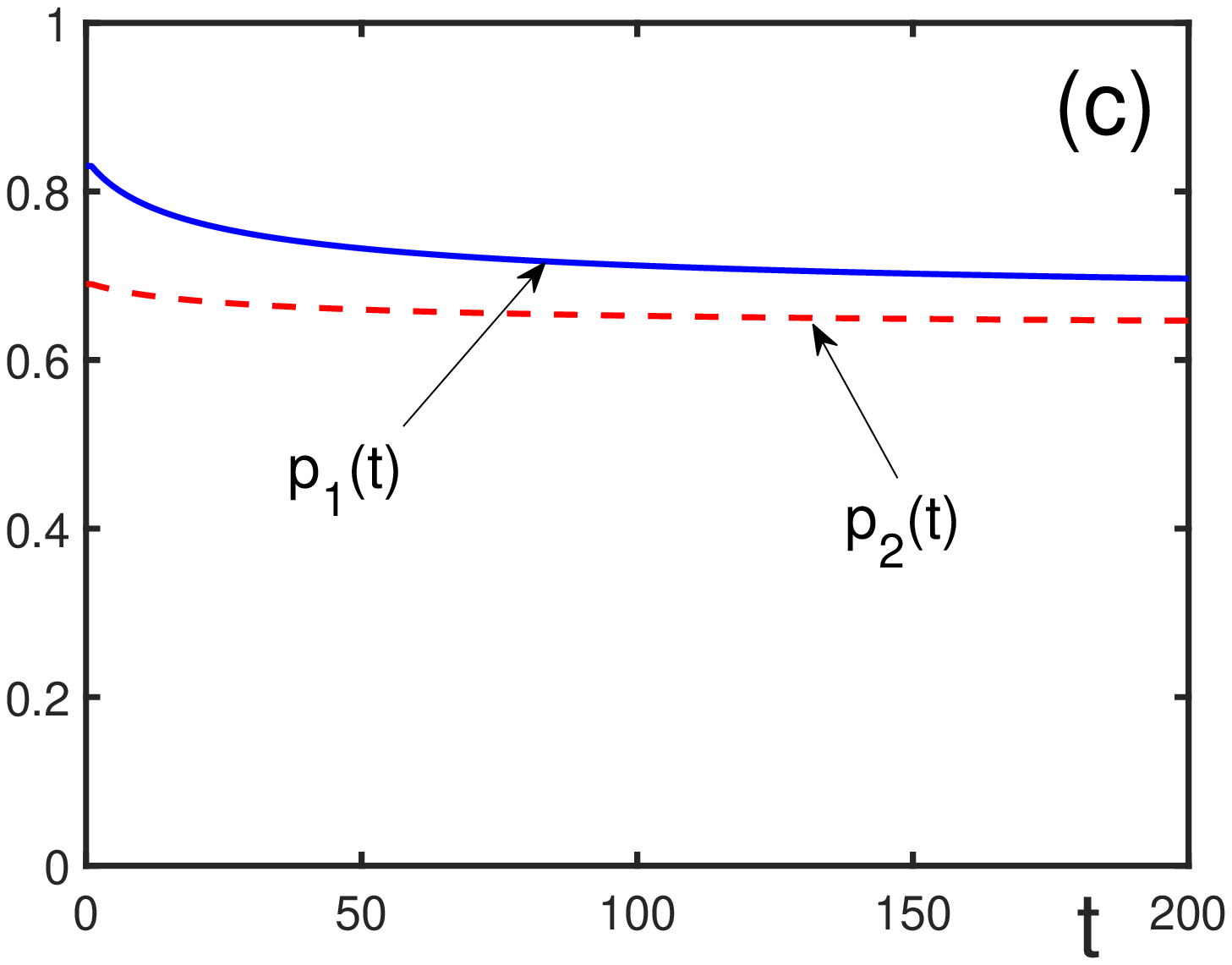} \hspace{1cm}
\includegraphics[width=7.5cm]{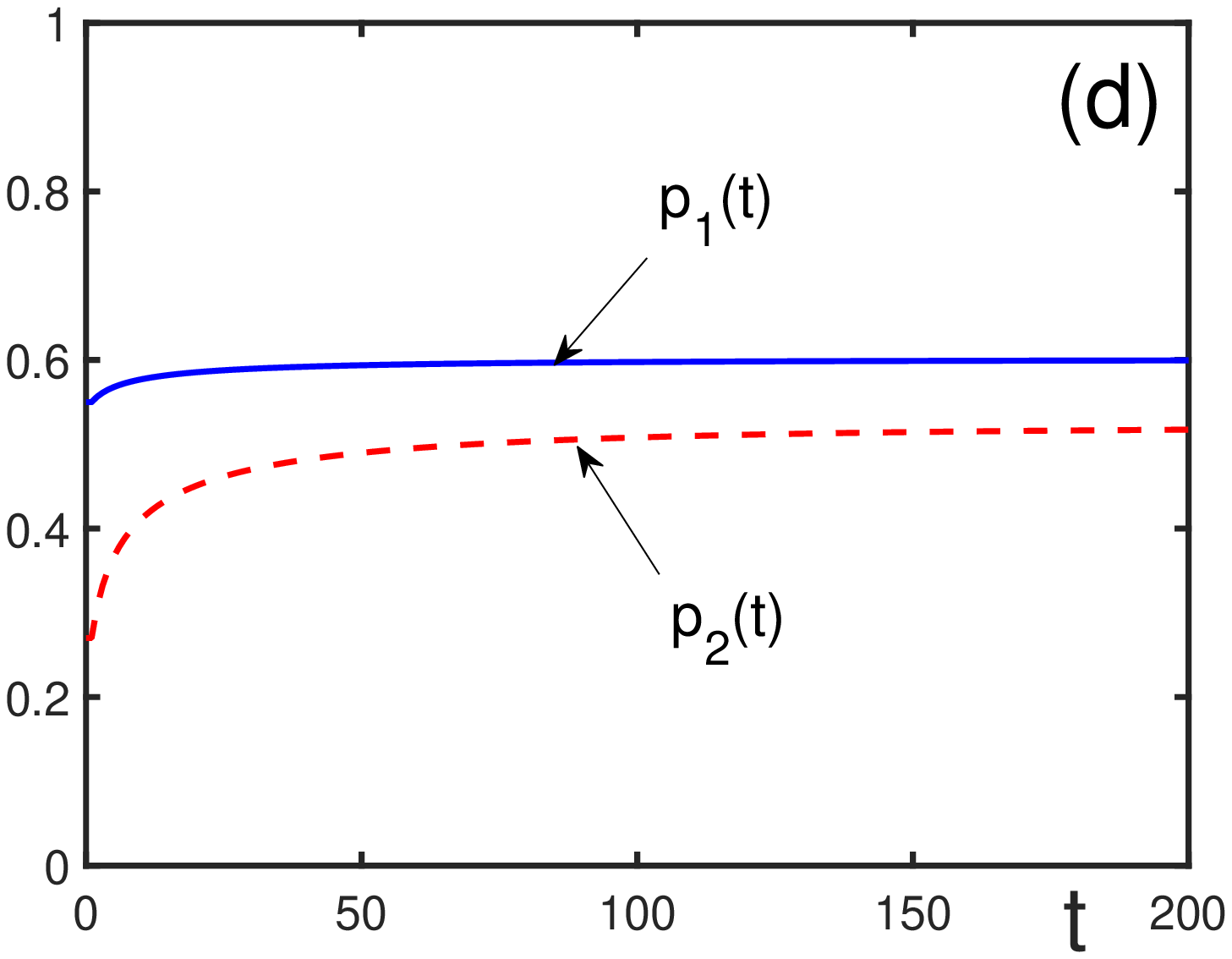} } }
\caption{
Long-term memory. Influence of the information field on 
 probabilities $p_1(t)$ (solid line) and $p_2(t)$  (dashed line)
approaching each other, in the case of 
the rational initial preference and the same initial conditions $f_1=0.7$, 
$f_2=0.5$, $q_1=0.2$, and $q_2=-0.3$: 
(a) $\ep_1=\ep_2=0$. The initial probabilities are $p_1(0)=0.9$ and $p_2(0)=0.2$. 
The limiting probabilities are $p_1(\infty)=f_1=0.7$ and $p_2(\infty)=f_2^*=0.5$.
(b) $\ep_1=0.2$ and $\ep_2=0.3$. The initial probabilities are $p_1(0)=0.76$ 
and $p_2(0)=0.41$. The limiting probabilities are $p_1(\infty)=p_1^*=0.66$ and 
$p_2(\infty)=p_2^*=0.56$.
(c) $\ep_1=0.1$ and $\ep_2=0.7$. The initial probabilities are $p_1(0)=0.83$ 
and $p_2(0)=0.69$. The limiting probabilities are $p_1(\infty)=p_1^*=0.68$ and 
$p_2(\infty)=p_2^*=0.64$. 
(d) $\ep_1=0.5$ and $\ep_2=0.1$. The initial probabilities are $p_1(0)=0.55$ 
and $p_2(0)=0.27$. The limiting probabilities are $p_1(\infty)=p_1^*=0.6$ 
and $p_2(\infty)=p_2^*=0.52$.
}
\label{fig:Fig.4}
\end{figure}

\newpage

\begin{figure}[ht]
\centerline{
\hbox{ \includegraphics[width=7.5cm]{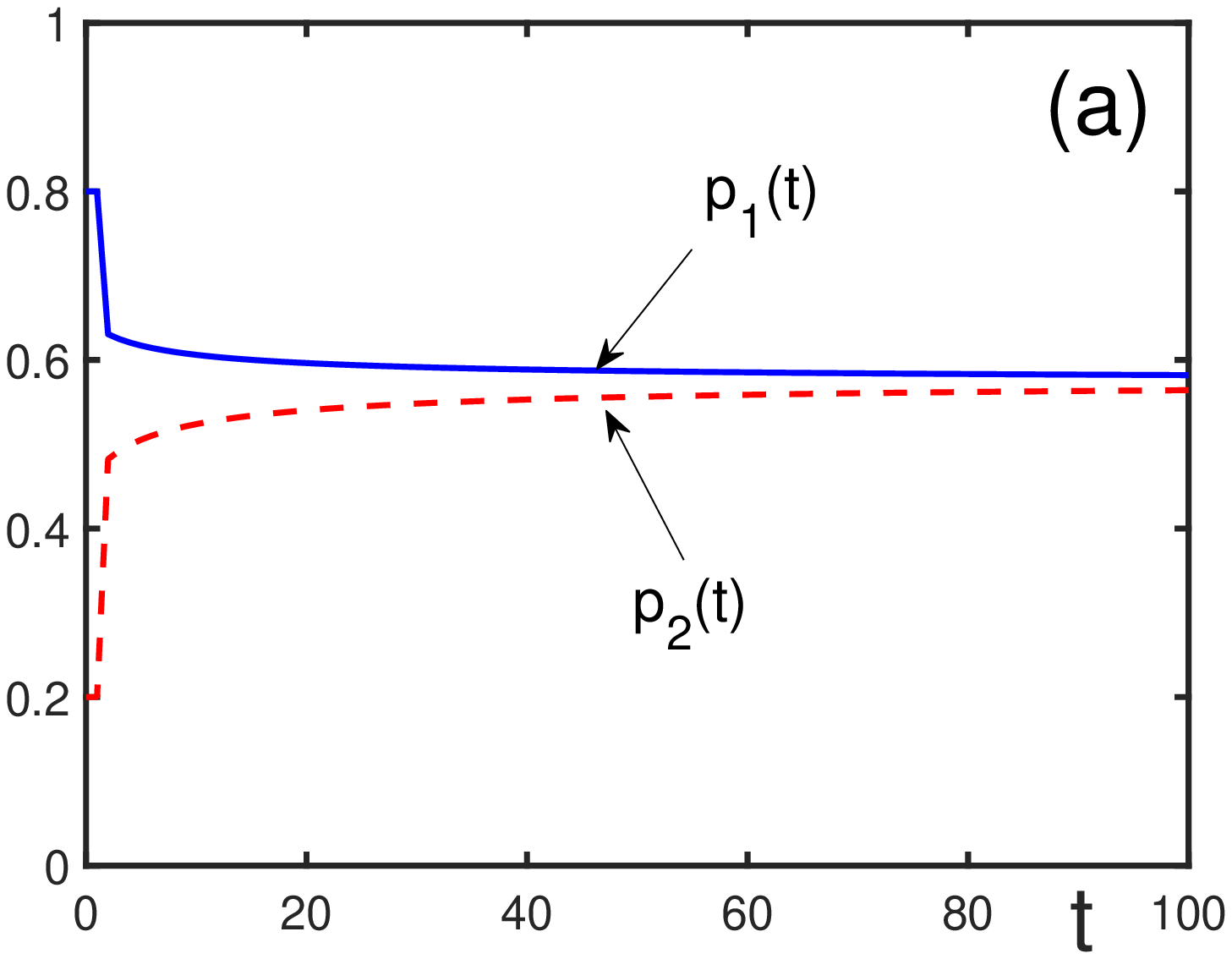} \hspace{1cm}
\includegraphics[width=7.5cm]{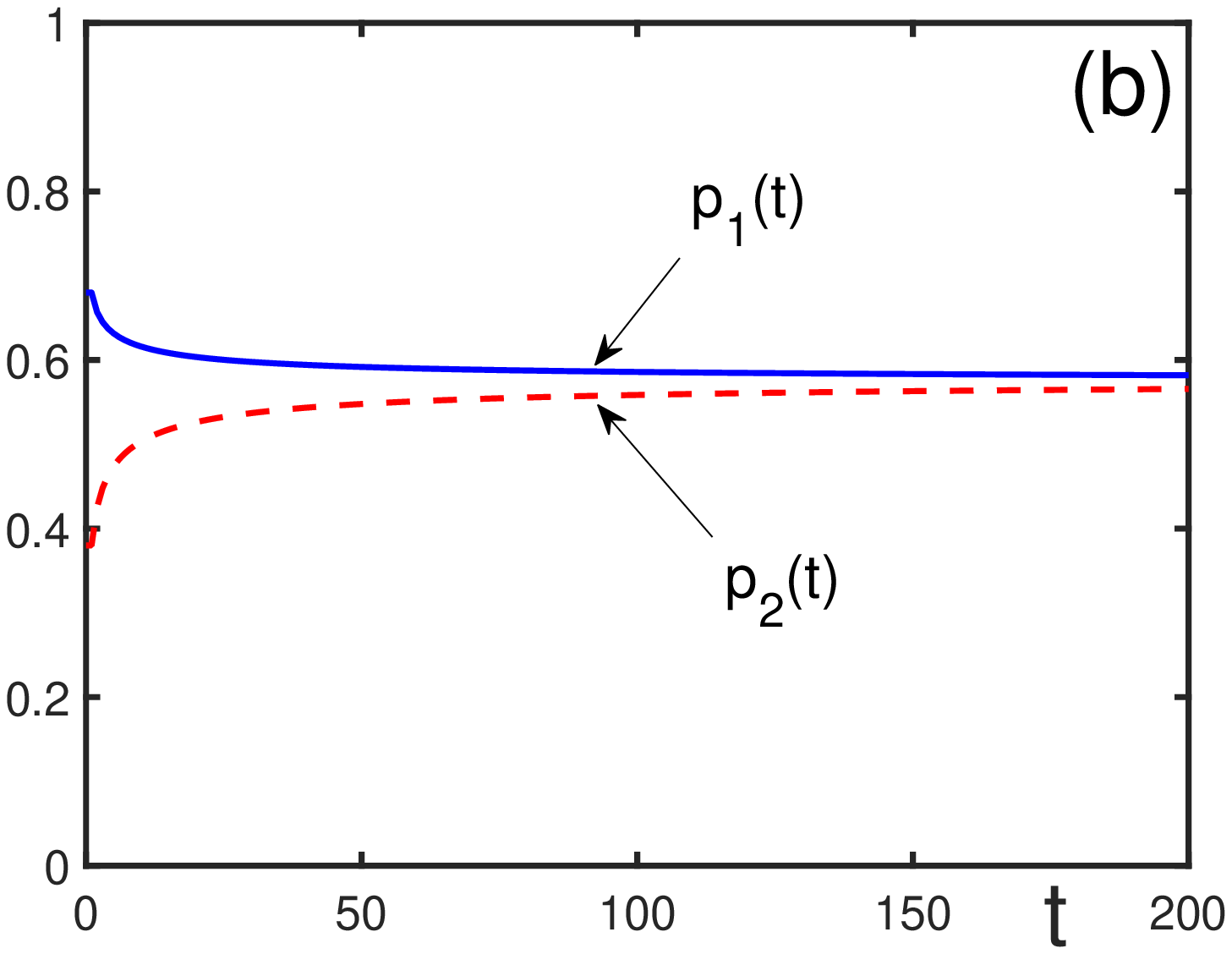}  } }
\vspace{12pt}
\centerline{
\hbox{ \includegraphics[width=7.5cm]{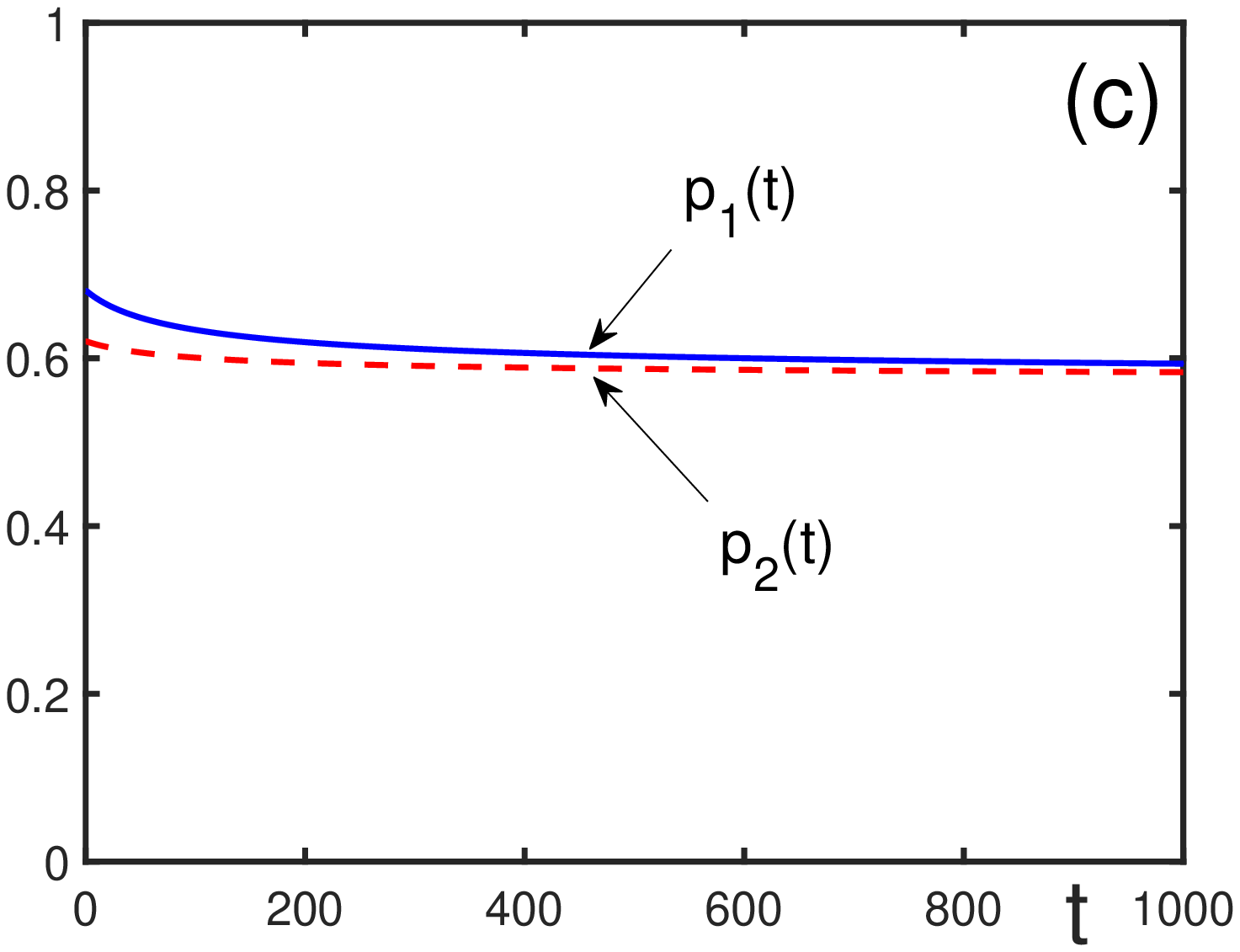} \hspace{1cm}
\includegraphics[width=7.5cm]{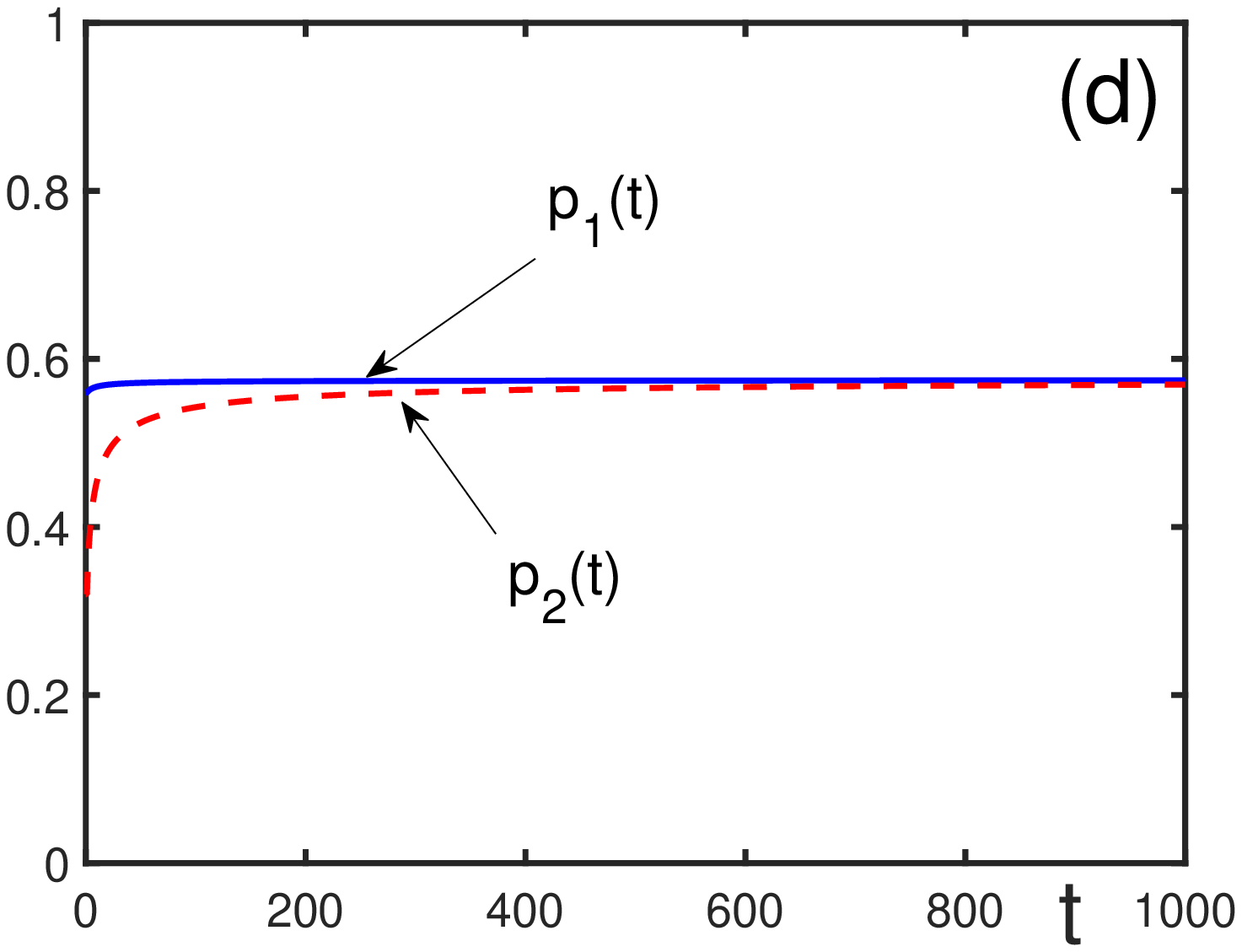} } }
\caption{
Long-term memory. Influence of the information field on 
 probabilities $p_1(t)$ (solid line) and $p_2(t)$  (dashed line)
approaching each other, in the 
case of the irrational initial preference and the same initial conditions 
$f_1=0.5$, $f_2=0.7$, $q_1=0.3$, and $q_2=-0.5$: 
(a) $\ep_1=\ep_2=0$. The initial probabilities are $p_1(0)=0.8$ and $p_2(0)=0.2$. 
The consensual limiting probabilities are $p_1(\infty)=p_2(\infty)=p^*=0.58$.
(b) $\ep_1=0.2$ and $\ep_2=0.3$. The initial probabilities are $p_1(0)=0.68$ 
and $p_2(0)=0.38$. The consensual limiting probabilities are 
$p_1(\infty)=p_2(\infty)=p^*=0.58$.
(c) $\ep_1=0.2$ and $\ep_2=0.7$. The initial probabilities are $p_1(0)= 0.68$ 
and $p_2(0)=0.62$. The consensual limiting probabilities are 
$p_1(\infty)=p_2(\infty)=p^*=0.57$.
(d) $\ep_1=0.4$ and $\ep_2=0.2$. The initial probabilities are $p_1(0)=0.56$ 
and  $p_2(0)=0.32$. The consensual limiting probabilities are 
$p_1(\infty)=p_2(\infty)=p^*=0.57$.
}
\label{fig:Fig.5}
\end{figure}

\newpage

\begin{figure}[ht]
\centerline{
\hbox{ \includegraphics[width=7.5cm]{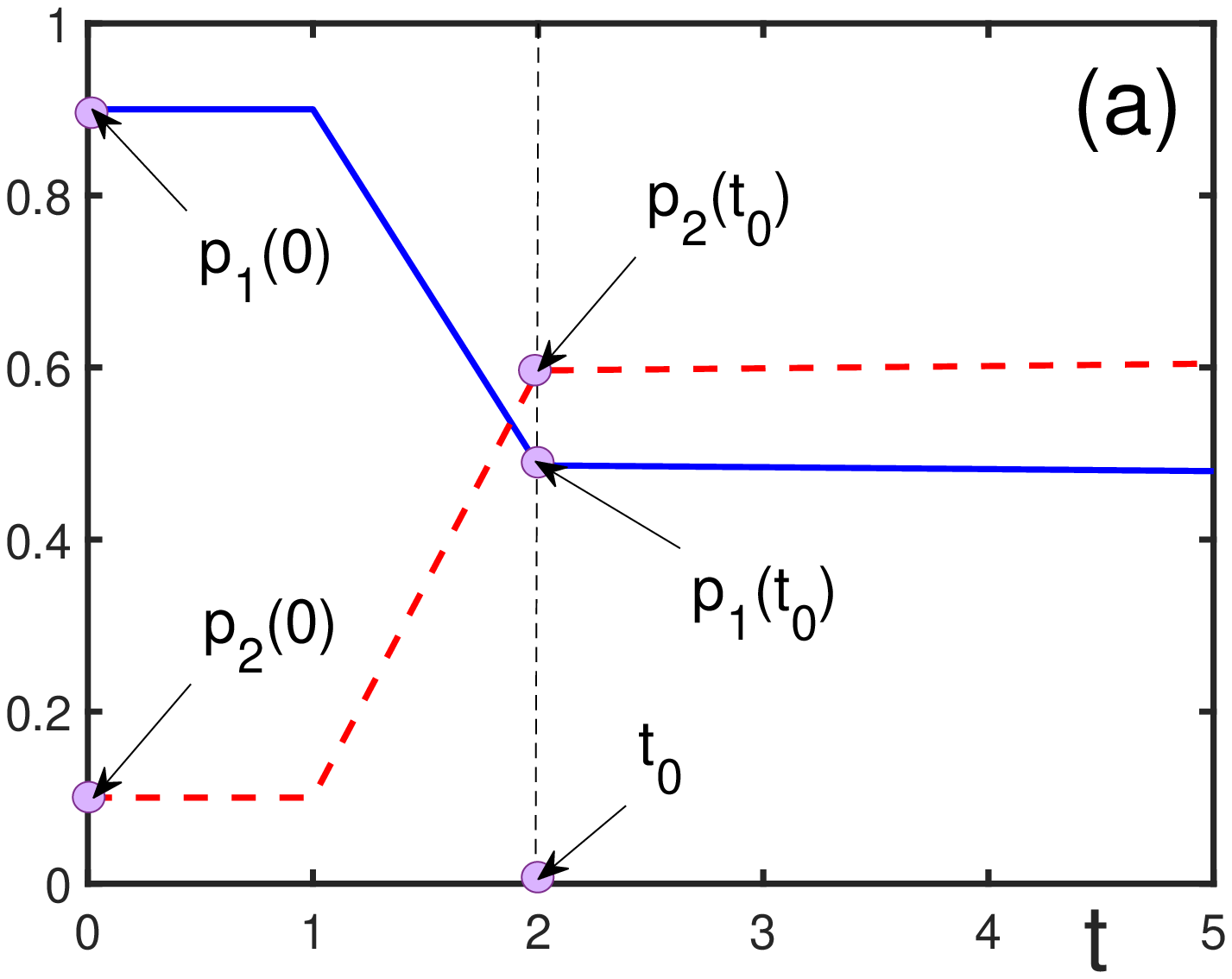} \hspace{1cm}
\includegraphics[width=7.5cm]{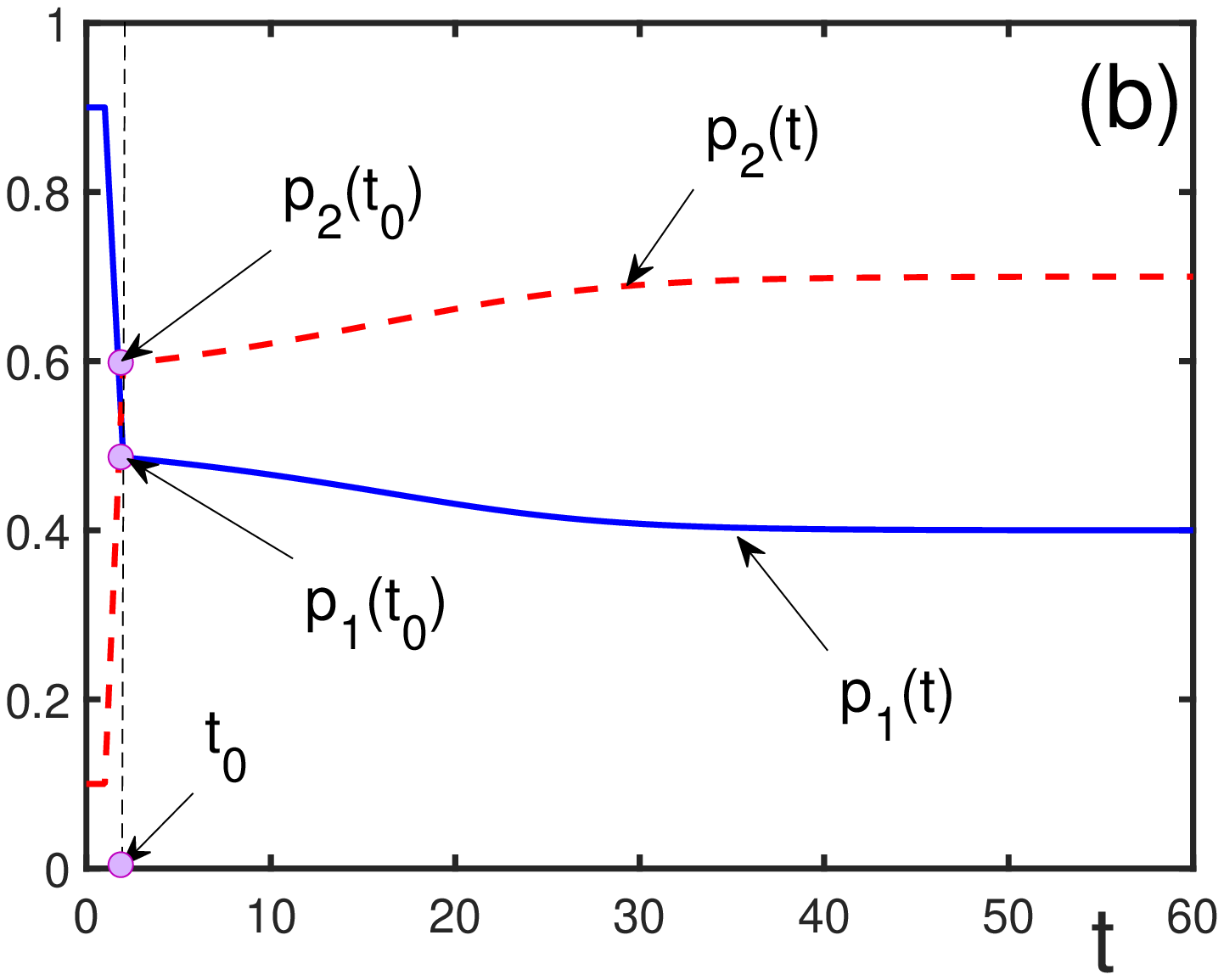}  } }
\vspace{12pt}
\centerline{
\hbox{ \includegraphics[width=7.5cm]{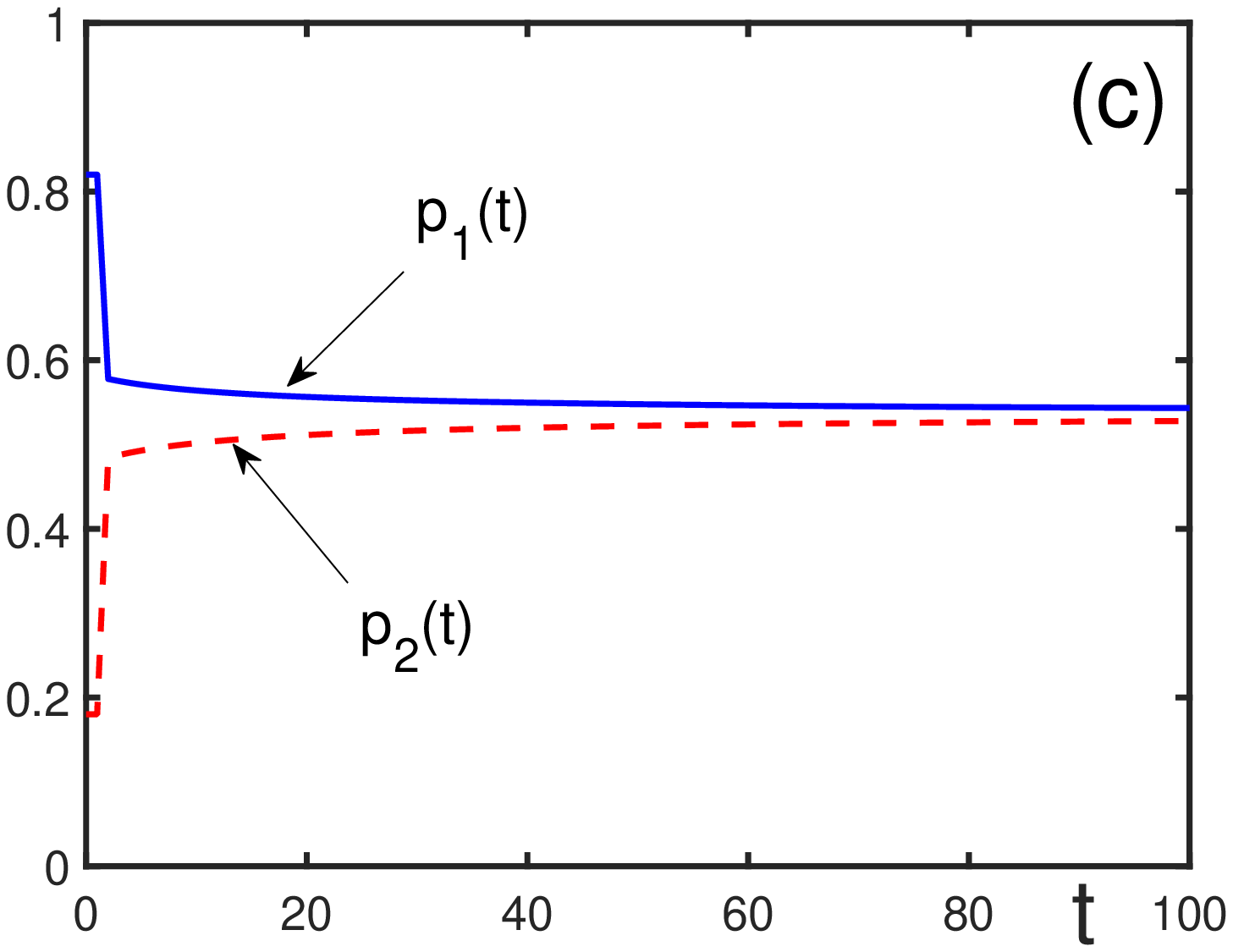} \hspace{1cm}
\includegraphics[width=7.5cm]{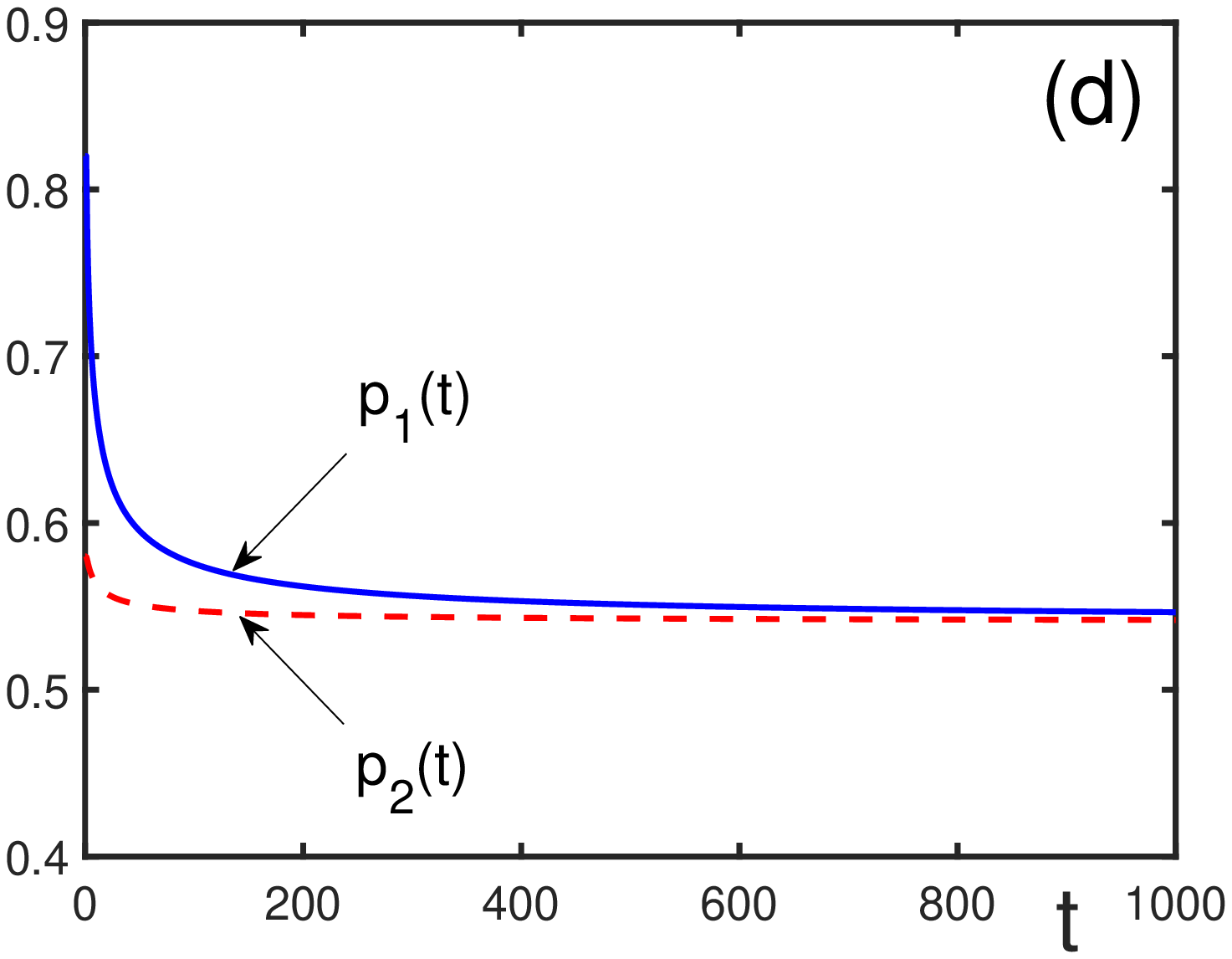} } }
\vspace{12pt}
\centerline{
\hbox{ 
\includegraphics[width=7.5cm]{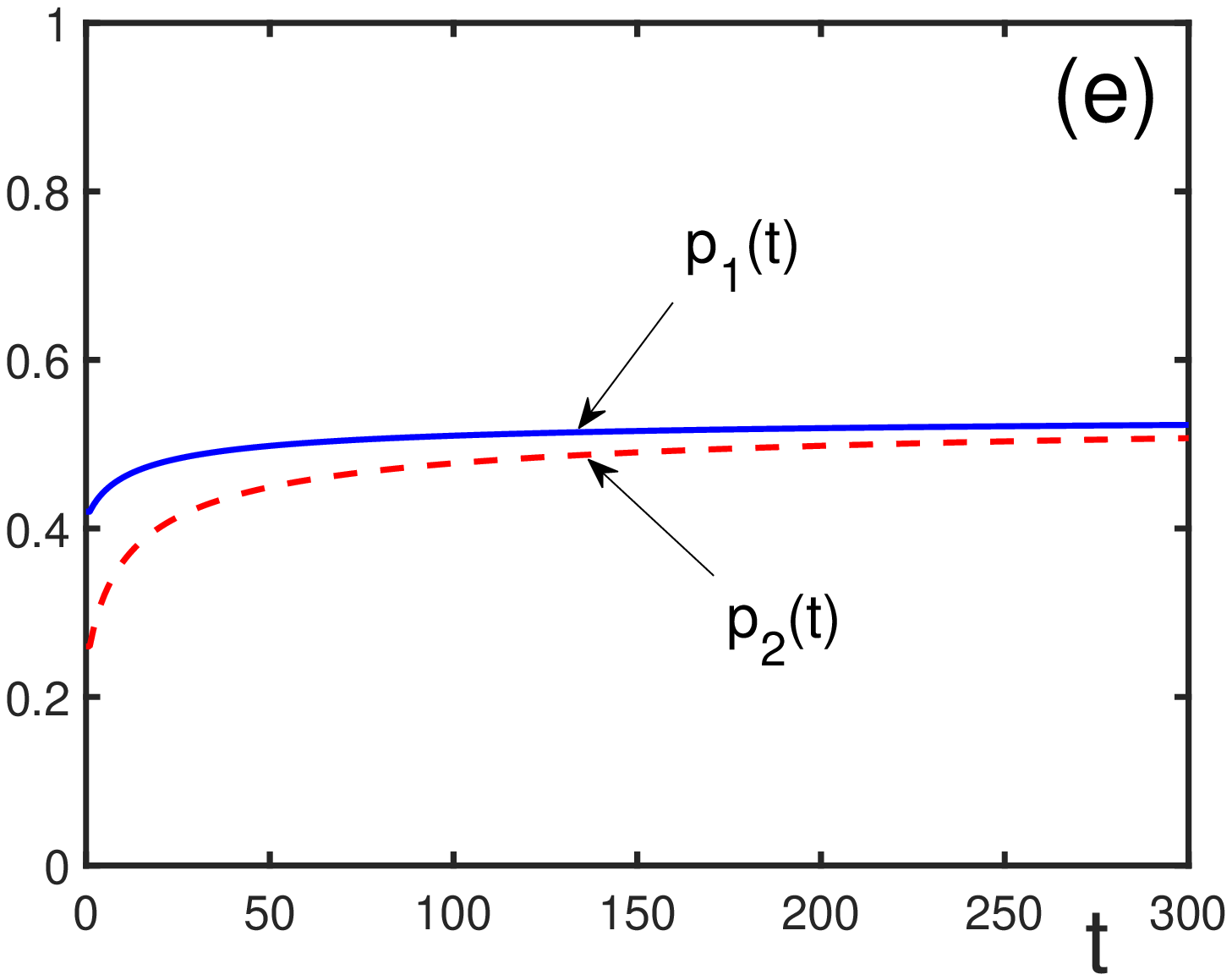} } }
\caption{
Long-term memory. Dynamic preference reversal under irrational initial 
preference. Probabilities $p_1(t)$ (solid line) and $p_2(t)$ (dashed line) 
for the initial conditions $f_1=0.4$, $f_2=0.7$, $q_1=0.5$, and $q_2=-0.6$. 
(a) The initial stage of the dynamic preference reversal for the time in the 
interval $0<t<5$, with $\ep_1=\ep_2=0$. The initial probabilities are 
$p_1(0)=0.9$ and  $p_2(0)=0.1$.
(b) Behavior of the probabilities after the preference reversal under 
$\ep_1=\ep_2=0$. The limiting probabilities are $p_1(\infty)=f_1=0.4$ and 
$p_2(\infty)=f_2=0.7$. 
(c) Switched on information field with $\ep_1=0.1$ and $\ep_2=0.1$. 
The consensual limiting probabilities are $p_j(t)=p^*=0.54$.
(d) Switched on information field with $\ep_1=0.1$ and $\ep_2=0.6$. 
The consensual limiting probabilities are $p_j(t)=p^*=0.54$.
(e) Switched on information field with $\ep_1=0.6$ and $\ep_2=0.2$. 
The consensual limiting probabilities are $p_j(t)=p^*=0.54$.
}
\label{fig:Fig.6}
\end{figure}

\newpage

\begin{figure}[ht]
\centerline{
\hbox{ \includegraphics[width=7.5cm]{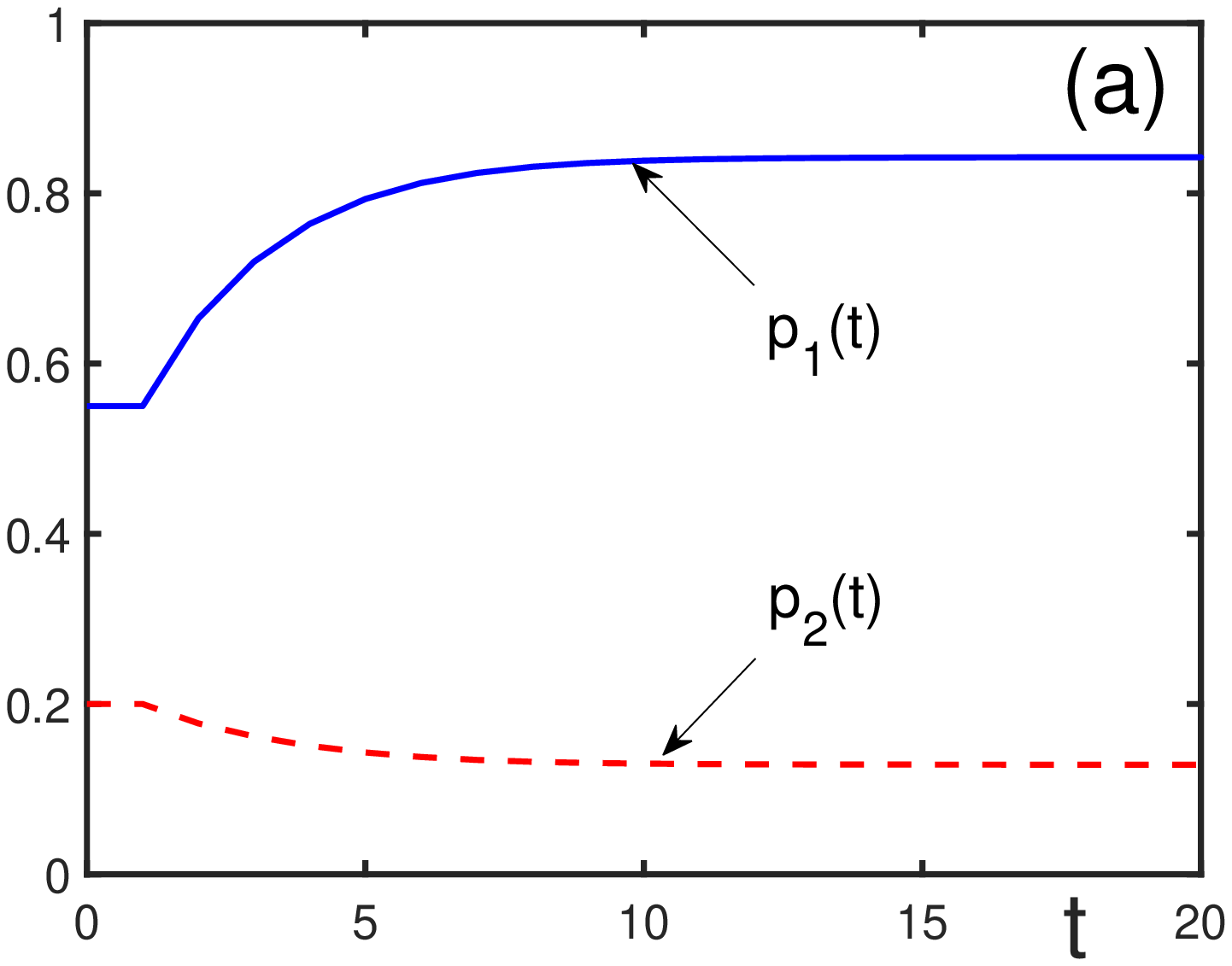} \hspace{1cm}
\includegraphics[width=7.5cm]{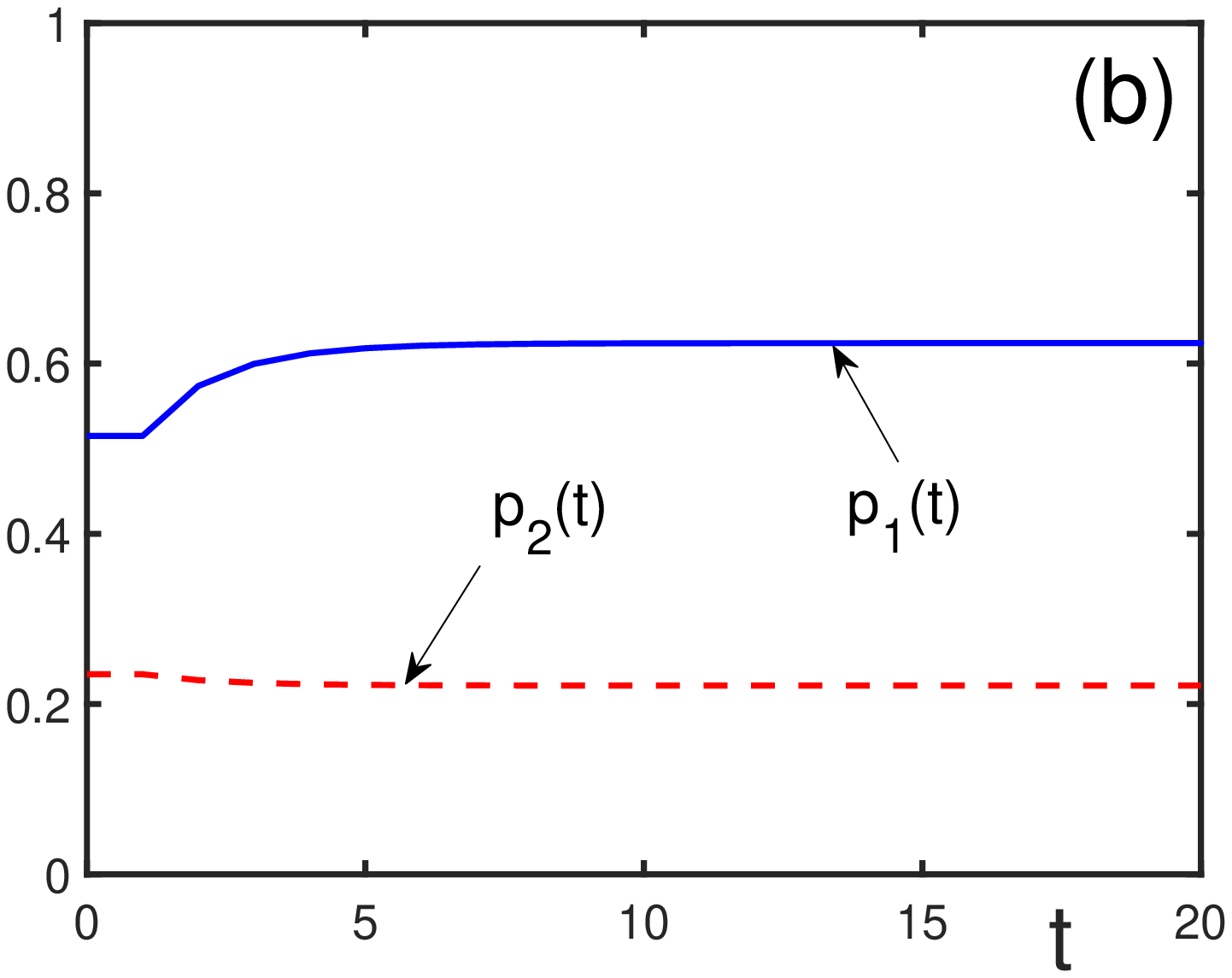}  } }
\vspace{12pt}
\centerline{
\hbox{ \includegraphics[width=7.5cm]{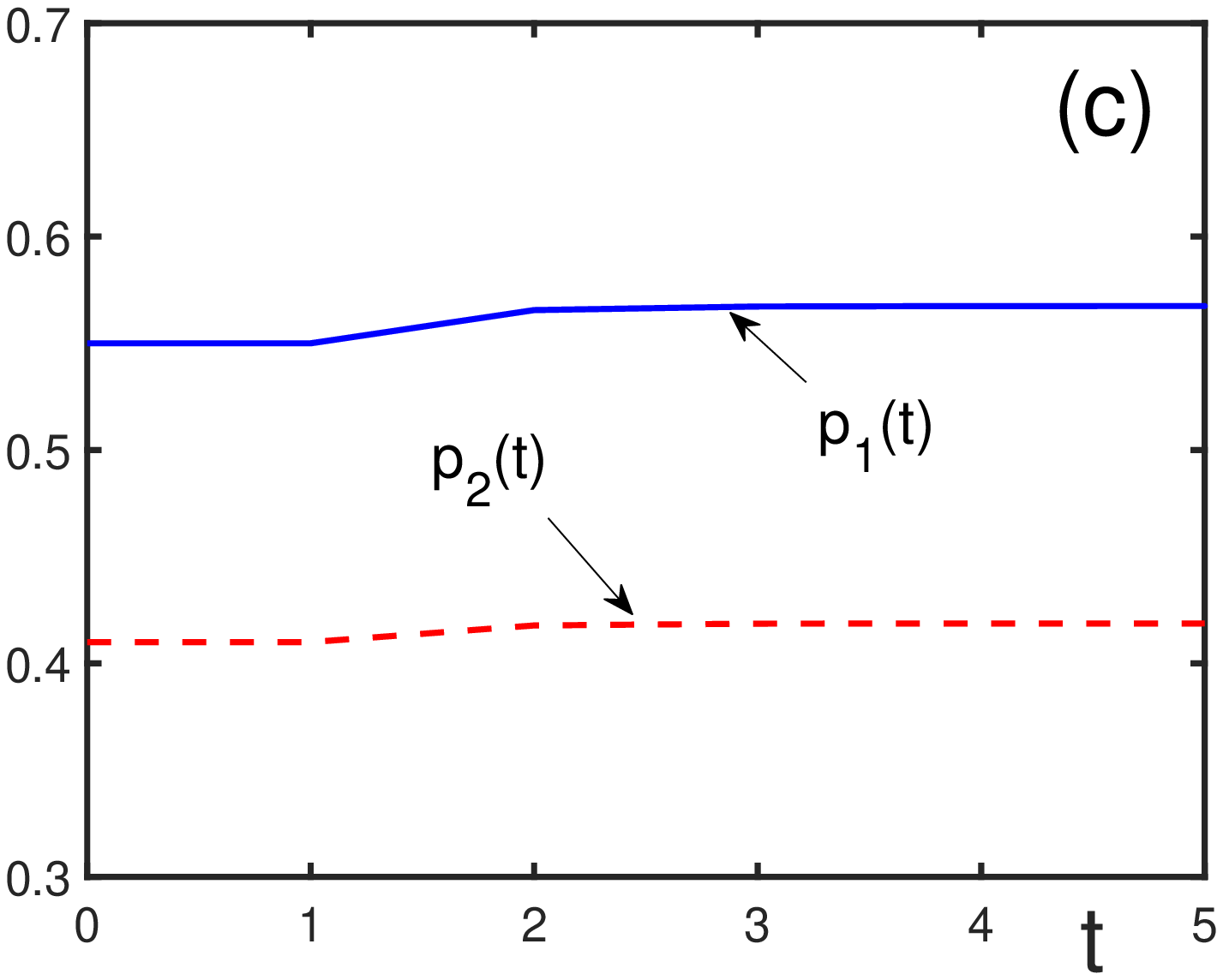} \hspace{1cm}
\includegraphics[width=7.5cm]{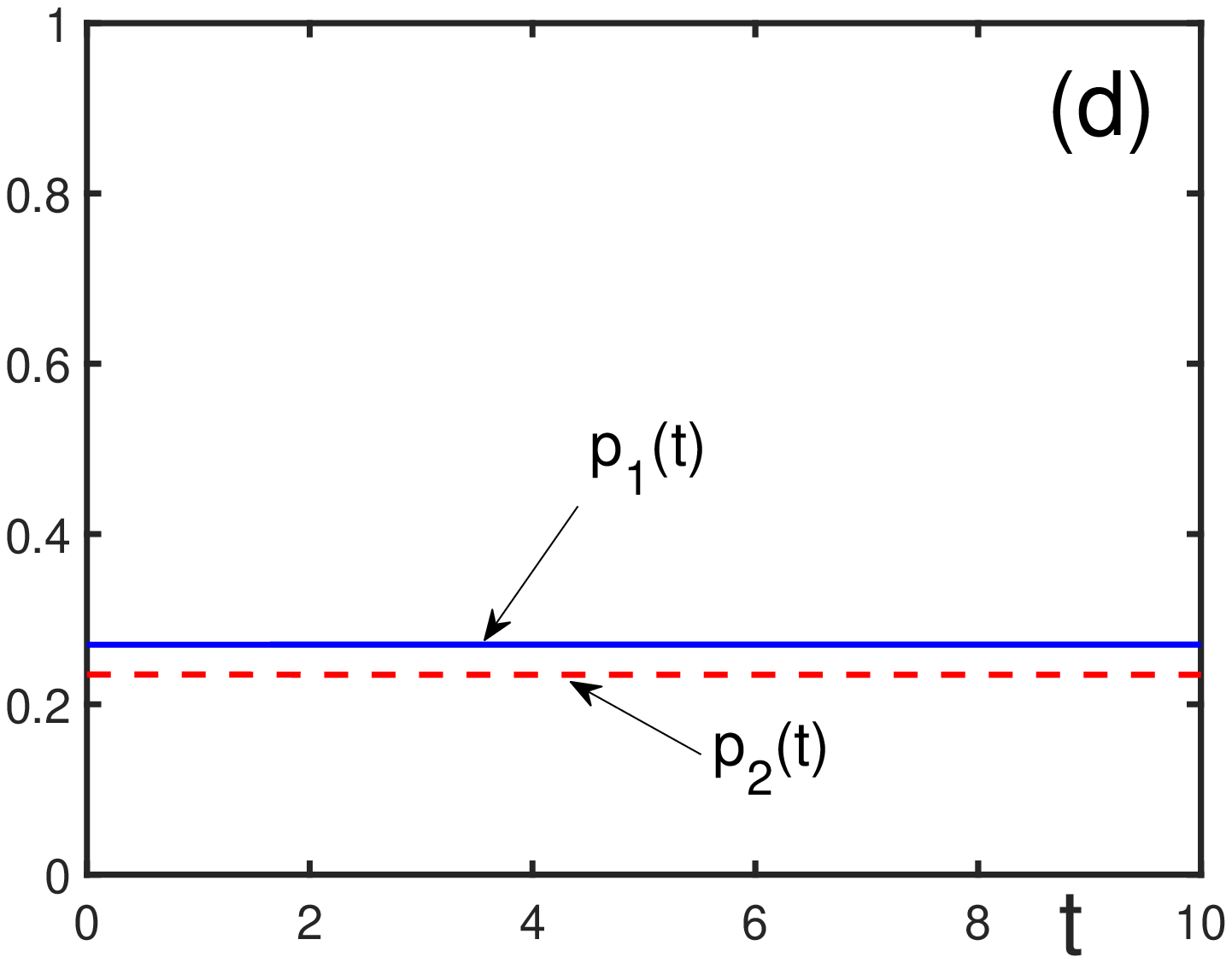} } }
\caption{
Short-term memory. Probabilities $p_1(t)$ (solid line) and $p_2(t)$ 
(dashed line). The initial conditions are $f_1=0.95$, $f_2=0.1$, $q_1=-0.4$, 
and $q_2=0.1$. 
(a) $\ep_1=\ep_2=0$. The initial probabilities are $p_1(0)=0.55$ and 
$p_2(0)=0.2$. The limiting probabilities are $p_1(\infty)=0.95$ and 
$p_2(\infty)=0.1$. 
(b) $\ep_1=\ep_2=0.1$. The initial probabilities are $p_1(0)=0.52$ and 
$p_2(0)=0.24$. The limiting probabilities are $p_1(\infty)=0.87$ and 
$p_2(\infty)=0.19$.
(c) $\ep_1=0$, $\ep_2=0.6$. The initial probabilities are $p_1(0)=0.52$ 
and $p_2(0)=0.41$. The limiting probabilities are $p_1(\infty)=0.87$ and 
$p_2(\infty)=0.61$.
(d) $\ep_1=0.8$, $\ep_2=0.1$. The initial probabilities are $p_1(0)=0.27$ 
and $p_2(0)=0.24$. The limiting probabilities are $p_1(\infty)=0.27$ and 
$p_2(\infty)=0.19$.
}
\label{fig:Fig.7}
\end{figure}

\newpage

\begin{figure}[ht]
\centerline{
\hbox{ \includegraphics[width=7.5cm]{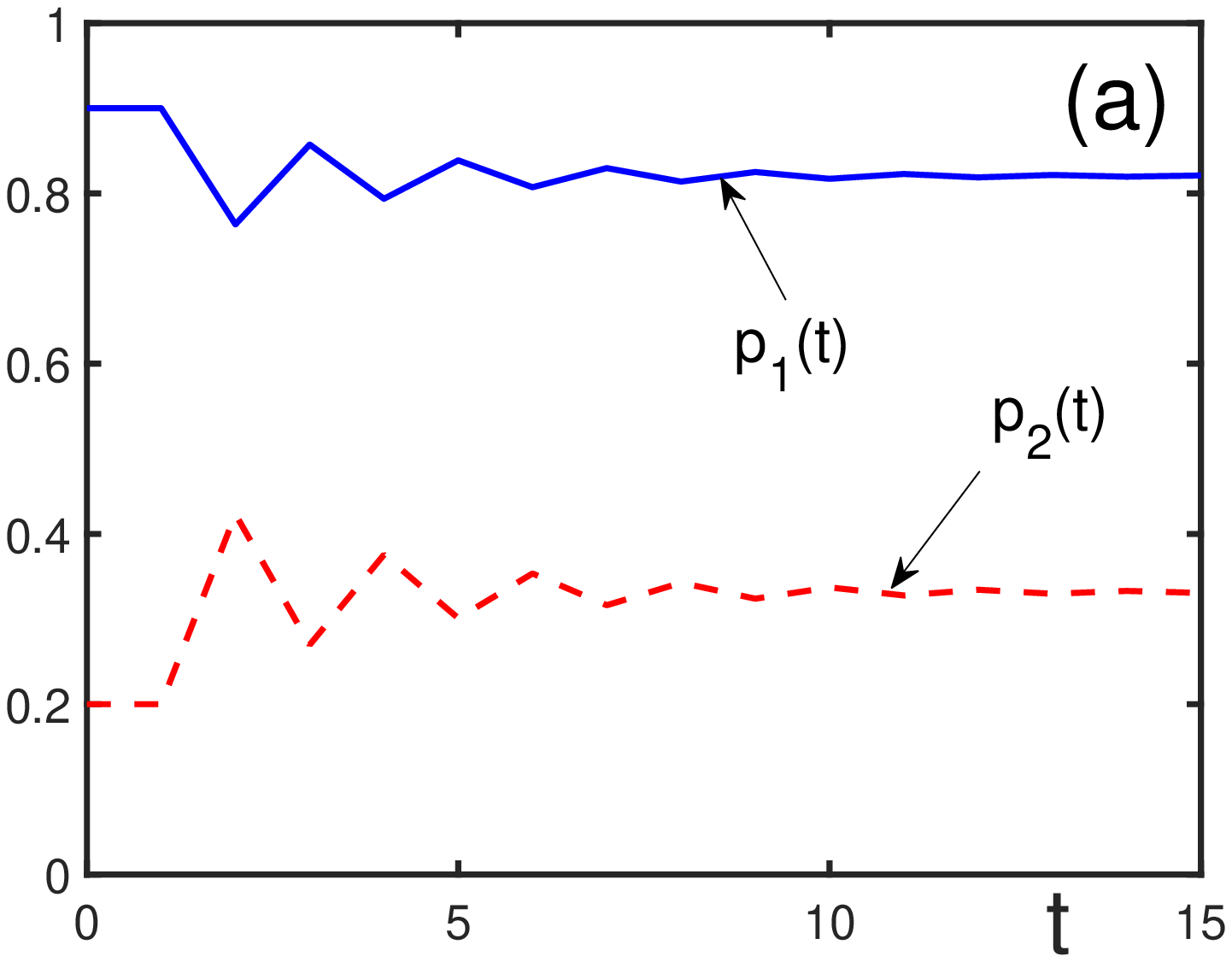} \hspace{1cm}
\includegraphics[width=7.5cm]{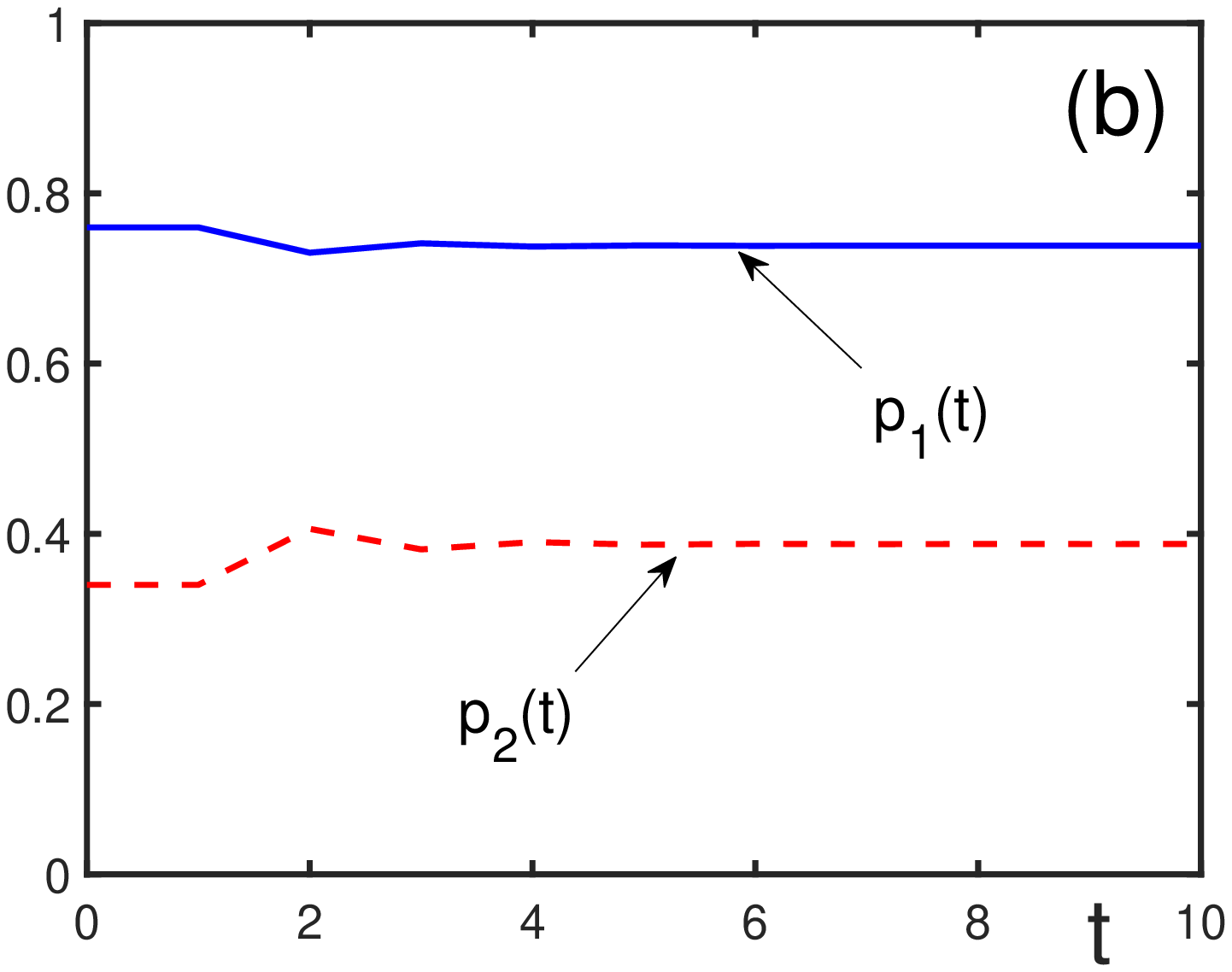}  } }
\vspace{12pt}
\centerline{
\hbox{ \includegraphics[width=7.5cm]{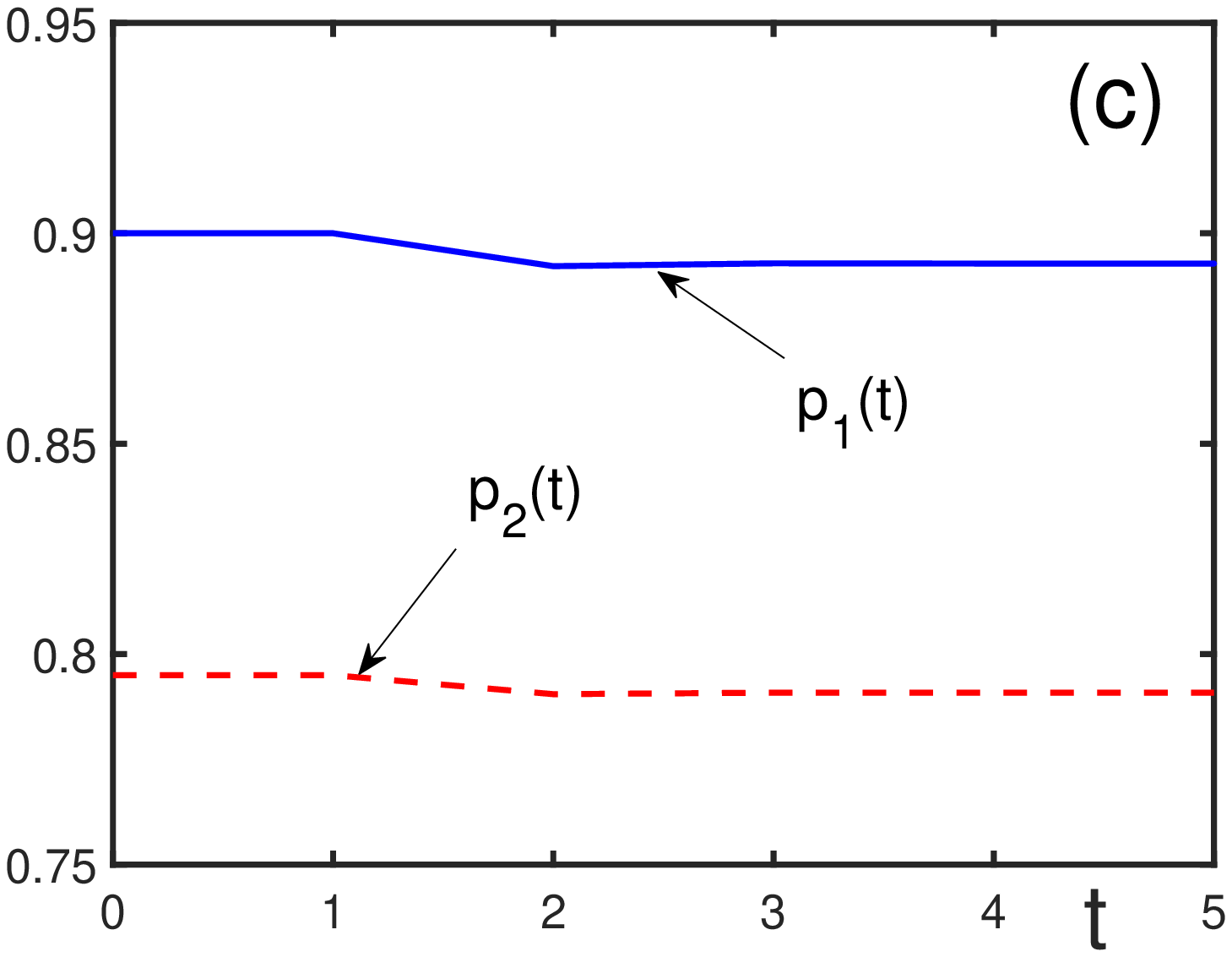} \hspace{1cm}
\includegraphics[width=7.5cm]{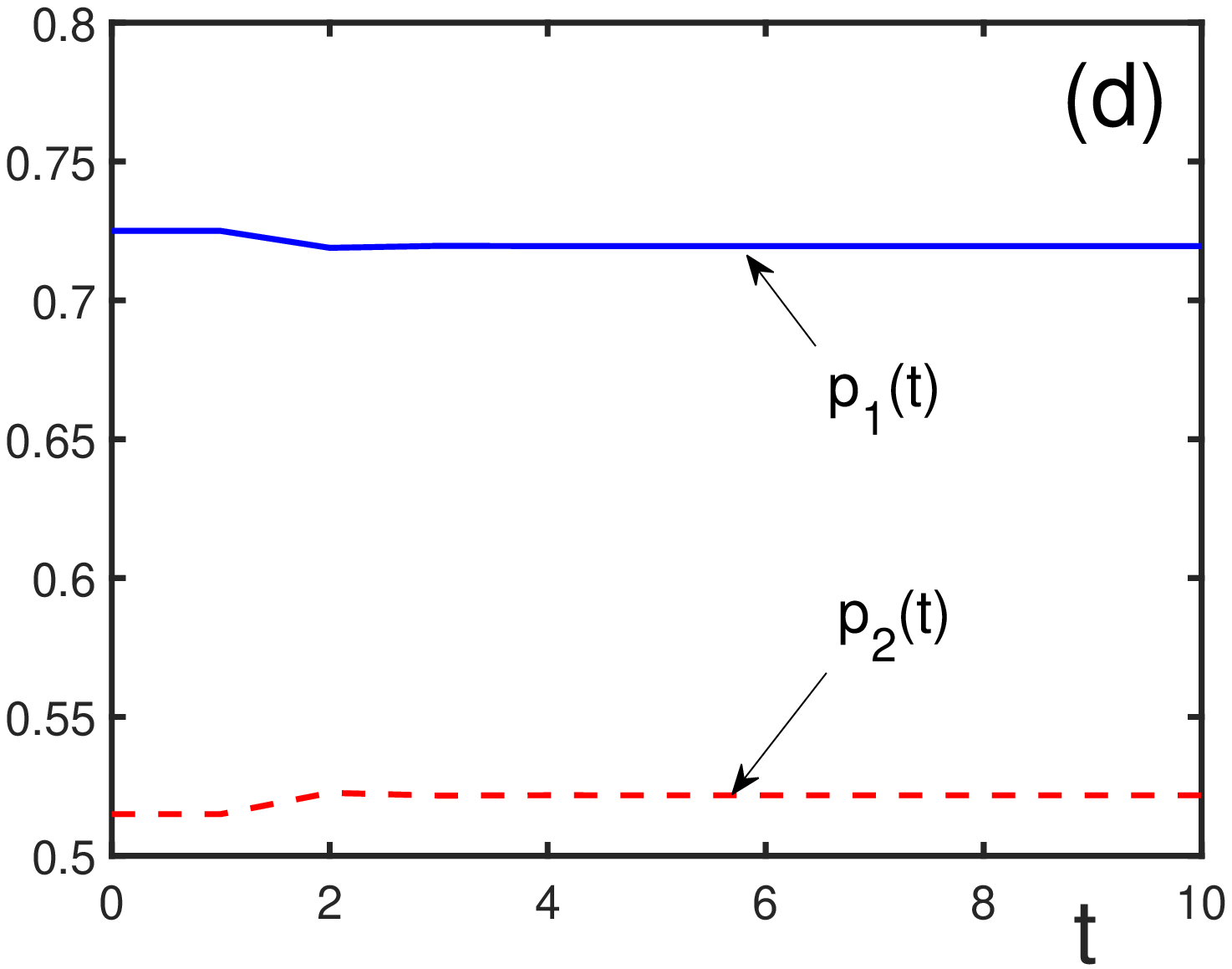} } }
\caption{
Short-term memory. Probabilities $p_1(t)$ (solid line) and $p_2(t)$ 
(dashed line). The initial conditions are $f_1=0.7$, $f_2=0.5$, 
$q_1=0.2$, and $q_2=-0.3$.
(a) $\ep_1=\ep_2=0$. The initial probabilities are $p_1(0)=0.9$ and 
$p_2(0)=0.2$. The limiting probabilities are $p_1(\infty)=0.82$ and 
$p_2(\infty)=0.33$. 
(b) $\ep_1=\ep_2=0.2$. The initial probabilities are $p_1(0)=0.76$ and 
$p_2(0)=0.34$. The limiting probabilities are $p_1(\infty)=0.74$ and 
$p_2(\infty)=0.39$.
(c) $\ep_1=0$, $\ep_2=0.85$. The initial probabilities are $p_1(0)=0.9$ 
and $p_2(0)=0.8$. The limiting probabilities are $p_1(\infty)=0.89$ and 
$p_2(\infty)=0.79$.
(d) $\ep_1=0.25$, $\ep_2=0.45$. The initial probabilities are $p_1(0)=0.73$ 
and $p_2(0)=0.52$. The limiting probabilities are $p_1(\infty)=0.65$ and 
$p_2(\infty)=0.59$.
}
\label{fig:Fig.8}
\end{figure}

\newpage

\begin{figure}[ht]
\centerline{
\hbox{ \includegraphics[width=7.5cm]{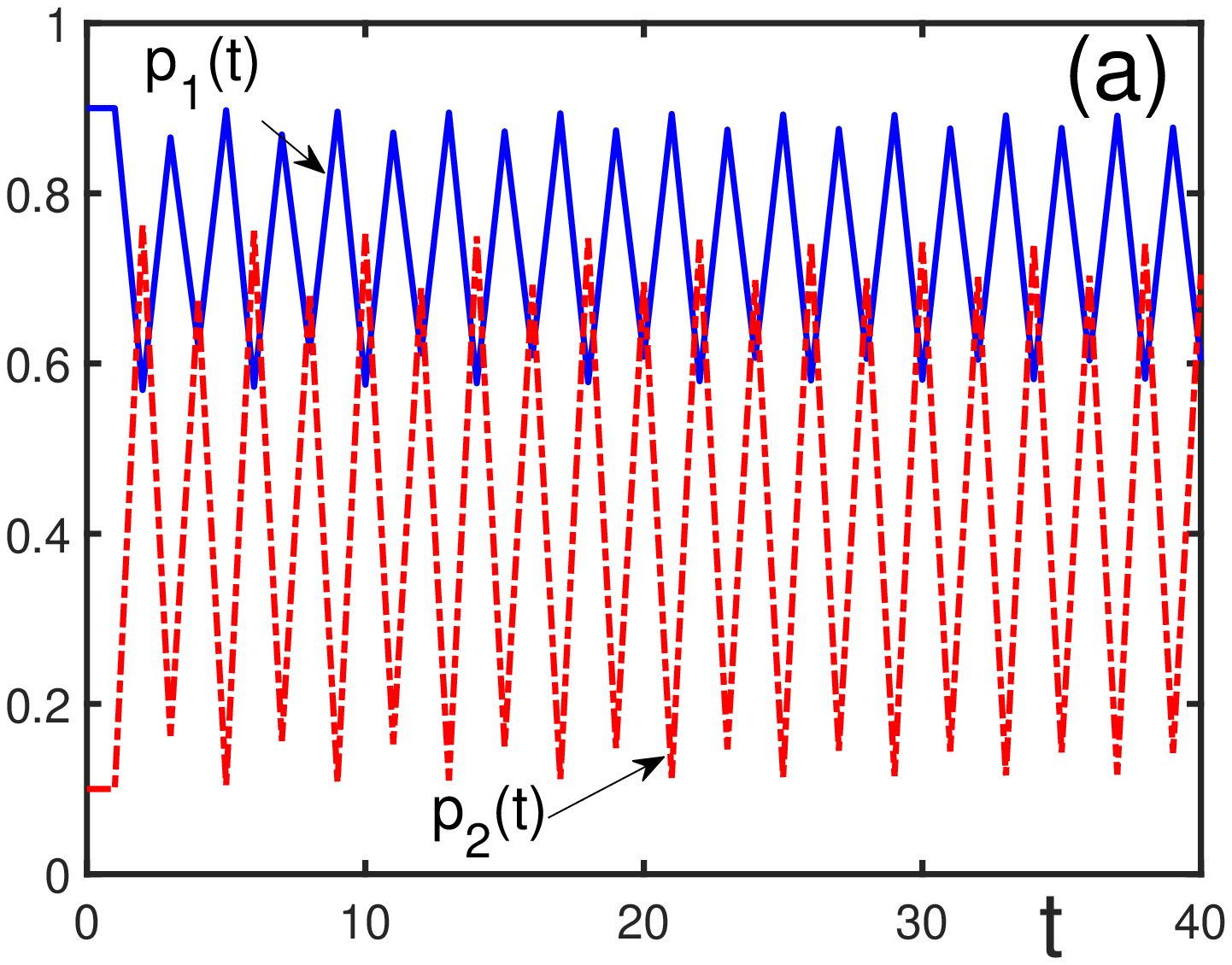} \hspace{1cm}
\includegraphics[width=7.5cm]{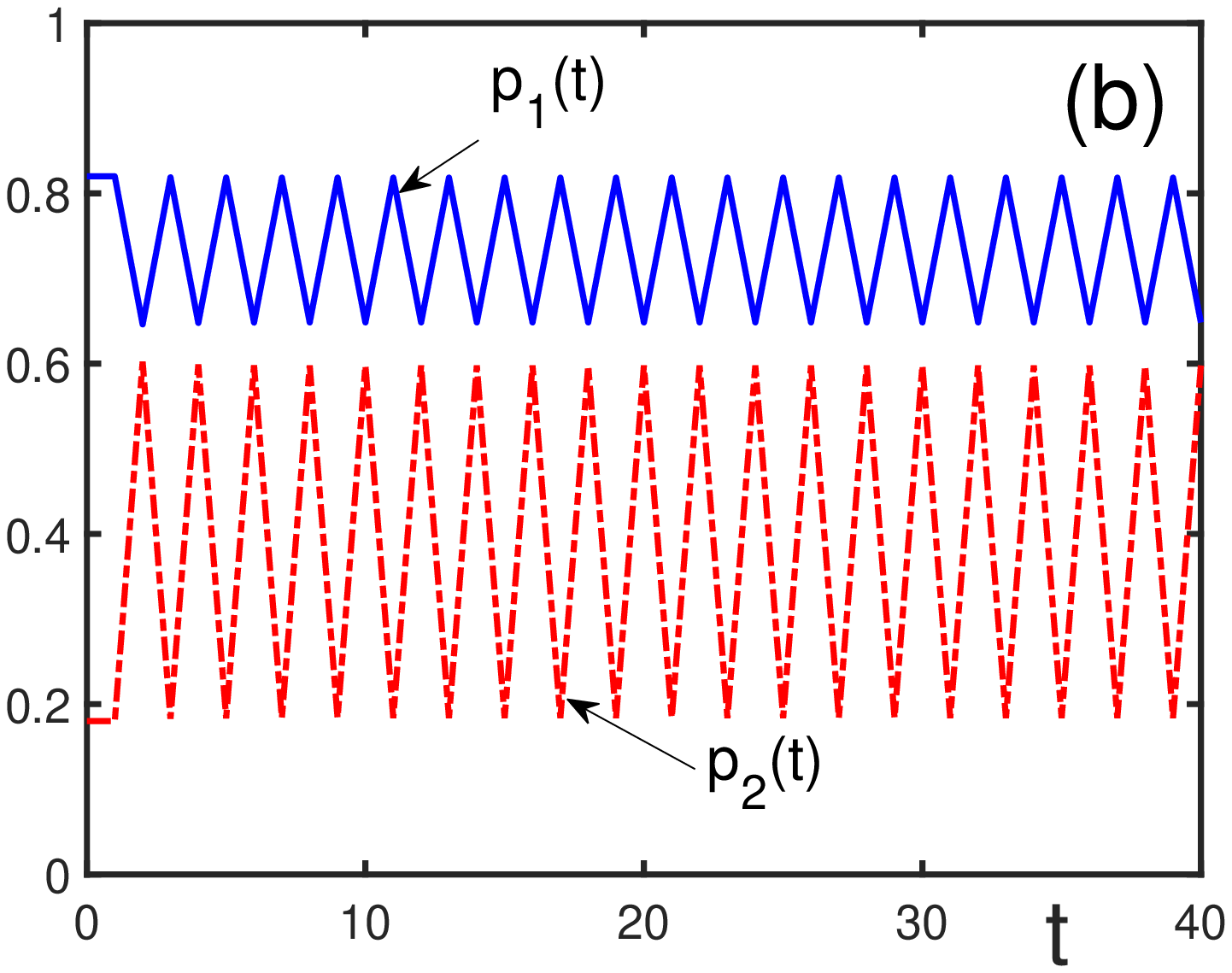}  } }
\vspace{12pt}
\centerline{
\hbox{ \includegraphics[width=7.5cm]{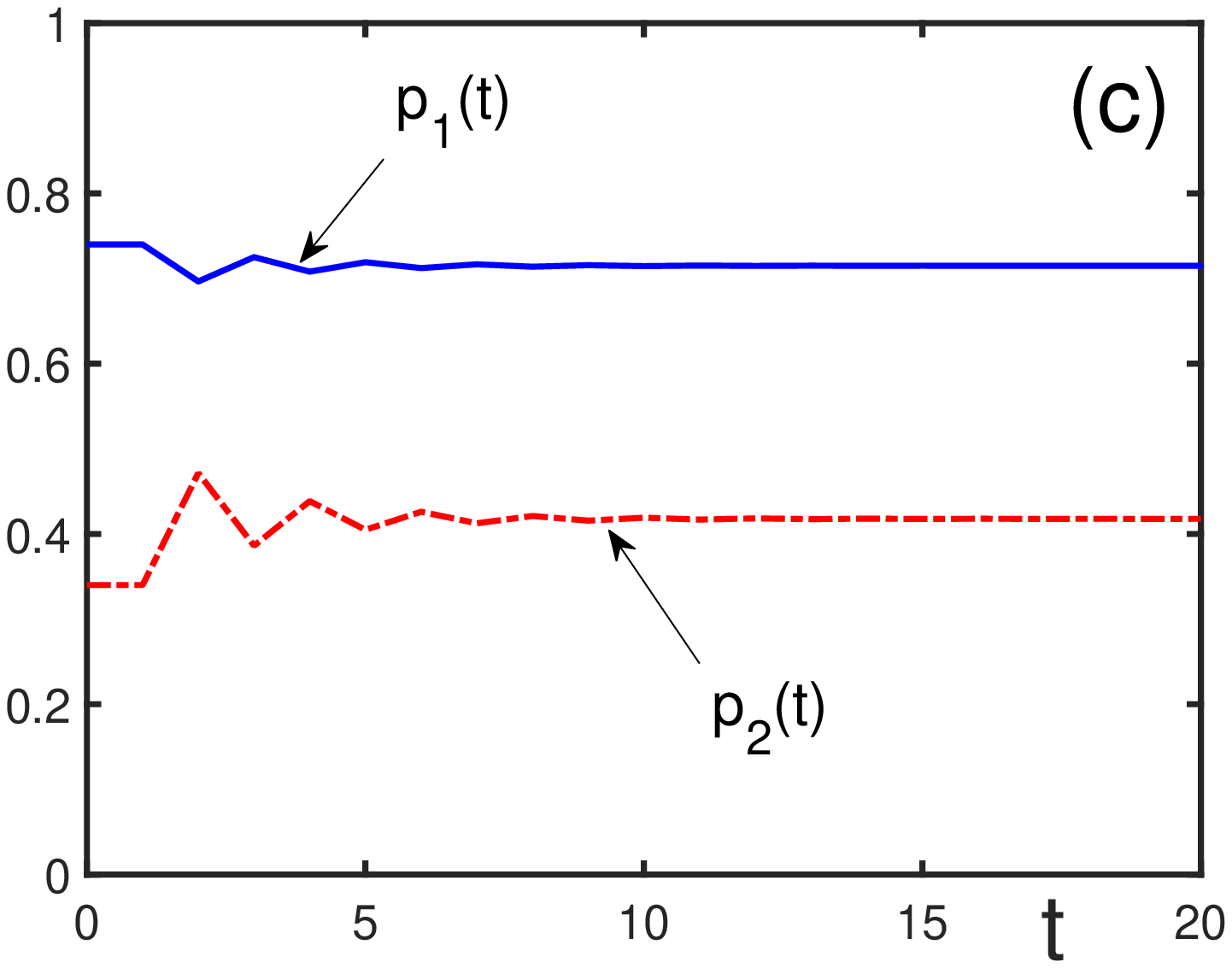} \hspace{1cm}
\includegraphics[width=7.5cm]{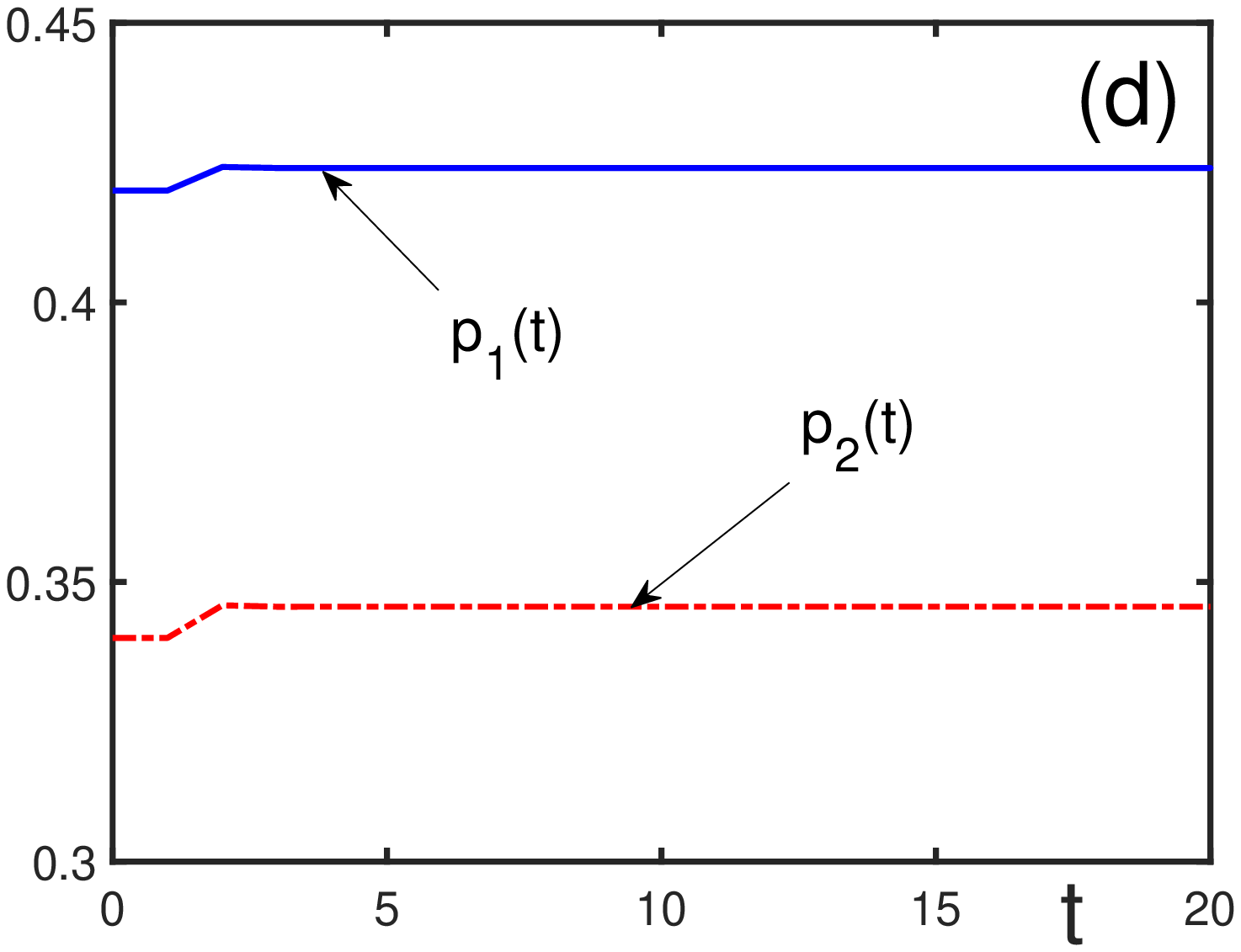} } }
\caption{
Short-term memory. Probabilities $p_1(t)$ (solid line) and $p_2(t)$ 
(dashed-dotted line). The initial conditions are $f_1=0.5$, $f_2=0.9$, 
$q_1=0.4$, and $q_2=-0.8$.
(a) Everlasting oscillations for $\ep_1=\ep_2=0$. The initial conditions 
are $p_1(0)=0.9$ and $p_2(0)=0.1$. 
(b) Everlasting oscillations with reduced amplitude for $\ep_1=\ep_2=0.1$. 
The initial conditions are $p_1(0)=0.82$ and $p_2(0)=0.18$. 
(c) Oscillatory tendency to limits for $\ep_1=0.2$, $\ep_2=0.3$. The initial 
conditions are $p_1(0)=0.74$ and $p_2(0)=0.34$. The limiting probabilities 
are $p_1(\infty)=0.72$ and $p_2(\infty)=0.42$.
(d) Monotonic tendency to limits for $\ep_1=0.6$, $\ep_2=0.3$. The initial 
conditions are $p_1(0)=0.42$ and $p_2(0)=0.34$. The limiting probabilities 
are $p_1(\infty)=0.424$ and $p_2(\infty)=0.346$.
}
\label{fig:Fig.9}
\end{figure}

\begin{figure}[ht]
\centerline{
\includegraphics[width=10cm]{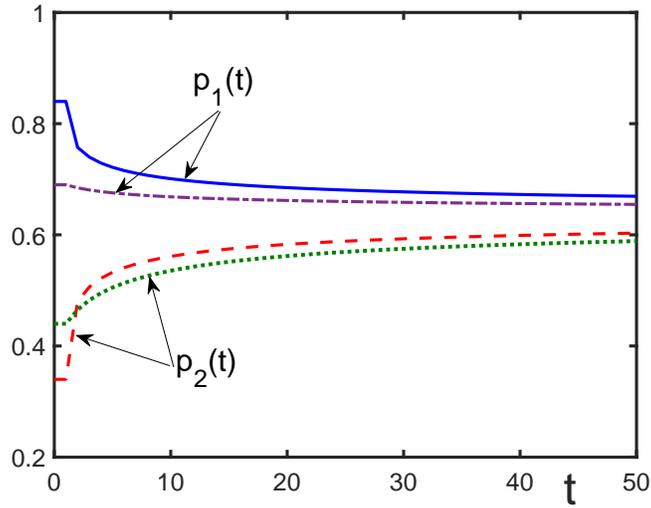}  }
\caption{Disjunction effect attenuation as a function of time and its reduction 
caused by common information. The initial conditions are $f_1=f_2=f=0.64$, 
$q_1=0.2$, and $q_2=-0.3$. When $\ep_1=\ep_2=0$, functions $p_1(t)$ (solid line) 
and $p_2(t)$ (dashed line), under initial conditions $p_1(0)=0.84$ and $p_2(0)=0.34$, 
monotonically tend, with increasing time $t\ra\infty$, to their common limit $f=0.64$. 
In the presence of the common information, with $\ep_1=0.3$ and $\ep_2=0.2$, the 
functions $p_1(t)$ (dashed-dotted line) and $p_2(t)$ (dotted line), under the initial 
conditions $p_1(0)=0.69$ and $p_2(0)=0.44$, monotonically tend to the same limit 
$f = 0.64$ as $t\ra\infty$. The common information reduces the disjunction effect 
from the beginning of the process.
}
\label{fig:Fig.10}
\end{figure}

\end{document}